\documentclass[fleqn,usenatbib]{mnras}

\usepackage{mathtools} 

\usepackage[T1]{fontenc}

\DeclareRobustCommand{\VAN}[3]{#2}
\let\VANthebibliography\thebibliography
\def\thebibliography{\DeclareRobustCommand{\VAN}[3]{##3}\VANthebibliography}

\usepackage{graphicx}	
\usepackage{amsmath}	
\usepackage{amssymb}	

\usepackage{multirow}
\usepackage{longtable}

\usepackage{txfonts}
\usepackage{hyperref} 
\usepackage{float} 
\usepackage{natbib}
\usepackage{numprint}
\usepackage{caption}

\newcommand\Pristine{\textit{Pristine}}
\newcommand\Espadons{ESPaDOnS}
\newcommand\Tstrut{\rule{0pt}{2.5ex}}         
\def\kpc{{\rm\,kpc}}

\title[The Pristine survey -- XV]{The Pristine survey -- XV: A CFHT ESPaDOnS view on the Milky Way halo and disc populations}

\author[Lucchesi et al.]{R. Lucchesi $^{1,2}$,
\thanks{E-mail: romain.lucchesi@epfl.ch}
C. Lardo$^{1,3}$,
P. Jablonka$^{1,4}$,
F. Sestito$^{5}$,
L. Mashonkina$^{6}$,
A. Arentsen$^{7}$,
W. Suter$^{1}$,
K. Venn$^{5}$,
\newauthor
N. Martin$^{7,8}$,
E. Starkenburg$^{9}$,
D. Aguado$^{10}$,
V. Hill$^{11}$,
G. Kordopatis$^{11}$,
J. F. Navarro$^{5}$,
\newauthor
J. I. Gonz\'alez Hern\'andez$^{12,13}$,
K. Malhan$^{14}$,
and Z. Yuan$^{7}$
\\
$^{1}$Institut de Physique, Laboratoire d'astrophysique, École Polytechnique Fédérale de Lausanne (EPFL), Observatoire, 1290 Versoix, Switzerland \\
$^{2}$ European Southern Observatory, Karl-Schwarzschild-str. 2, 85748 Garching bei München, Germany\\
$^{3}$ Dipartimento di Fisica e Astronomia, Universit\`a degli Studi di Bologna, Via Gobetti 93/2, I-40129 Bologna, Italy\\
$^{4}$ GEPI, Observatoire de Paris, Université PSL, CNRS, 5 Place Jules Janssen, 92190 Meudon, France\\
$^{5}$ Department of Physics and Astronomy, University of Victoria, PO Box 3055, STN CSC, Victoria, BC V8W 3P6, Canada\\
$^{6}$ Institute of Astronomy of the Russian Academy of Sciences, Pyatnitskaya st. 48, 119017, Moscow, Russia \\
$^{7}$ Universit\'e de Strasbourg, CNRS, Observatoire astronomique de Strasbourg, UMR 7550, F-67000 Strasbourg, France \\
$^{8}$ Max-Planck-Institut f\"ur Astronomie, K\"onigstuhl 17, D-69117, Heidelberg, Germany\\
$^{9}$ Kapteyn Astronomical Institute, University of Groningen, Postbus 800, 9700 AV, Groningen, the Netherlands \\
$^{10}$ Institute of Astronomy, University of Cambridge, Madingley Road, Cambridge CB3 0HA, UK \\
$^{11}$ Universit\'e C\^ote d’Azur, Observatoire de la C\^ote d’Azur, CNRS, Laboratoire Lagrange, Nice, France \\
$^{12}$ Instituto de Astrofísica de Canarias, Vía Lactea, E-38205 La Laguna, Tenerife, Spain\\
$^{13}$ Departamento de Astrofísica, Universidad de La Laguna, E-38206 La Laguna, Tenerife, Spain\\
$^{14}$ The  Oskar  Klein  Centre  for  Cosmoparticle  Physics,  Department  of Physics,  Stockholm  University,  AlbaNova,  10691  Stockholm,  Sweden\\
}

\date{Accepted XXX. Received YYY; in original form ZZZ}

\pubyear{2021}

\begin{document}
\label{firstpage}
\pagerange{\pageref{firstpage}--\pageref{lastpage}}
\maketitle

\begin{abstract}
We present a one-dimensional, local thermodynamic equilibrium (1D-LTE) homogeneous analysis of 132 stars observed at
high-resolution with ESPaDOnS. This represents the largest sample observed at high
resolution (R$\sim$40,000) from the \Pristine\ survey. This sample is based on
the first version of the \Pristine\ catalog and covers the full range of
metallicities from [Fe/H]$\sim -3$ to $\sim +0.25$, with nearly half of our
sample (58 stars) composed of very metal-poor stars ([Fe/H]~$\le$~$-$2). This wide range of metallicities provides the opportunity of a new detailed study of the Milky Way 
stellar population. Because it includes both dwarf and giant stars, it also
enables the analysis of any potential bias induced by the \Pristine\ selection
process. Based on Gaia EDR3, the orbital analysis of this \Pristine\--\Espadons\ sample shows that it is composed of 65 halo stars and 67 disc stars. After a general assessment of the sample chemical properties with the
$\alpha$-elements Mg and Ca, we focus on the abundance of carbon and the
neutron capture elements Ba and Sr.  While most of our very metal-poor subsample is
carbon normal, we also find that 14 stars out of the 38 stars with [Fe/H] $\leq$ --2 and measured carbon abundances turn out to be carbon enhanced metal-poor (CEMP)
stars. We show that these CEMP stars are nearly exclusively (i.e. 12 stars out of 14) in the regime of low
luminosity, unevolved, dwarf stars, which we interpret as the consequence of bias of the
\Pristine\ filter against C-rich giants. Among the very metal-poor (VMP) stars, we identify 2 CEMP stars with no enhancement
in neutron-capture process elements (CEMP-no) and another one enriched in
s-process element (CEMP-s). Finally, one VMP star is found with a very low [Sr/Fe]
abundance ratio for its metallicity, as expected if it had been accreted from an
ultra-faint dwarf galaxy.
\end{abstract}




\section{Introduction}

The most metal-poor stars in the Galaxy and its close satellites are witnesses of the early
stages of star formation in the Universe \citep[e.g.,][and references therein]{Pagel1997,Bromm2004,Heger2010,Frebel2015}. Their formation follows the explosions
of a few population III (Pop III) supernovae only.  Hence, their elemental abundances reflect the physical conditions and the nucleosynthesis of the
primordial chemical evolution \citep[e.g.,][]{Beers1992,Cayrel2004,Beers2005,Keller2007,Christlieb2008,Caffau2013,Yong2013,Roederer2014,Jacobson2015,Frebel2018}.

The detection of very metal-poor (VMP; [Fe/H]$\leq$ --2.0), extremely metal-poor (EMP; [Fe/H]$\leq$ --3.0), and ultra metal-poor (UMP; [Fe/H]$\leq$ --4.0) stars is a challenging task that requires surveying large volumes of the sky\footnote{In a high Galactic latitude field towards the anticentre direction 
only $\sim$1/2000 stars in the magnitude range between 14 $<$ V $<$ 18 are expected to have metallicity less than [Fe/H] $\leq$ --3; \citep{Youakim2017}.}. Indeed, many observational efforts have been devoted to the search and identification of such key stellar population --e.g., the HK objective-prism survey \citep{Beers1992}; the RAdial Velocity Experiment survey \citep{Steinmetz2006}, the Hamburg/ESO objective-prism survey \citep{Christlieb2008}; the Sloan Extension for Galactic Understanding and Exploration Survey \citep{ Yanny2009}; the LAMOST Experiment for Galactic Understanding and Exploration \citep{Deng2012}; the Apache Point Observatory Galactic Evolution Experiment \citep{Majewski2016}; the SkyMapper Southern Sky Survey \citep{Keller2007}.

Along these lines, \Pristine\ is a photometric survey designed to efficiently pre-select very metal-poor star candidates  \citep{Starkenburg2017,Youakim2017,Aguado2019}. It
takes advantage of a narrow band filter centered on
the Ca H\&K spectral lines and of the large field of view of MegaCam at the
Canada-France-Hawaii Telescope (CFHT). Briefly, the  \Pristine\ selection method combines information
from the metallicity-sensitive Ca H\&K filter with broad band photometry from
large-field, multi-band photometric surveys; e.g. the Sloan Digital Survey
(SDSS;\citealp{York2000,Eisenstein2011,Blanton2017}) and Gaia DR2
\citep{GaiaCollaboration2016,GaiaCollaboration2018}.

Spectroscopic follow-up of EMP star candidates is on-going. These observations are conducted
both at medium and high- spectroscopic resolution
\citep{Caffau2017,Starkenburg2018,Bonifacio2019,Venn2020,Youakim2017,Aguado2019,Caffau2020,Kielty2021}.
The detailed chemical analysis of individual stars allows us to address three
main threads of open issues related  to stellar evolution  and
galaxy formation: {\em (i)} the nature and properties of the first stars, {\em (ii)} how and
when the different components of the structure of the Milky Way assemble, and, finally,
{\em (iii)} in the hierarchical galaxy formation paradigm, the mass and the degree of
chemical evolution of the dwarf galaxy building blocks.

The existence of $\alpha$-poor stars ([(Mg+Ca)/Fe]~$\le$~0) in some of the
\Pristine\ subsample was reported by \citet{Caffau2020} in their ESO/FORS2 medium resolution
spectroscopic follow-up. These stars were found in a metallicity regime that is
more metal-poor ([Fe/H]$< -1.2$) than the sample of \citet{Nissen2010}, which is
interpreted as the result of quiescent star formation forming the Milky Way
thin disc \citep{Khoperskov2021}. Such metal deficient and $\alpha$-poor stars were
also identified in other studies \citep[e.g.,][]{Ivans2003,Cohen2013}. Their origin is
still unclear and could be heterogeneous, such as a formation from pockets
of interstellar medium enhanced in Type Ia supernovae (SNeIa) products, biased initial mass function (IMF) sampling, or accretion of merging dwarf systems \citep{Sakari2019, Xing2019}. The present \Pristine\ sample is large enough to shed some light on the fraction of $\alpha$-poor stars in the Milky Way halo.

Carbon-enhanced metal-poor (CEMP; i.e. stars having [C/Fe]$\geq$1.0 according to the definition given in \citealp{Beers2005}) represent an increased fraction of the halo component of the Milky Way with decreasing metallicity \citep[see][for a complete discussion]{Frebel2015}. For three metallicity bins in the range between --4.5 $<$[Fe/H] $<$--3.0, \citet{Yong2013} found that the fraction of CEMP stars was increasing from 0.22 to 0.32, and 0.33 with decreasing metallicity, up to 1.00 for  
[Fe/H] $<$ --4.5. Higher fractions were derived by \citet{Placco2014} when correcting for internal mixing effects depleting surface carbon abundance with stellar evolution --e.g., they derive a fraction of 
0.43, 0.60, 0.70, and 1.00, respectively, for the same metallicity bins defined in \citet{Yong2013}.
The exact origin of this increase in CEMP stars towards lower metallicities has yet to be
unveiled, however this result suggest that significant amounts of carbon were
produced in the early Universe.  This production could be a necessary condition
for the transition from massive pop III to low-mass stars \citep{Bromm2003}. 
However, the discoveries of SDSS J102915+172927 at
[Fe/H]=$-4.73$ \citep{Caffau2011} and Pristine 221.8781$+$9.7844 at $-4.66$
\citep{Starkenburg2018}, two stars with a significant low enrichment of carbon, nitrogen and oxygen suggest that there must have been more than one formation channel of low-mass stars in the early Universe. Unexpectedly \Pristine\ has found two contradictory results on this matter. While \citet{Aguado2019} reported a normal fraction of CEMP stars, \citet{Caffau2020} found a fraction of carbon-enhanced MP stars much lower than those provided by \citet{Placco2014}, thereby suggesting some sensitivity of the \Pristine\ filter to carbon abundance. This issue can be addressed in the present study.

In the following, we present the analysis of the 132 bright (V~$<$~15.5) metal-poor candidates from the original 1000 deg$^2$ of the \Pristine\ survey, calibrated using the original SDSS gri photometry and observed at the CFHT with the high-resolution spectrograph \Espadons. Out of this full sample \cite{Venn2020} presented the detailed abundances of 10 elements (Na, Mg, Ca, Sc, Ti, Cr, Fe, Ni, Y, and Ba) for the 28 very-metal poor stars identified at the time, as well as the analysis of their orbital properties.
Because the full sample comes from the first stages of the
\Pristine\ calibration, the confirmation of very metal-deficient stars does not
reach a success rate as high as in the later stages. Nevertheless, near
half of the present sample (58 stars) is composed of very metal-poor stars ([Fe/H]~$\le$~$-$2). The more metal-rich stars offer us the opportunity of a new
and detailed study of the Milky Way halo stellar population.

This paper is organised as follows: in \S~ \ref{observations} we first discuss
observations and data reduction. The abundance analysis is presented in
\S~\ref{analysis} and the discussion of the abundances of C, Mg, Ca, Sr, and Ba
takes place in \S~\ref{Results}. In Section \S~\ref{orbits}, we look into the orbits of our sample stars. Finally, Section \ref{conclusion} summarizes our results
and conclusions.

\section{Observational material}\label{observations}

\subsection{Source catalog and sample selection}\label{sample}

The targets were selected from the \Pristine\ 
diagnostics originally presented in \citet{Starkenburg2017}.  Stars
were selected upon their probability to be very metal-poor, in the bright (V
$\lesssim$ 15.5) regime of the original $\sim$1000~deg$^2$ footprint of 
\Pristine.  

\begin{figure}
    \centering
    \includegraphics[width=1.0\columnwidth]{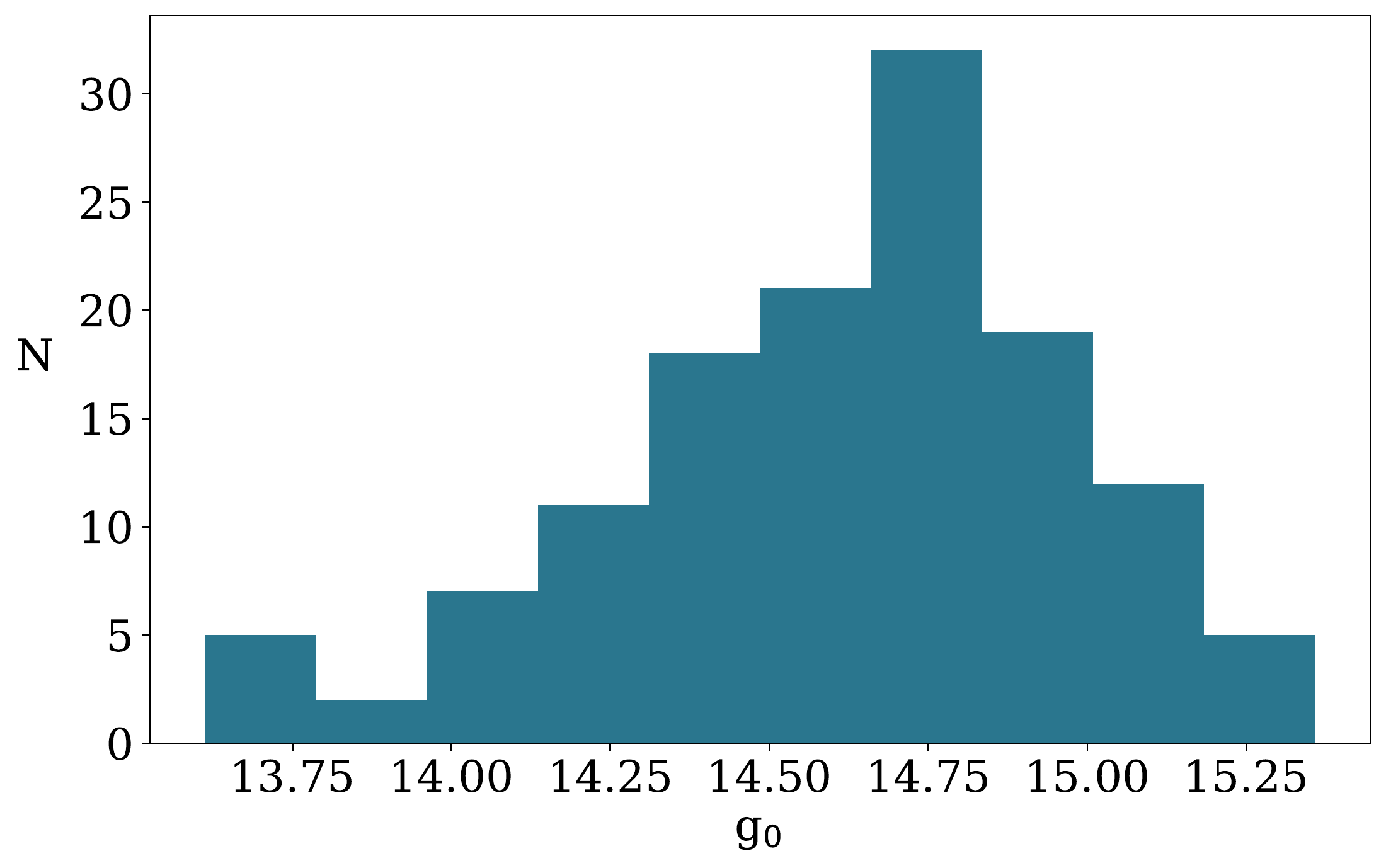}
    \caption{Histogram of the SDSS dereddened g magnitudes ($g_0$) of the 132 stars analysed in this work.}
    \label{Fig:magnitues}
\end{figure}

The final sample consists in 132 stars and includes~:
\begin{enumerate}
    \item   112 stars which were introduced in \citet{Venn2020}.  [Fe/H] was spectroscopically derived only for a subset of those (86) using the equivalent widths of six selected iron lines (4 \ion{Fe}{I} lines and 2 \ion{Fe}{II} lines). In the following, we adopt the same nomenclature as \citet{Venn2020} and refer to this metallicity estimate as the quick six (Q6) one. 
    Only stars with metallicity estimate [Fe/H]~$\leq$~$-$2.5 were then retained for further spectroscopic chemical analysis (28 stars). 
    
    \item 20 new stars from the CFHT \Espadons\ programs 16BF10, 17AF09, and 17BF18.
\end{enumerate}

Figure~\ref{Fig:magnitues} shows the distribution of the full sample in the
dereddened $g_{0}$ SDSS magnitude, with E(B--V) values taken from the galactic
reddening maps of \citet{Schlegel1998}. Stellar magnitudes range from
g$_0$~=~13.6 to 15.3 with a peak around $g_{0}$~=~14.7. Table~\ref{tab:parameters}
provides the coordinates, dereddened {\em $g_{0}$} and {\em $i_{0}$} SDSS
magnitudes along with the corresponding E(B--V) values, and de-reddened \Pristine\ CaH\&K
magnitudes for all stars analysed in this study.

The full dataset is analysed in a  homogeneous way. The data reduction has
been improved compared to \citet{Venn2020}; all stellar atmospheric parameters
are now spectroscopically derived and we provide detailed chemical abundances
for C, Mg, Ca, Fe, Sr, and Ba.

\begin{figure*}[!ht]
    \centering
    \includegraphics[width=0.9\textwidth]{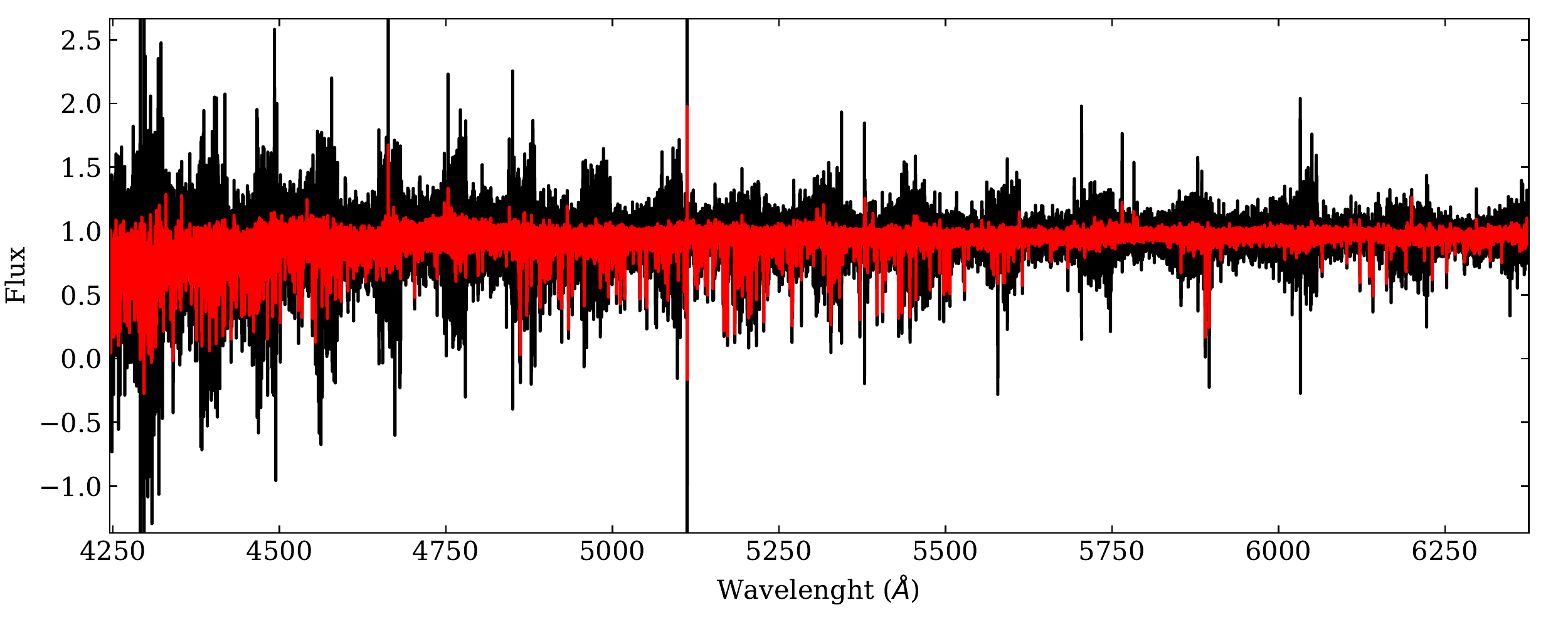}\vspace{-0.6cm}
    \includegraphics[width=0.9\textwidth]{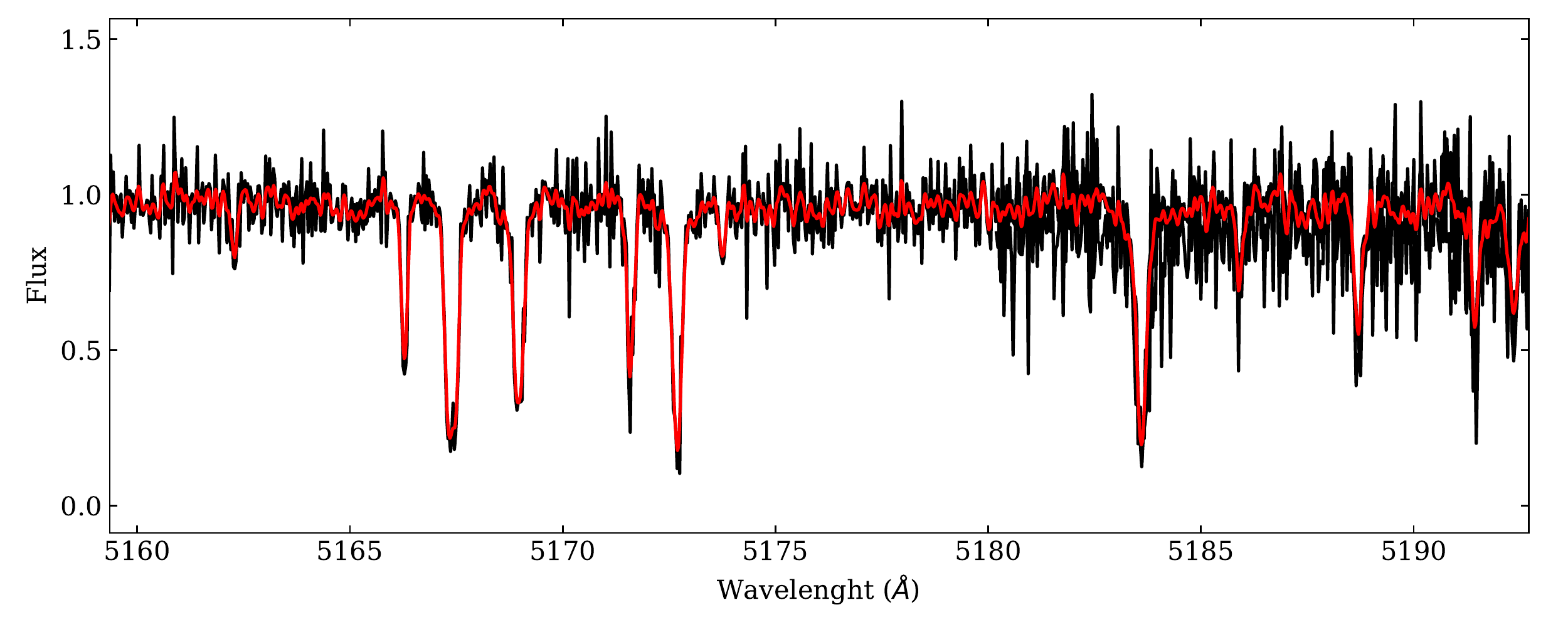}
    \caption{The top panel shows the spectrum of Pr\_236.1077+10.5311 in the wavelength range between 4250 and 6300~\AA. The bottom panel shows the spectrum of the same star over a limited wavelength range around the magnesium triplet spectral features.
    The raw ESPaDOnS spectrum from the Libre-ESpRIT pipeline is plotted in black. The red line shows the spectrum degraded to a resolution of R=40\ 000 and with optimal order merging applied to data (see text).}
    \label{fig:orders}
\end{figure*}

\subsection{Observations and data reduction}

The observations were performed during the periods 2016A and 2018B with the
high-resolution spectrograph \Espadons\ at the CFHT \citep{Donati2006}.  To enable a good sky subtraction, ESPaDOnS was used in the ``star+sky'' mode, providing a
high-resolution (R=68,000) spectrum from 4\ 000 to 10\ 000~\AA.  Exposure times
range from $\sim$12 minutes for the brightest targets to $\sim$120 minutes for
the faintest ones.

The data reduction was performed with the dedicated pipeline Libre-ESpRIT \footnote{\url{https://www.cfht.hawaii.edu/Instruments/Spectroscopy/Espadons/Espadons\_esprit.html}}. This
includes bias subtraction, flat fielding, wavelength calibration, and spectral
extraction.  ESPaDOnS records 40 orders and each of them is curved. Libre-ESpRIT
proceeds in two steps.  First, the pipeline performs a geometrical analysis from
a sequence of calibration exposures. The position and shape of each order is
derived from a mean flat field image. The details of the wavelength-to-pixel
relationship along and across each spectral order is measured from a thorium
lamp exposure. Second, Libre-ESpRIT performs an optimal extraction of each
object spectrum, using the geometrical information found in the previous step.
It computes the intensity spectra with error bars, and applies corrections to
compensate for Earth's motion.

\setlength{\tabcolsep}{4pt}
\begin{table*}

\caption{Right Ascension, Declination, dereddened SDDS g and i magnitudes, averaged radial velocities, uncertainties on radial velocities, spectroscopic T$\mathrm{_{eff}}$ log$g$ micro-turbulence velocities and metallicities for the 132 stars analyzed. Stars marked with * were rejected by the new \Pristine\ photometric selection. Stars marked with $\dagger$ are three fast rotators and discarded from the analysis.}
\label{tab:parameters}
\begin{tabular}{lccccccrccccl}
\hline\hline\Tstrut
Star & RA & Dec & g$_0$ & i$_0$ & CaHK$_0$ & E(B$-$V) & RV & $\sigma_{RV}$ & T$\mathrm{_{eff}}$ & logg & v$_{\mathrm t}$ & [Fe/H] \\
     & [$\deg$] & [$\deg$] &  &  &  &  & [km~s$^{-1}$] &  &  [K]  & [cgs]  &  [km~s$^{-1}$] & [dex] \\
\hline\Tstrut

Pr\_134.3232+17.6970 & 134.3232 & 17.6970 & 14.671 & 14.310 & 15.002 &  0.033 & $+115.27$ (5) & 1.78 & 6350 & 4.40 & 1.12 & $-2.63$ \\ 
Pr\_180.0090+3.7165  & 182.5090 & 03.7165 & 13.613 & 13.040 & 14.227 &  0.018 & $ -17.30$ (1) & 1.32 & 5200 & 4.40 & 1.12 & $-0.18$* \\ 
Pr\_180.2206+9.5683  & 180.2206 & 09.5683 & 15.186 & 14.370 & 15.890 &  0.020 & $ +22.12$ (6) & 0.46 & 5050 & 1.86 & 1.55 & $-2.96$ \\ 
Pr\_180.3790+0.9470  & 180.3790 & 00.9470 & 14.353 & 13.776 & 14.987 &  0.019 & $ -57.47$ (1) & 0.90 & 5684 & 3.70 & 1.05 & $-0.74$ \\ 
Pr\_180.7918+3.4084  & 180.7918 & 03.4084 & 13.748 & 12.456 & 14.884 &  0.027 & $ +21.20$ (1) & 0.81 & 4800 & 3.70 & 1.26 & $-0.08$* \\ 
Pr\_180.9118+11.3258 & 180.9118 & 11.3258 & 14.244 & 13.630 & 14.855 &  0.026 & $ +33.98$ (1) & 0.95 & 5800 & 3.80 & 1.15 & $-0.49$* \\ 
Pr\_181.2243+7.4160  & 181.2243 & 07.4160 & 15.057 & 14.694 & 15.450 &  0.014 & $-146.70$ (2) & 0.77 & 6455 & 3.81 & 1.24 & $-2.92$ \\ 
Pr\_181.3119+11.6850 & 181.3119 & 11.6850 & 14.270 & 13.488 & 15.006 &  0.032 & $  +1.49$ (2) & 0.02 & 5300 & 4.60 & 1.08 & $-1.84$* \\ 
Pr\_181.3473+11.6698 & 181.3473 & 11.6698 & 14.414 & 13.722 & 15.006 &  0.032 & $ +11.90$ (2) & 0.02 & 5900 & 4.45 & 1.11 & $-0.49$ \\ 
Pr\_181.3708+11.7636 & 181.3708 & 11.7636 & 14.443 & 13.689 & 15.119 &  0.033 & $ +79.06$ (1) & 3.03 & 5494 & 3.18 & 1.36 & $-1.48$ $\dagger$\\ 
Pr\_181.4395+1.6294  & 181.4395 & 01.6294 & 14.969 & 14.052 & 15.795 &  0.021 & $+206.58$ (1) & 2.00 & 4935 & 1.80 & 1.80 & $-2.50$ \\ 
Pr\_181.6954+13.8076 & 181.6954 & 13.8076 & 14.714 & 14.055 & 15.384 &  0.030 & $ +78.01$ (1) & 1.06 & 5608 & 3.60 & 1.05 & $-0.80$ \\ 
Pr\_182.1670+3.4771  & 182.1670 & 03.4771 & 14.209 & 13.107 & 14.967 &  0.020 & $ -19.09$ (1) & 0.64 & 5350 & 4.45 & 1.11 & $-0.35$* \\ 
Pr\_182.5364+0.9431  & 182.5364 & 00.9431 & 15.053 & 14.645 & 15.476 &  0.029 & $+222.71$ (2) & 0.34 & 6270 & 3.75 & 1.25 & $-1.74$ \\ 
Pr\_182.8521+14.1594 & 182.8521 & 14.1594 & 14.409 & 13.462 & 15.333 &  0.045 & $ +78.56$ (1) & 1.35 & 4959 & 1.75 & 1.60 & $-1.79$ \\ 
Pr\_183.6850+4.8619  & 183.6850 & 04.8619 & 15.038 & 14.718 & 15.374 &  0.017 & $ +41.00$ (2) & 0.30 & 6491 & 4.44 & 1.11 & $-3.16$ \\ 
Pr\_185.4112+7.4778  & 185.4112 & 07.4778 & 14.847 & 14.412 & 15.234 &  0.020 & $+175.18$ (2) & 0.40 & 6304 & 4.53 & 1.09 & $-1.85$* \\ 
Pr\_187.8517+13.4560 & 187.8517 & 13.4560 & 13.914 & 13.382 & 14.269 &  0.025 & $  +1.99$ (1) & 0.43 & 5700 & 4.10 & 1.18 & $-0.45$* \\ 
Pr\_187.9786+8.7294  & 187.9786 & 08.7294 & 15.190 & 14.532 & 15.775 &  0.018 & $ -54.11$ (2) & 0.06 & 5618 & 3.66 & 1.27 & $-0.50$ \\ 
Pr\_188.1264+8.7740  & 188.1264 & 08.7740 & 14.528 & 13.780 & 15.228 &  0.017 & $ -48.83$ (1) & 1.11 & 5600 & 3.60 & 1.00 & $-1.05$ \\ 
Pr\_189.9449+11.5535 & 189.9449 & 11.5535 & 14.427 & 14.033 & 14.812 &  0.037 & $ +30.35$ (1) & 6.00 & 6491 & 3.83 & 1.23 & $-2.57$ \\ 
Pr\_190.2669+11.1092 & 190.2669 & 11.1092 & 14.076 & 13.854 & 14.759 &  0.027 & $ -15.71$ (1) & 1.18 & 5800 & 4.20 & 1.16 & $-0.34$* \\ 
Pr\_190.5813+12.8577 & 190.5813 & 12.8577 & 14.327 & 13.632 & 14.779 &  0.030 & $  +4.42$ (1) & 0.70 & 5800 & 4.20 & 1.16 & $-0.20$* \\ 
Pr\_190.6313+8.5138  & 190.6313 & 08.5138 & 15.114 & 14.511 & 16.060 &  0.021 & $ -64.16$ (2) & 0.02 & 5500 & 3.80 & 1.24 & $-0.24$ \\ 
Pr\_192.2121+15.9263 & 192.2121 & 15.9263 & 14.843 & 13.809 & 15.837 &  0.025 & $  -4.56$ (1) & 0.76 & 5600 & 4.30 & 1.14 & $+0.25$ \\ 
Pr\_192.4285+15.9119 & 192.4285 & 15.9119 & 14.238 & 13.795 & 15.031 &  0.026 & $ -62.24$ (1) & 0.62 & 5450 & 4.65 & 1.07 & $-0.49$* \\ 
Pr\_192.8540+15.8199 & 192.8540 & 15.8199 & 14.277 & 13.748 & 14.763 &  0.023 & $  +0.98$ (1) & 0.39 & 5700 & 4.05 & 1.19 & $-0.20$* \\ 
Pr\_192.9068+6.8314  & 192.9068 & 06.8314 & 13.766 & 13.017 & 14.490 &  0.034 & $ +15.64$ (1) & 1.52 & 5800 & 4.25 & 1.15 & $-0.35$ \\ 
Pr\_193.1159+8.0557  & 193.1159 & 08.0557 & 14.660 & 14.224 & 15.129 &  0.026 & $ +42.13$ (1) & 4.43 & 6100 & 4.55 & 1.09 & $-1.85$ \\ 
Pr\_193.1501+15.7966 & 193.1501 & 15.7966 & 13.757 & 13.536 & 14.373 &  0.025 & $  +9.57$ (1) & 0.93 & 5800 & 4.00 & 1.20 & $-0.31$* \\ 
Pr\_193.5542+11.5036 & 193.5542 & 11.5036 & 14.460 & 13.681 & 15.087 &  0.032 & $ +16.20$ (1) & 0.66 & 5679 & 4.15 & 1.17 & $-0.30$ \\ 
Pr\_193.8390+11.4150 & 193.8390 & 11.4150 & 15.072 & 13.983 & 16.070 &  0.030 & $  +2.53$ (2) & 0.07 & 4650 & 1.22 & 1.76 & $-2.91$ \\ 
Pr\_196.3755+8.5138  & 196.3755 & 08.5138 & 14.914 & 14.051 & 15.656 &  0.029 & $ -72.24$ (1) & 3.59 & 5012 & 1.73 & 1.65 & $-2.77$ \\ 
Pr\_196.4126+14.3177 & 196.4126 & 14.3177 & 14.884 & 14.003 & 15.634 &  0.029 & $ -49.36$ (2) & 0.00 & 5702 & 4.55 & 1.09 & $-0.44$ \\ 
Pr\_196.5323+8.7716  & 196.5323 & 08.7716 & 14.157 & 13.874 & 14.491 &  0.031 & $-100.40$ (2) & 1.59 & 6483 & 4.55 & 1.09 & $-2.55$ \\ 
Pr\_196.5453+12.1211 & 196.5453 & 12.1211 & 14.745 & 14.468 & 15.090 &  0.028 & $ +54.47$ (2) & 0.24 & 5950 & 3.40 & 1.32 & $-2.51$ \\ 
Pr\_196.6013+15.6768 & 196.6013 & 15.6768 & 14.624 & 14.001 & 15.438 &  0.027 & $-103.91$ (2) & 0.17 & 5600 & 4.60 & 1.08 & $-0.69$ \\ 
Pr\_197.5045+15.6970 & 197.5045 & 15.6970 & 13.980 & 13.807 & 14.490 &  0.022 & $ -59.46$ (1) & 0.98 & 5920 & 4.05 & 1.19 & $-0.80$* \\ 
Pr\_197.9861+12.3578 & 197.9861 & 12.3578 & 15.074 & 14.626 & 15.647 &  0.028 & $+114.83$ (2) & 0.35 & 6050 & 3.80 & 1.24 & $-1.22$ \\ 
Pr\_198.5288+12.1493 & 198.5288 & 12.1493 & 14.232 & 14.111 & 14.840 &  0.024 & $ -26.75$ (1) & 1.23 & 6120 & 4.25 & 1.15 & $-0.59$* \\ 
Pr\_198.5495+11.4125 & 198.5495 & 11.4125 & 14.357 & 13.988 & 14.772 &  0.022 & $ +74.45$ (1) & 5.31 & 6494 & 4.35 & 1.13 & $-2.20$ \\ 
Pr\_199.9269+8.3816  & 199.9269 & 08.3816 & 14.100 & 13.423 & 15.092 &  0.022 & $  +3.17$ (1) & 0.54 & 5300 & 4.55 & 1.09 & $-0.15$* \\ 
Pr\_200.0999+13.7229 & 200.0999 & 13.7229 & 15.282 & 14.223 & 16.288 &  0.024 & $+189.83$ (4) & 0.56 & 4750 & 1.24 & 1.75 & $-2.48$ \\ 
Pr\_200.5298+8.9768  & 200.5298 & 08.9768 & 15.358 & 14.664 & 16.306 &  0.027 & $ +96.79$ (2) & 0.14 & 5363 & 3.45 & 1.20 & $-1.02$ \\ 
Pr\_200.7620+9.4376  & 200.7620 & 09.4376 & 14.695 & 14.096 & 15.680 &  0.023 & $ +33.26$ (1) & 0.68 & 5650 & 4.35 & 1.13 & $-0.15$ \\ 
Pr\_201.1159+15.4382 & 201.1159 & 15.4382 & 14.548 & 13.859 & 15.586 &  0.021 & $  -5.77$ (1) & 0.70 & 5589 & 4.40 & 1.12 & $-0.19$ \\ 
Pr\_202.3435+13.2291 & 202.3435 & 13.2291 & 14.398 & 13.840 & 14.795 &  0.021 & $+100.81$ (2) & 0.20 & 5950 & 2.80 & 1.50 & $-0.75$ \\ 
Pr\_203.2831+13.6326 & 203.2831 & 13.6326 & 15.116 & 14.249 & 15.833 &  0.024 & $-139.55$ (1) & 3.42 & 5008 & 1.95 & 1.45 & $-2.70$ \\ 
Pr\_204.9008+10.5513 & 204.9008 & 10.5513 & 14.894 & 14.617 & 15.171 &  0.031 & $-246.62$ (1) & 6.02 & 6718 & 4.22 & 1.16 & $-2.66$ \\ 
Pr\_205.1342+13.8234 & 205.1342 & 13.8234 & 14.803 & 14.188 & 15.402 &  0.022 & $+126.90$ (1) & 3.05 & 5462 & 2.90 & 1.42 & $-2.12$ \\ 
Pr\_205.8132+15.3832 & 205.8132 & 15.3832 & 14.678 & 14.371 & 15.076 &  0.032 & $+121.42$ (1) & 5.22 & 6718 & 4.23 & 1.15 & $-2.13$ \\ 
Pr\_206.3487+9.3099  & 206.3487 & 09.3099 & 14.365 & 14.027 & 14.726 &  0.026 & $+157.19$ (1) & 4.17 & 6522 & 3.94 & 1.21 & $-1.80$ \\ 
Pr\_207.9961+1.1795  & 207.9961 & 01.1795 & 14.306 & 13.493 & 14.823 &  0.030 & $ +36.17$ (1) & 0.51 & 5450 & 4.40 & 1.12 & $-0.49$* \\ 
Pr\_208.0799+4.4267  & 208.0799 & 04.4267 & 14.686 & 14.132 & 15.151 &  0.026 & $-129.96$ (1) & 5.27 & 5572 & 2.97 & 1.41 & $-2.77$ \\ 
Pr\_209.2123+1.5275  & 209.2123 & 01.5275 & 14.562 & 14.030 & 15.100 &  0.035 & $ +13.53$ (1) & 2.67 & 5540 & 3.30 & 1.15 & $-1.90$ \\ 
Pr\_209.7189+10.8613 & 209.7189 & 10.8613 & 14.642 & 14.250 & 15.069 &  0.024 & $-136.51$ (2) & 0.33 & 6358 & 4.40 & 1.12 & $-1.98$* \\ 
Pr\_209.9364+15.9251 & 209.9364 & 15.9251 & 14.903 & 14.611 & 15.281 &  0.021 & $ -91.43$ (2) & 0.37 & 6664 & 3.96 & 1.21 & $-2.25$ \\ 
Pr\_210.0175+14.6289 & 210.0175 & 14.6289 & 14.789 & 14.079 & 15.464 &  0.018 & $ -74.32$ (2) & 0.67 & 5150 & 2.37 & 1.53 & $-2.67$ \\ 
Pr\_210.0316+14.0027 & 210.0322 & 14.0036 & 14.590 & 14.019 & 15.247 &  0.016 & $ +22.68$ (2) & 0.03 & 5400 & 3.55 & 1.29 & $-0.98$* \\ 
Pr\_210.7513+12.7744 & 210.7513 & 12.7744 & 14.881 & 13.714 & 16.127 &  0.028 & $ +40.58$ (1) & 1.65 & 4652 & 1.35 & 1.60 & $-2.12$ \\ 
Pr\_210.8633+8.1798  & 210.8633 & 08.1798 & 14.675 & 14.067 & 15.237 &  0.024 & $ -12.33$ (3) & 0.79 & 5542 & 3.31 & 1.34 & $-1.95$ \\ 
Pr\_211.2766+10.3280 & 211.2766 & 10.3280 & 14.825 & 14.393 & 15.348 &  0.022 & $ +44.59$ (1) & 1.79 & 5740 & 3.75 & 1.25 & $-1.39$* \\ 
\hline

\end{tabular}
\end{table*}

\setlength{\tabcolsep}{4pt}
\begin{table*}

\contcaption{}
\label{tab:continued1}
\begin{tabular}{lccccccrccccl}
\hline\hline\Tstrut
Star & RA & Dec & g$_0$ & i$_0$ & CaHK$_0$ & E(B$-$V) & RV & $\sigma_{RV}$ & T$\mathrm{_{eff}}$ & logg & v$_{\mathrm t}$ & [Fe/H] \\
     & [$\deg$] & [$\deg$] &  &  &  &  & [km~s$^{-1}$] &  &  [K]  & [cgs]  &  [km~s$^{-1}$] & [dex] \\
\hline\Tstrut

Pr\_211.7184+15.5516 & 211.7184 & 15.5516 & 14.885 & 13.876 & 15.899 &  0.017 & $-111.27$ (2) & 0.19 & 4750 & 1.31 & 1.74 & $-2.42$ \\ 
Pr\_212.5834+10.5365 & 212.5834 & 10.5365 & 14.553 & 14.082 & 15.036 &  0.023 & $-125.50$ (1) & 3.96 & 6222 & 4.55 & 1.09 & $-1.79$ \\ 
Pr\_213.2814+14.8983 & 213.2814 & 14.8983 & 14.643 & 14.283 & 15.006 &  0.019 & $ -11.28$ (1) & 5.46 & 6002 & 3.55 & 1.29 & $-1.95$ $\dagger$\\ 
Pr\_213.7878+8.4232  & 213.7878 & 08.4232 & 14.989 & 14.254 & 15.573 &  0.030 & $-106.36$ (1) & 3.40 & 5289 & 2.45 & 1.51 & $-2.45$* \\ 
Pr\_214.5557+7.4670  & 214.5557 & 07.4670 & 14.713 & 14.350 & 15.106 &  0.029 & $ +47.24$ (1) & 5.51 & 6482 & 4.10 & 1.18 & $-2.14$ \\ 
Pr\_215.6129+15.0163 & 215.6129 & 15.0163 & 14.419 & 13.688 & 15.164 &  0.023 & $-120.21$ (1) & 0.81 & 5180 & 2.62 & 1.40 & $-1.92$ \\ 
Pr\_215.6783+7.6929  & 215.6783 & 07.6929 & 14.530 & 13.849 & 15.019 &  0.028 & $  +9.50$ (1) & 0.54 & 5527 & 4.00 & 1.20 & $-0.35$* \\ 
Pr\_216.1245+10.2135 & 216.1245 & 10.2135 & 14.683 & 13.999 & 15.341 &  0.027 & $+106.05$ (1) & 3.65 & 5412 & 2.88 & 1.30 & $-2.21$ \\ 
Pr\_217.3862+15.1651 & 217.3862 & 15.1651 & 14.675 & 14.349 & 15.125 &  0.025 & $-161.83$ (1) & 4.11 & 5700 & 3.30 & 1.34 & $-1.97$ \\ 
Pr\_217.5786+14.0379 & 217.5786 & 14.0379 & 14.765 & 13.867 & 15.489 &  0.029 & $ -17.17$ (3) & 0.40 & 4968 & 1.64 & 1.67 & $-2.66$ \\ 
Pr\_217.6444+15.9634 & 217.6444 & 15.9634 & 14.897 & 14.562 & 15.277 &  0.031 & $ -19.95$ (2) & 0.85 & 6550 & 4.17 & 1.17 & $-1.82$ \\ 
Pr\_218.4256+7.5213  & 218.4256 & 07.5213 & 14.661 & 13.992 & 15.322 &  0.024 & $ +11.45$ (2) & 0.15 & 5500 & 3.80 & 1.24 & $-0.60$* \\ 
Pr\_218.4622+10.3683 & 218.4622 & 10.3683 & 14.998 & 14.595 & 15.363 &  0.023 & $-125.40$ (2) & 1.22 & 5923 & 3.50 & 1.30 & $-2.40$ \\ 
Pr\_218.4977+15.7251 & 218.4977 & 15.7251 & 14.022 & 13.144 & 15.178 &  0.023 & $ -32.61$ (2) & 0.00 & 5150 & 4.40 & 1.12 & $-0.16$* \\ 
Pr\_223.5283+11.1353 & 223.5283 & 11.1353 & 14.612 & 13.537 & 15.772 &  0.033 & $+123.90$ (1) & 1.31 & 4540 & 1.05 & 1.50 & $-2.30$ \\ 
Pr\_227.2895+1.3378  & 227.2895 & 01.3378 & 13.917 & 13.304 & 14.570 &  0.051 & $  +4.91$ (1) & 0.49 & 5750 & 4.30 & 1.14 & $-0.35$* \\ 
Pr\_228.4607+8.3553  & 228.4607 & 08.3553 & 14.894 & 14.587 & 15.252 &  0.030 & $  +8.08$ (2) & 0.01 & 6525 & 4.32 & 1.14 & $-2.20$ \\ 
Pr\_228.6558+9.0914  & 228.6558 & 09.0914 & 14.832 & 14.530 & 15.166 &  0.032 & $-147.32$ (2) & 0.02 & 6695 & 4.26 & 1.15 & $-2.26$* \\ 
Pr\_228.8159+0.2222  & 228.8159 & 00.2222 & 14.751 & 14.366 & 15.129 &  0.052 & $  +5.61$ (2) & 0.12 & 6520 & 4.36 & 1.13 & $-2.02$ \\ 
Pr\_229.0409+10.3020 & 229.0409 & 10.3020 & 14.728 & 13.727 & 15.790 &  0.038 & $  -8.15$ (2) & 0.21 & 5400 & 4.55 & 1.09 & $-0.10$ \\ 
Pr\_229.1219+0.9089  & 229.1219 & 00.9089 & 14.747 & 14.403 & 15.118 &  0.048 & $-223.25$ (2) & 0.64 & 6385 & 3.70 & 1.26 & $-2.25$ \\ 
Pr\_229.8911+0.1106  & 229.8911 & 00.1106 & 14.430 & 13.577 & 15.016 &  0.064 & $  -2.05$ (1) & 0.47 & 5600 & 4.00 & 1.20 & $+0.10$* \\ 
Pr\_230.4663+6.5252  & 230.4663 & 06.5252 & 14.521 & 13.737 & 15.098 &  0.040 & $ -26.63$ (1) & 1.09 & 5348 & 3.45 & 1.31 & $-1.15$* \\ 
Pr\_231.0318+6.4867  & 231.0318 & 06.4867 & 14.626 & 13.893 & 15.183 &  0.041 & $ +21.45$ (1) & 0.63 & 5468 & 3.90 & 1.22 & $-0.35$* \\ 
Pr\_232.6956+8.3392  & 232.6956 & 08.3392 & 15.076 & 14.462 & 15.661 &  0.044 & $ -37.84$ (2) & 0.20 & 5641 & 3.40 & 1.32 & $-2.22$ \\ 
Pr\_232.8039+6.1178  & 232.8039 & 06.1178 & 15.016 & 14.268 & 15.867 &  0.052 & $-212.58$ (1) & 3.92 & 5280 & 2.30 & 1.54 & $-2.26$* \\ 
Pr\_233.5730+6.4702  & 233.5730 & 06.4702 & 14.774 & 13.869 & 15.545 &  0.044 & $ -80.66$ (1) & 4.16 & 4991 & 1.90 & 1.62 & $-2.74$ \\ 
Pr\_233.9312+9.5596  & 233.9312 & 09.5596 & 14.970 & 14.368 & 15.557 &  0.037 & $ -99.38$ (2) & 0.29 & 5505 & 3.50 & 1.20 & $-2.20$ \\ 
Pr\_234.4403+13.3742 & 234.4403 & 13.3742 & 14.817 & 13.506 & 15.815 &  0.046 & $ -66.48$ (2) & 0.05 & 5600 & 4.51 & 0.90 & $-0.40$* \\ 
Pr\_235.1448+8.7464  & 235.1448 & 08.7464 & 14.649 & 14.151 & 15.086 &  0.042 & $-156.26$ (1) & 4.79 & 6167 & 3.64 & 1.27 & $-2.54$ \\ 
Pr\_235.7578+9.0000  & 235.7578 & 09.0000 & 14.937 & 14.601 & 15.330 &  0.042 & $ -58.70$ (2) & 0.37 & 6654 & 4.20 & 1.16 & $-1.81$ \\ 
Pr\_235.9710+9.1864  & 235.9710 & 09.1864 & 14.631 & 14.257 & 15.069 &  0.042 & $ -32.49$ (1) & 3.69 & 6300 & 3.70 & 1.26 & $-1.83$* \\ 
Pr\_236.1077+10.5311 & 236.1077 & 10.5311 & 14.808 & 13.753 & 15.809 &  0.048 & $ -41.51$ (1) & 1.79 & 4650 & 1.47 & 1.71 & $-2.55$ \\ 
Pr\_236.4855+10.6903 & 236.4855 & 10.6903 & 14.575 & 14.005 & 14.933 &  0.052 & $ -36.42$ (1) & 0.93 & 5850 & 3.70 & 1.26 & $-0.27$* \\ 
Pr\_236.7138+9.6084  & 236.7138 & 09.6084 & 14.529 & 13.466 & 15.271 &  0.045 & $ +54.25$ (1) & 0.92 & 5450 & 3.45 & 1.31 & $-0.85$* \\ 
Pr\_237.8246+10.1427 & 237.8246 & 10.1427 & 15.216 & 14.501 & 15.776 &  0.046 & $-165.90$ (2) & 1.08 & 5405 & 2.80 & 1.44 & $-3.23$ \\ 
Pr\_237.8353+10.5902 & 237.8353 & 10.5902 & 14.734 & 13.964 & 15.370 &  0.056 & $-237.67$ (2) & 0.31 & 5250 & 2.70 & 1.46 & $-2.32$* \\ 
Pr\_237.9609+15.4023 & 237.9609 & 15.4023 & 14.543 & 14.215 & 14.916 &  0.046 & $-267.22$ (2) & 0.66 & 6557 & 4.09 & 1.18 & $-1.90$ \\ 
Pr\_238.7217+6.1945  & 238.7217 & 06.1945 & 14.770 & 14.462 & 15.103 &  0.039 & $-195.14$ (2) & 0.23 & 6551 & 4.20 & 1.16 & $-2.06$ \\ 
Pr\_240.0348+13.8279 & 240.0348 & 13.8279 & 14.785 & 13.835 & 15.721 &  0.052 & $  +3.55$ (1) & 1.80 & 4760 & 1.52 & 1.70 & $-2.30$ \\ 
Pr\_240.4216+9.6761  & 240.4216 & 09.6761 & 14.944 & 14.164 & 15.528 &  0.040 & $ +37.59$ (3) & 0.23 & 5204 & 2.60 & 1.47 & $-2.98$ \\ 
Pr\_241.1186+9.4156  & 241.1186 & 09.4156 & 14.532 & 14.070 & 14.999 &  0.044 & $ -51.47$ (1) & 4.89 & 6299 & 3.95 & 1.21 & $-1.92$ \\ 
Pr\_241.7900+14.0920 & 241.7900 & 14.0920 & 14.680 & 14.351 & 15.078 &  0.038 & $-127.80$ (2) & 0.91 & 6485 & 3.90 & 1.22 & $-2.51$ \\ 
Pr\_242.3556+7.9425  & 242.3556 & 07.9425 & 14.851 & 14.461 & 15.249 &  0.046 & $-181.11$ (2) & 1.00 & 6326 & 3.90 & 1.22 & $-1.95$ \\ 
Pr\_243.8390+6.9966  & 243.8390 & 06.9966 & 14.748 & 14.242 & 15.192 &  0.063 & $ -17.94$ (3) & 0.32 & 5878 & 3.52 & 1.30 & $-1.95$ \\ 
Pr\_244.4872+16.8936 & 244.4872 & 16.8936 & 14.379 & 13.953 & 14.701 &  0.043 & $-192.08$ (2) & 0.46 & 6214 & 3.90 & 1.22 & $-2.10$ \\ 
Pr\_245.1096+8.8947  & 245.1096 & 08.8947 & 14.463 & 13.739 & 15.033 &  0.064 & $ -30.90$ (1) & 0.59 & 5567 & 4.45 & 1.11 & $-0.35$* \\ 
Pr\_245.4387+8.9954  & 245.4387 & 08.9954 & 14.780 & 14.354 & 15.199 &  0.055 & $ -82.89$ (2) & 0.60 & 6464 & 4.10 & 1.18 & $-1.67$ \\ 
Pr\_245.5747+6.8844  & 245.5747 & 06.8844 & 15.041 & 14.323 & 15.579 &  0.062 & $-188.96$ (4) & 0.79 & 5424 & 2.95 & 1.40 & $-3.17$ \\ 
Pr\_245.8364+13.8778 & 245.8364 & 13.8778 & 14.012 & 13.264 & 14.641 &  0.045 & $-176.99$ (1) & 4.42 & 5150 & 2.28 & 1.40 & $-3.06$ \\ 
Pr\_246.8588+12.3193 & 246.8588 & 12.3193 & 14.763 & 13.932 & 15.500 &  0.056 & $ -86.02$ (2) & 0.05 & 5070 & 2.21 & 1.50 & $-2.25$ \\ 
Pr\_248.4394+7.9230  & 248.4394 & 07.9230 & 14.340 & 13.892 & 14.767 &  0.081 & $ -15.66$ (3) & 0.85 & 6350 & 4.40 & 1.12 & $-1.72$ \\ 
Pr\_248.4959+15.0776 & 248.4959 & 15.0776 & 14.717 & 13.830 & 15.454 &  0.062 & $ -74.03$ (2) & 0.25 & 5069 & 1.86 & 1.63 & $-2.63$ \\ 
Pr\_248.5263+8.9342  & 248.5263 & 08.9342 & 15.089 & 14.279 & 15.829 &  0.069 & $-109.73$ (2) & 0.63 & 5300 & 2.45 & 1.51 & $-2.07$ \\ 
Pr\_250.6971+8.3743  & 250.6971 & 08.3743 & 14.615 & 13.696 & 20.337 &  0.085 & $  -4.08$ (1) & 3.84 & 5020 & 1.80 & 1.64 & $-2.66$ \\ 
Pr\_250.8797+12.1101 & 250.8797 & 12.1101 & 14.261 & 13.069 & 15.327 &  0.053 & $  +4.67$ (1) & 0.52 & 5600 & 4.40 & 1.12 & $+0.05$* \\ 
Pr\_251.4082+12.3657 & 251.4082 & 12.3657 & 15.062 & 14.076 & 15.785 &  0.056 & $  -4.43$ (2) & 0.03 & 4919 & 1.54 & 1.69 & $-3.22$ \\ 
Pr\_252.1648+15.0648 & 252.1648 & 15.0648 & 14.621 & 13.615 & 15.518 &  0.068 & $-157.06$ (1) & 2.24 & 4770 & 1.35 & 1.50 & $-2.43$ \\ 
Pr\_252.4208+12.6477 & 252.4208 & 12.6477 & 14.486 & 13.739 & 15.172 &  0.053 & $ +35.27$ (2) & 0.05 & 5500 & 3.60 & 1.28 & $-1.30$* \\ 
Pr\_252.4917+15.2984 & 252.4917 & 15.2984 & 14.349 & 13.443 & 15.056 &  0.070 & $ +62.65$ (1) & 0.93 & 5900 & 3.70 & 1.26 & $-0.60$* \\ 
Pr\_252.6179+16.0546 & 252.6179 & 16.0546 & 14.178 & 13.208 & 15.057 &  0.074 & $-113.39$ (2) & 0.04 & 4810 & 1.60 & 1.68 & $-2.52$ \\ 
Pr\_253.8582+15.7240 & 253.8582 & 15.7240 & 14.973 & 14.037 & 15.720 &  0.088 & $ -98.39$ (3) & 0.44 & 5077 & 2.25 & 1.50 & $-2.58$ \\ 
Pr\_254.0662+14.2694 & 254.0662 & 14.2694 & 13.658 & 13.079 & 14.172 &  0.071 & $ -29.70$ (1) & 0.46 & 5500 & 3.80 & 1.24 & $-0.56$* \\ 
\hline

\end{tabular}
\end{table*}

\setlength{\tabcolsep}{4pt}
\begin{table*}
\contcaption{}
\label{tab:continued2}
\begin{tabular}{lccccccrccccl}
\hline\hline\Tstrut
Star & RA & Dec & g$_0$ & i$_0$ & CaHK$_0$ & E(B$-$V) & RV & $\sigma_{RV}$ & T$\mathrm{_{eff}}$ & logg & v$_{\mathrm t}$ & [Fe/H] \\
     & [$\deg$] & [$\deg$] &  &  &  &  & [km~s$^{-1}$] &  &  [K]  & [cgs]  &  [km~s$^{-1}$] & [dex] \\
\hline\Tstrut
Pr\_254.3844+12.9653 & 254.3844 & 12.9653 & 14.969 & 13.978 & 15.743 &  0.059 & $-386.12$ (2) & 0.47 & 5300 & 2.80 & 1.44 & $-2.45$ $\dagger$\\  
Pr\_254.5215+15.4969 & 254.5215 & 15.4969 & 14.045 & 13.246 & 14.807 &  0.096 & $ +25.01$ (1) & 0.52 & 5640 & 4.30 & 1.14 & $-0.09$ \\ 
Pr\_254.5478+10.9129 & 254.5478 & 10.9129 & 14.447 & 13.687 & 15.096 &  0.077 & $ +94.21$ (1) & 2.72 & 5460 & 3.35 & 1.33 & $-2.15$ \\ 
Pr\_254.7768+13.8208 & 254.7768 & 13.8208 & 14.018 & 13.239 & 14.479 &  0.077 & $ -16.41$ (1) & 0.51 & 5600 & 4.05 & 1.19 & $-0.35$* \\ 
Pr\_255.2679+14.9714 & 255.2679 & 14.9714 & 14.332 & 13.867 & 14.700 &  0.083 & $ +27.92$ (1) & 5.11 & 6479 & 3.88 & 1.22 & $-2.09$ \\ 
Pr\_255.5564+10.8613 & 255.5564 & 10.8613 & 14.782 & 14.125 & 15.306 &  0.075 & $-372.85$ (2) & 0.43 & 5495 & 2.86 & 1.43 & $-2.55$ \\ 
Pr\_255.8043+10.8443 & 255.8043 & 10.8443 & 14.246 & 13.859 & 14.629 &  0.082 & $-266.01$ (2) & 1.12 & 5600 & 3.20 & 1.36 & $-3.00$* \\ 

\hline

\end{tabular}
\end{table*}

\normalsize

The echelle orders were merged using a procedure developed by the authors of
this paper. In short, the script isolates the different echelle orders to
remove, in the overlapping wavelength regions, the part with the lowest
signal-to-noise ratio (SNR). The orders are then combined again with a sigma
clipping routine.

Spectra with multiple exposures were corrected for radial velocity (RV)
shifts before combination. Radial velocities were measured with the {\tt DAOSPEC} package \citep{Stetson2008} using the {\tt  4DAO} wrapper \citep{Mucciarelli2013}.  The spectra were then degraded to R~=~40,000 to increase the SNR and allow for the automatic equivalent width (EW) measurement under the assumption that the line profile has a Gaussian shape. The final RV corrected and combined spectra are normalized using {\tt DAOSPEC}
in 3 wavelength ranges (4000--4800~\AA, 4800--5800~\AA\ and 5800--6800~\AA), using a 40 to 60 order polynomial.
The final RV measurements and their associated errors, along with the number of exposures for each stars, are presented in Table~\ref{tab:parameters}.

Figure~\ref{fig:orders} illustrates how the quality of the reduced spectra is improved by this procedure. 
It shows a portion of the original ESPaDOnS spectrum of the star Pr\_236.1077+10.5311 
and its spectrum after our optimal merging procedure, degraded to a resolution of R~=~40,000.

\section{Atmospheric parameters and chemical abundances}\label{analysis}

The chemical abundances of iron, carbon, the $\alpha$- Mg and Ca, and neutron
($n$)-capture (Sr and Ba) elements are calculated in 1D local
thermodynamic equilibrium (LTE) with {\tt Turbospectrum} \citep{Plez2012}
combined with MARCS model
atmospheres\footnote{\url{https://marcs.astro.uu.se/}}.

The equivalent widths of unblended spectral lines with Gaussian shape together
with their associated uncertainties were measured with {\tt DAOSPEC}
\citep{Stetson2008} which was launched iteratively with the code 4DAO
\citep{Mucciarelli2013}.\\ A $\chi ^{2}$~minimization between the observed and synthetic
spectra was applied for the strong and blended lines, and for the elements
presenting hyper fine splitting (HFS) of their energy levels such as barium and
carbon. The latter was estimated from the CH molecular feature at 4300~\AA . All
synthetic spectra were convoluted to the instrumental resolution, then rebinned at
the same pixel step as the observed spectra.

The abundance analysis is carried out with our own code. It enables the
interpolation of the stellar atmosphere models, allows the derivations 
of the atmospheric parameters and chemical abundances from EW measurements, 
as well as enables a spectral synthesis
$\chi ^{2}$~minimization for a set of chosen lines and elements. Spectral
synthesis is typically done over small wavelength ranges, centered on
the line of interest. The abundance of an element X is varied between
$-$2.0~$\le$~[X/Fe]~$\le$~$+$2.0~dex in steps of 0.1~dex, and refined in a
second iteration with smaller steps.

The linelist used in the calculations is the same as in \citet{Lucchesi2020}. It
combines the list from \citet{Jablonka2015}, \citet{Tafelmeyer2010}, and
\citet{VanderSwaelmen2013}.  The data for the selected atomic and molecular
transitions are taken from the VALD database
\citep{Piskunov1995,Ryabchikova1997,Kupka2000}. The solar abundances are taken
from \citet{Asplund2009}.

\subsection{Atmospheric parameters and metallicities} \label{Sec:optimization}

The stellar atmospheric parameters, i.e. effective temperature
(T$_{\mathrm{eff}}$), surface gravity ($\log$~g), micro-turbulence velocity
(v$_{\mathrm t}$), and metallicity ([Fe/H]), were adjusted spectroscopically
using the classical EW method.  Only Fe lines with EW~$\geq$~25~m\AA\ were
considered, in order to exclude the weak and noisy ones. Lines with
EW~$\geq$~110~m\AA\ were also excluded from the EW analysis, in order to avoid the flat part of the
curve of growth that is less sensitive to the abundance. These lines are highly
sensitive to the microturbulent velocity and the velocity fields. Moreover the
Gaussian approximation of the line profile starts to fail. Lines with excitation
potential $\chi_{ex}$~>~1.4~eV were also rejected in order to minimise non-local
thermodynamic (NLTE) effects. Finally, the \ion{Fe}{I} lines at wavelength
shorter than $\lambda$~$\leq$~4500~\AA\ were excluded as the consequence of the
low SNR at the blue end of the spectra.

The stellar atmospheric parameters have been optimized iteratively as follows:

\begin{itemize}
\item The effective temperatures were derived  by minimizing the slopes between iron abundance and excitation potential, allowing the slope to deviate from zero by less than about twice the uncertainty on the slope;

\item the surface gravities  were obtained from ionization equilibrium between  \ion{Fe}{II} and \ion{Fe}{I}. However, since NLTE  impacts the abundances derived from the \ion{Fe}{I} lines at low metallicities \citep[e.g.][]{Mashonkina2017}, we tolerated a difference in abundance $\Delta$(\ion{Fe}{II}--\ion{Fe}{I})~=~+0.15~dex for stars with [Fe/H]~$\le$~$-$2.5.  

\item the initial microturbulent velocities were obtained from the empirical relation $\log$~g~: v$_t = 2.0 - 0.2*\log~g$ as in \citet{Theler2020}. Convergence to the final value was reached by minimizing the slope between the neutral iron lines and their EWs.

\item The model's metallicity was adjusted to the derived \ion{Fe}{I} abundance after each iteration until they agreed within $\pm$0.05~dex.

\end{itemize}

\begin{figure}
    \centering
    \includegraphics[width=1.0\columnwidth]{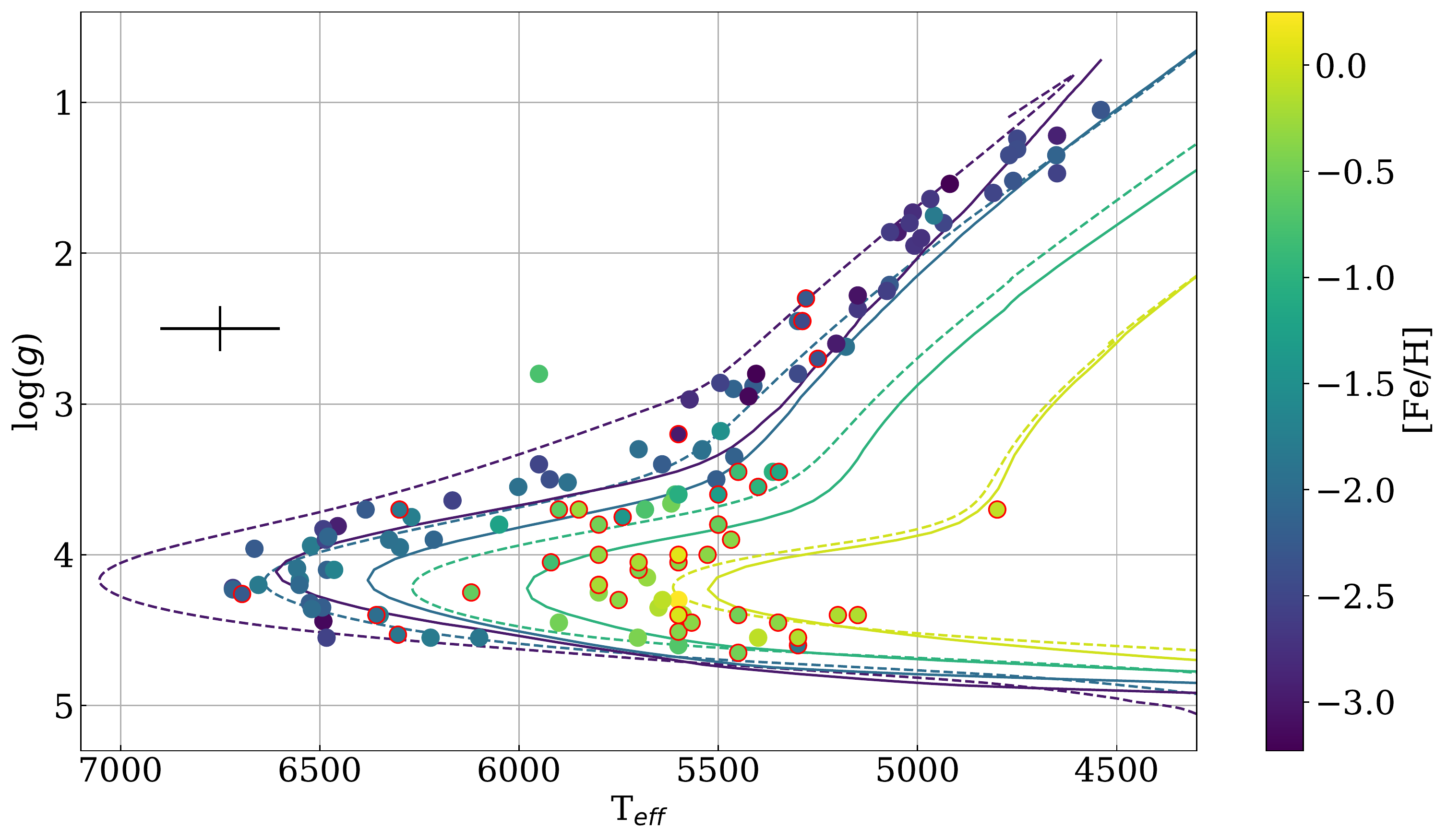}
    \caption{T$_{\mathrm{eff}}$ vs. $\log$~g diagram. Stars are colour-coded according to their metallicities. The Yonsei-Yale stellar isochrones (solid lines) are shown for the age of 13~Gyr and metallicities [Fe/H]~=~$[-3,-2,-1,0]$ from left to right respectively, with $\alpha$-enhancement  [$\alpha$/Fe]~=~+0.3 at [Fe/H]$<-1$. The MIST/MESA isochrones (dashed lines) with same parameters are also shown, at constant [$\alpha$/Fe]~=~0. Stars denoted with an open red circle have been rejected from the later versions of the \Pristine\ catalogues (see text). The black cross indicates the typical uncertainty of 150~K on T$_{\mathrm{eff}}$ and 0.15~dex on $\log$~g.}
    \label{Fig:CMD}
\end{figure}

Figure~\ref{Fig:CMD} presents the distribution of our sample in the
T$_{\mathrm{eff}}$ vs. $\log$~g diagram. Stars are colour-coded according to
their metallicity. We consider two sets of isochrones both taken at 13 Gyr and
at metallicities [Fe/H]=--3,--2,--1, and 0. The Yonsei-Yale (YY) stellar
isochrones \citep{Demarque2004} (solid lines) assumes an enhancement in $\alpha$-elements
of [$\alpha$/Fe]~=~+0.3 at [Fe/H]~<~$-$1. The MESA/MIST isochrones are currently
only available for solar-scaled abundances
\citep[e.g.][]{Choi2016,Dotter2016}.  The YY isochrones seem to better fit our
spectroscopic results.  Indeed, the turn-off region of the MIST isochrones is
significantly warmer ($\sim$500~K) than the corresponding YY isochrone at
[Fe/H]~=~$-$3 and none of our sample stars reach this extremely hot
temperature.  At higher metallicities, the difference between the two sets of
models decreases, reaching differences in temperature smaller than
$\Delta$T$_{\rm eff}$~<~100~K at [Fe/H]~=~0, well within the errors of the
spectroscopic T$_{\rm eff}$. As to red giant branch stars, the largest
difference between the two sets of isochrones is again observed at low
metallicities (e.g., $\Delta$(T$_{\rm{eff}})_{\rm{MIST-YY}}$~=~--100~K and
$\Delta(\log{\rm g})_{\rm{MIST-YY}}$~=~+0.2 and at [Fe/H]~=~--3). However, their
difference stays well within the typical spectroscopic uncertainties (150~K in
T$_{\mathrm{eff}}$ and 0.15 in $\log$~g). Thus, both sets of isochrones
represent a good fit to data in this region of stellar evolution.

Table~\ref{tab:parameters} lists the final stellar atmospheric parameters adopted in the rest of our analysis.  Our
sample includes both unevolved and evolved stars with effective temperatures and
surface gravities in the interval 
T$_{\mathrm{eff}}$~=~4540--6720~K and $\log$~g~=~1.05 to 4.65.  It covers a wide
range in metallicity, from [Fe/H]~=~$-3.2$ to +0.25.  As also seen in
Fig~\ref{Fig:CMD}, all the RGB stars selected by \Pristine\, are confirmed
very metal-poor stars.

One star stands out from the rest of the sample, Pr\_202.3435+13.2291 at
T$_{\mathrm{eff}}$~=~5950~K, $\log$g~=~2.80, and [Fe/H]~=~$-0.75$. It is most
likely an horizontal branch (HB) star. It was removed from further chemical analysis.

Three stars have very broaden lines, one originates from the new sample and two from the Q6 sample. They are fast rotators (identified with the $\dagger$ symbol in
Table~\ref{tab:parameters}) and their spectroscopic metallicity is poorly
constrained. As a consequence, they were also removed from further chemical
analysis. Abundances are provided for a total of 128 stars in Table~\ref{tab:abundances}.


\setlength{\tabcolsep}{4pt}
\begin{table*}
\caption{Elemental abundances of the 129 stars kept for the chemical analysis. The number of lines used is indicated in parentheses for each element and each star. The quoted errors correspond to the uncertainties resulting from the EW analysis or spectral fitting (see text). }
\label{tab:abundances}

\begin{tabular}{lllllllll}
  
\hline\hline\Tstrut
Star & [Fe/H] & log(\ion{Fe}{I}) $\pm$ $\sigma$ (N) & log(\ion{Fe}{II}) $\pm$ $\sigma$ (N) & [C/Fe] $\pm$ $\sigma$ & [Mg/Fe] $\pm$ $\sigma$ (N) & [Ca/Fe] $\pm$ $\sigma$ (N) & [Sr/Fe] $\pm$ $\sigma$ (N) & [Ba/Fe] $\pm$ $\sigma$ (N) \\
\hline\Tstrut

134.3232  & $-2.63$ & 4.87 $\pm$ 0.04 ( 17) & 4.63 $\pm$ 0.09 ( 3) & $ +1.17$ $\pm$ 0.24 & $ +0.19$ $\pm$ 0.08 ( 2) &            --            & $ +0.03$ $\pm$ 0.12 ( 1) &            --            \\
180.0090  & $-0.18$ & 7.32 $\pm$ 0.01 ( 47) & 7.30 $\pm$ 0.04 (10) & $ -0.47$ $\pm$ 0.19 & $ +0.17$ $\pm$ 0.12 ( 1) & $ +0.14$ $\pm$ 0.08 ( 2) &            --            & $ -0.05$ $\pm$ 0.09 ( 4) \\
180.2206  & $-2.96$ & 4.54 $\pm$ 0.02 ( 34) & 4.58 $\pm$ 0.04 ( 7) & $ +0.86$ $\pm$ 0.22 & $ +0.40$ $\pm$ 0.09 ( 2) & $ +0.25$ $\pm$ 0.09 ( 2) & $<-2.05$  ~~~~~~~   ( 1) & $ -1.26$ $\pm$ 0.12 ( 1) \\
180.3790  & $-0.74$ & 6.76 $\pm$ 0.02 ( 95) & 6.78 $\pm$ 0.02 (16) & $ +0.02$ $\pm$ 0.20 & $ +0.22$ $\pm$ 0.04 ( 4) & $ +0.16$ $\pm$ 0.02 (15) & $ -0.19$ $\pm$ 0.08 ( 1) & $ +0.05$ $\pm$ 0.06 ( 4) \\
180.7918  & $-0.08$ & 7.42 $\pm$ 0.03 ( 32) & 7.40 $\pm$ 0.04 (15) & $ -0.08$ $\pm$ 0.20 & $ +0.18$ $\pm$ 0.12 ( 1) & $ -0.05$ $\pm$ 0.07 ( 3) &            --            & $ +0.18$ $\pm$ 0.09 ( 3) \\
180.9118  & $-0.49$ & 7.01 $\pm$ 0.03 (100) & 7.04 $\pm$ 0.03 (20) & $ -0.30$ $\pm$ 0.25 & $ +0.14$ $\pm$ 0.06 ( 3) & $ +0.11$ $\pm$ 0.04 (11) & $ +0.20$ $\pm$ 0.10 ( 1) & $ +0.13$ $\pm$ 0.06 ( 4) \\
181.2243  & $-2.92$ & 4.58 $\pm$ 0.14 (  3) & 4.37 $\pm$ 0.14 ( 3) & $<+2.07$            & $ +0.71$ $\pm$ 0.17 ( 2) & $<+1.00$  ~~~~~~~   ( 4) &            --            & $ +0.71$ $\pm$ 0.24 ( 1) \\
181.3119  & $-1.84$ & 5.66 $\pm$ 0.01 ( 80) & 5.67 $\pm$ 0.04 ( 7) & $ -0.07$ $\pm$ 0.21 & $ +0.34$ $\pm$ 0.07 ( 1) & $ +0.38$ $\pm$ 0.02 (15) & $ -0.19$ $\pm$ 0.07 ( 1) & $ -0.08$ $\pm$ 0.05 ( 3) \\
181.3473  & $-0.49$ & 7.01 $\pm$ 0.01 ( 94) & 7.03 $\pm$ 0.02 (18) & $ +0.25$ $\pm$ 0.21 & $ +0.11$ $\pm$ 0.08 ( 1) & $ +0.07$ $\pm$ 0.02 (18) & $ -0.06$ $\pm$ 0.08 ( 1) & $ +0.08$ $\pm$ 0.05 ( 3) \\
181.4395  & $-2.50$ & 5.00 $\pm$ 0.02 ( 70) & 5.04 $\pm$ 0.06 (10) & $<-0.12$            & $ +0.58$ $\pm$ 0.07 ( 4) & $ +0.35$ $\pm$ 0.04 ( 7) & $ -0.09$ $\pm$ 0.16 ( 1) & $ -0.79$ $\pm$ 0.06 ( 3) \\
181.6954  & $-0.80$ & 6.70 $\pm$ 0.03 (102) & 6.75 $\pm$ 0.04 (11) & $ -0.03$ $\pm$ 0.20 & $ +0.25$ $\pm$ 0.07 ( 2) & $ +0.29$ $\pm$ 0.03 (14) &            --            & $ +0.03$ $\pm$ 0.05 ( 4) \\
182.1670  & $-0.35$ & 7.15 $\pm$ 0.01 ( 58) & 7.20 $\pm$ 0.03 (17) & $ +0.11$ $\pm$ 0.20 & $ +0.15$ $\pm$ 0.08 ( 1) & $ +0.15$ $\pm$ 0.03 ( 9) & $ +0.24$ $\pm$ 0.08 ( 1) & $ +0.17$ $\pm$ 0.07 ( 3) \\
182.5364  & $-1.74$ & 5.76 $\pm$ 0.03 ( 53) & 5.78 $\pm$ 0.04 (10) & $<+1.06$            & $ +0.24$ $\pm$ 0.08 ( 4) & $ +0.12$ $\pm$ 0.06 ( 3) & $ +0.81$ $\pm$ 0.10 ( 1) & $ +0.07$ $\pm$ 0.07 ( 2) \\
182.8521  & $-1.79$ & 5.71 $\pm$ 0.02 ( 88) & 5.81 $\pm$ 0.03 (13) & $ -0.35$ $\pm$ 0.20 & $ +0.26$ $\pm$ 0.06 ( 4) & $ +0.20$ $\pm$ 0.03 (14) & $ +0.37$ $\pm$ 0.11 ( 1) & $ +0.26$ $\pm$ 0.07 ( 3) \\
183.6850  & $-3.16$ & 4.34 $\pm$ 0.05 (  2) &        --            & $<+2.30$            & $ +0.01$ $\pm$ 0.05 ( 2) &            --            &            --            &            --            \\
185.4112  & $-1.85$ & 5.65 $\pm$ 0.02 ( 39) & 5.71 $\pm$ 0.04 ( 4) & $ +0.85$ $\pm$ 0.22 & $ +0.16$ $\pm$ 0.06 ( 2) & $ +0.42$ $\pm$ 0.03 ( 9) & $ +0.24$ $\pm$ 0.09 ( 1) & $ +0.05$ $\pm$ 0.12 ( 2) \\
187.8517  & $-0.45$ & 7.05 $\pm$ 0.01 (100) & 7.13 $\pm$ 0.01 (22) & $ -0.08$ $\pm$ 0.20 & $ +0.05$ $\pm$ 0.07 ( 1) & $ +0.14$ $\pm$ 0.02 (13) & $ -0.19$ $\pm$ 0.07 ( 1) & $ +0.03$ $\pm$ 0.05 ( 3) \\
187.9786  & $-0.50$ & 7.00 $\pm$ 0.01 ( 73) & 6.99 $\pm$ 0.03 (15) & $ -0.02$ $\pm$ 0.20 & $ +0.30$ $\pm$ 0.07 ( 2) & $ +0.13$ $\pm$ 0.03 (16) &            --            & $ +0.02$ $\pm$ 0.05 ( 4) \\
188.1264  & $-1.05$ & 6.45 $\pm$ 0.02 (112) & 6.50 $\pm$ 0.04 (15) & $ +0.24$ $\pm$ 0.20 & $ +0.18$ $\pm$ 0.14 ( 1) & $ +0.21$ $\pm$ 0.04 (10) & $ -0.25$ $\pm$ 0.14 ( 1) & $ -0.46$ $\pm$ 0.10 ( 2) \\
189.9449  & $-2.57$ & 4.93 $\pm$ 0.05 (  8) & 4.87 $\pm$ 0.10 ( 4) & $<+1.44$            & $ +0.20$ $\pm$ 0.14 ( 1) & $ +0.46$ $\pm$ 0.09 ( 3) &            --            & $<-0.03$  ~~~~~~~   ( 1) \\
190.2669  & $-0.34$ & 7.16 $\pm$ 0.03 ( 88) & 7.21 $\pm$ 0.03 (23) & $ +0.09$ $\pm$ 0.20 & $ +0.26$ $\pm$ 0.10 ( 1) & $ +0.08$ $\pm$ 0.04 (10) &            --            & $ +0.28$ $\pm$ 0.06 ( 3) \\
190.5813  & $-0.20$ & 7.30 $\pm$ 0.01 ( 79) & 7.31 $\pm$ 0.02 (20) & $ +0.41$ $\pm$ 0.20 & $ -0.10$ $\pm$ 0.06 ( 3) & $ +0.12$ $\pm$ 0.03 (13) & $ +0.09$ $\pm$ 0.09 ( 1) & $ -0.02$ $\pm$ 0.05 ( 4) \\
190.6313  & $-0.24$ & 7.26 $\pm$ 0.02 ( 67) & 7.30 $\pm$ 0.03 (20) & $ +0.15$ $\pm$ 0.19 & $ +0.16$ $\pm$ 0.08 ( 1) & $ +0.06$ $\pm$ 0.03 (12) &            --            & $ -0.01$ $\pm$ 0.05 ( 3) \\
192.2121  & $+0.25$ & 7.75 $\pm$ 0.02 ( 44) & 7.71 $\pm$ 0.03 (17) & $ -0.02$ $\pm$ 0.18 & $ -0.31$ $\pm$ 0.05 ( 2) & $ -0.00$ $\pm$ 0.03 ( 6) & $ -0.80$ $\pm$ 0.07 ( 1) & $ +0.11$ $\pm$ 0.09 ( 3) \\
192.4285  & $-0.49$ & 7.01 $\pm$ 0.01 ( 65) & 7.11 $\pm$ 0.03 (15) & $ +0.32$ $\pm$ 0.19 & $ +0.09$ $\pm$ 0.07 ( 1) & $ +0.07$ $\pm$ 0.03 (11) &            --            & $ +1.06$ $\pm$ 0.05 ( 3) \\
192.8540  & $-0.20$ & 7.30 $\pm$ 0.01 ( 82) & 7.31 $\pm$ 0.01 (22) & $ +0.10$ $\pm$ 0.20 & $ -0.01$ $\pm$ 0.04 ( 3) & $ +0.15$ $\pm$ 0.03 ( 6) & $ -0.23$ $\pm$ 0.07 ( 1) & $ +0.12$ $\pm$ 0.05 ( 4) \\
192.9068  & $-0.35$ & 7.15 $\pm$ 0.02 ( 78) & 7.17 $\pm$ 0.02 (18) & $ +0.12$ $\pm$ 0.20 & $ -0.01$ $\pm$ 0.05 ( 3) & $ +0.13$ $\pm$ 0.04 ( 7) & $ +0.17$ $\pm$ 0.09 ( 1) & $ -0.03$ $\pm$ 0.07 ( 4) \\
193.1159  & $-1.85$ & 5.65 $\pm$ 0.03 ( 52) & 5.60 $\pm$ 0.05 ( 7) & $ +0.82$ $\pm$ 0.22 & $ +0.37$ $\pm$ 0.07 ( 3) & $ +0.27$ $\pm$ 0.07 ( 4) & $ +0.35$ $\pm$ 0.12 ( 1) & $ -0.02$ $\pm$ 0.12 ( 1) \\
193.1501  & $-0.31$ & 7.19 $\pm$ 0.01 ( 91) & 7.34 $\pm$ 0.03 (20) & $ -0.03$ $\pm$ 0.20 & $ -0.08$ $\pm$ 0.07 ( 2) & $ +0.10$ $\pm$ 0.03 (13) &            --            & $ +0.15$ $\pm$ 0.05 ( 4) \\
193.5542  & $-0.30$ & 7.20 $\pm$ 0.01 ( 72) & 7.19 $\pm$ 0.02 (18) & $ -0.04$ $\pm$ 0.20 & $ +0.03$ $\pm$ 0.05 ( 2) & $ +0.20$ $\pm$ 0.02 (10) & $ +0.31$ $\pm$ 0.07 ( 1) & $ +0.12$ $\pm$ 0.05 ( 4) \\
193.8390  & $-2.91$ & 4.59 $\pm$ 0.01 ( 64) & 4.64 $\pm$ 0.04 (10) & $ +0.00$ $\pm$ 0.25 & $ +0.49$ $\pm$ 0.06 ( 3) & $ +0.25$ $\pm$ 0.04 ( 4) & $ +0.40$ $\pm$ 0.10 ( 1) & $<-2.00$  ~~~~~~~   ( 1) \\
196.3755  & $-2.77$ & 4.73 $\pm$ 0.02 ( 47) & 4.74 $\pm$ 0.04 ( 4) & $ +0.60$ $\pm$ 0.24 & $ +0.35$ $\pm$ 0.04 ( 3) & $ +0.36$ $\pm$ 0.04 ( 3) &            --            & $ -0.63$ $\pm$ 0.05 ( 3) \\
196.4126  & $-0.44$ & 7.06 $\pm$ 0.01 ( 81) & 7.01 $\pm$ 0.02 (18) & $ +0.26$ $\pm$ 0.20 & $ +0.24$ $\pm$ 0.05 ( 3) & $ +0.21$ $\pm$ 0.03 (11) & $ -0.15$ $\pm$ 0.09 ( 1) & $ -0.25$ $\pm$ 0.08 ( 3) \\
196.5323  & $-2.55$ & 4.95 $\pm$ 0.04 ( 24) & 4.74 $\pm$ 0.14 ( 3) &         --          & $ +0.03$ $\pm$ 0.10 ( 3) & $ +0.27$ $\pm$ 0.17 ( 1) & $ +0.50$ $\pm$ 0.17 ( 1) & $<-0.04$  ~~~~~~~   ( 1) \\
196.5453  & $-2.51$ & 4.99 $\pm$ 0.03 ( 18) & 5.08 $\pm$ 0.06 ( 6) & $<+0.84$            & $ +0.43$ $\pm$ 0.10 ( 2) & $ +0.45$ $\pm$ 0.06 ( 5) & $ +0.41$ $\pm$ 0.14 ( 1) & $<-0.45$  ~~~~~~~   ( 1) \\
196.6013  & $-0.69$ & 6.81 $\pm$ 0.01 ( 76) & 6.85 $\pm$ 0.03 (10) & $ +0.16$ $\pm$ 0.20 & $ +0.34$ $\pm$ 0.09 ( 1) & $ +0.12$ $\pm$ 0.03 ( 8) & $ -0.04$ $\pm$ 0.09 ( 1) & $ +0.06$ $\pm$ 0.07 ( 2) \\
197.5045  & $-0.80$ & 6.70 $\pm$ 0.02 (105) & 6.67 $\pm$ 0.03 (17) & $ +0.37$ $\pm$ 0.21 & $ +0.11$ $\pm$ 0.07 ( 2) & $ +0.22$ $\pm$ 0.03 (15) & $ +0.07$ $\pm$ 0.09 ( 1) & $ +0.06$ $\pm$ 0.06 ( 4) \\
197.9861  & $-1.22$ & 6.28 $\pm$ 0.03 ( 72) & 6.30 $\pm$ 0.07 (14) & $<+0.67$            & $ +0.42$ $\pm$ 0.14 ( 1) & $ +0.45$ $\pm$ 0.06 ( 7) &            --            &            --            \\
198.5288  & $-0.59$ & 6.91 $\pm$ 0.02 ( 91) & 6.88 $\pm$ 0.03 (19) & $ +0.29$ $\pm$ 0.23 & $ +0.12$ $\pm$ 0.05 ( 3) & $ +0.00$ $\pm$ 0.03 (12) & $ -0.55$ $\pm$ 0.08 ( 1) & $ +0.31$ $\pm$ 0.05 ( 3) \\
198.5495  & $-2.20$ & 5.30 $\pm$ 0.03 ( 33) & 5.23 $\pm$ 0.04 ( 6) & $<+1.12$            & $ +0.36$ $\pm$ 0.09 ( 2) & $<+0.24$  ~~~~~~~   ( 4) & $ +1.06$ $\pm$ 0.07 ( 2) & $<-0.35$  ~~~~~~~   ( 1) \\
199.9269  & $-0.15$ & 7.35 $\pm$ 0.01 ( 56) & 7.47 $\pm$ 0.02 (19) & $ +0.11$ $\pm$ 0.19 & $ -0.01$ $\pm$ 0.07 ( 1) & $ +0.12$ $\pm$ 0.03 ( 7) &            --            & $ +0.14$ $\pm$ 0.05 ( 3) \\
200.0999  & $-2.48$ & 5.02 $\pm$ 0.04 ( 57) & 5.27 $\pm$ 0.07 ( 8) & $<+0.39$            & $ +0.26$ $\pm$ 0.14 ( 1) & $ +0.32$ $\pm$ 0.09 ( 3) &            --            & $ +0.42$ $\pm$ 0.14 ( 1) \\
200.5298  & $-1.02$ & 6.48 $\pm$ 0.03 ( 98) & 6.49 $\pm$ 0.03 (12) & $ -0.05$ $\pm$ 0.19 & $ +0.42$ $\pm$ 0.08 ( 2) & $ +0.38$ $\pm$ 0.04 (11) &            --            & $ +0.69$ $\pm$ 0.07 ( 3) \\
200.7620  & $-0.15$ & 7.35 $\pm$ 0.02 ( 63) & 7.31 $\pm$ 0.03 (15) & $ +0.14$ $\pm$ 0.20 & $ -0.04$ $\pm$ 0.10 ( 1) & $ +0.13$ $\pm$ 0.04 ( 8) &            --            & $ +0.12$ $\pm$ 0.10 ( 4) \\
201.1159  & $-0.19$ & 7.31 $\pm$ 0.02 ( 63) & 7.19 $\pm$ 0.02 (18) & $ -0.01$ $\pm$ 0.20 & $ +0.06$ $\pm$ 0.09 ( 1) & $ +0.10$ $\pm$ 0.03 (11) & $ -0.02$ $\pm$ 0.09 ( 1) & $ -0.04$ $\pm$ 0.05 ( 4) \\
203.2831  & $-2.70$ & 4.80 $\pm$ 0.03 ( 54) & 4.79 $\pm$ 0.04 ( 9) & $ +0.35$ $\pm$ 0.23 & $ +0.20$ $\pm$ 0.08 ( 4) & $ +0.28$ $\pm$ 0.04 ( 7) &            --            & $ -0.65$ $\pm$ 0.07 ( 3) \\
204.9008  & $-2.66$ & 4.84 $\pm$ 0.12 (  4) & 4.83 $\pm$ 0.17 ( 2) & $<+2.25$            & $ +0.32$ $\pm$ 0.14 ( 3) &            --            & $<-0.49$  ~~~~~~~   ( 1) &            --            \\
205.1342  & $-2.12$ & 5.38 $\pm$ 0.03 ( 66) & 5.56 $\pm$ 0.05 ( 6) & $ -0.01$ $\pm$ 0.25 & $ +0.17$ $\pm$ 0.07 ( 4) & $ +0.30$ $\pm$ 0.05 ( 5) & $ +0.85$ $\pm$ 0.12 ( 1) & $ +1.30$ $\pm$ 0.10 ( 3) \\
205.8132  & $-2.13$ & 5.37 $\pm$ 0.04 ( 19) & 5.39 $\pm$ 0.07 ( 6) & $<+1.43$            & $ +0.43$ $\pm$ 0.11 ( 2) &            --            & $ +1.03$ $\pm$ 0.15 ( 1) & $<+0.13$  ~~~~~~~   ( 1) \\
206.3487  & $-1.80$ & 5.70 $\pm$ 0.02 ( 31) & 5.79 $\pm$ 0.04 ( 9) & $<+1.17$            &            --            & $ +0.43$ $\pm$ 0.04 (10) & $ -0.11$ $\pm$ 0.10 ( 1) & $ -0.38$ $\pm$ 0.08 ( 1) \\
207.9961  & $-0.49$ & 7.01 $\pm$ 0.01 ( 71) & 7.09 $\pm$ 0.02 (17) & $ -0.03$ $\pm$ 0.19 & $ +0.27$ $\pm$ 0.08 ( 1) & $ +0.22$ $\pm$ 0.03 (10) & $ -0.22$ $\pm$ 0.08 ( 1) & $ -0.10$ $\pm$ 0.05 ( 4) \\
208.0799  & $-2.77$ & 4.73 $\pm$ 0.03 ( 29) & 4.80 $\pm$ 0.05 ( 6) & $<+0.59$            & $ +0.26$ $\pm$ 0.07 ( 2) & $<+0.39$  ~~~~~~~   ( 5) &            --            & $<-0.10$  ~~~~~~~   ( 1) \\
209.2123  & $-1.90$ & 5.60 $\pm$ 0.03 ( 71) & 5.56 $\pm$ 0.04 ( 9) & $ +0.29$ $\pm$ 0.25 & $ +0.30$ $\pm$ 0.04 ( 5) & $ +0.39$ $\pm$ 0.04 ( 8) & $ +0.69$ $\pm$ 0.10 ( 1) & $ -0.36$ $\pm$ 0.06 ( 3) \\
209.7189  & $-1.98$ & 5.52 $\pm$ 0.03 ( 22) & 5.56 $\pm$ 0.06 ( 5) & $<+0.72$            & $ +0.06$ $\pm$ 0.10 ( 2) & $ +0.20$ $\pm$ 0.07 ( 4) & $ +0.88$ $\pm$ 0.14 ( 1) & $<-0.18$  ~~~~~~~   ( 3) \\
209.9364  & $-2.25$ & 5.25 $\pm$ 0.03 ( 20) & 5.27 $\pm$ 0.07 ( 6) & $ +2.18$ $\pm$ 0.23 & $ +0.22$ $\pm$ 0.07 ( 3) & $ +0.37$ $\pm$ 0.09 ( 2) & $ +0.62$ $\pm$ 0.12 ( 1) & $<+0.91$  ~~~~~~~   ( 2) \\
210.0175  & $-2.67$ & 4.83 $\pm$ 0.01 ( 46) & 4.95 $\pm$ 0.04 ( 6) & $ +0.54$ $\pm$ 0.22 & $ +0.35$ $\pm$ 0.04 ( 4) & $ +0.32$ $\pm$ 0.03 ( 7) & $ +0.21$ $\pm$ 0.09 ( 1) & $ +0.79$ $\pm$ 0.05 ( 4) \\
210.0316  & $-0.98$ & 6.52 $\pm$ 0.01 ( 93) & 6.46 $\pm$ 0.01 (16) & $ -0.08$ $\pm$ 0.19 & $ +0.38$ $\pm$ 0.04 ( 2) & $ +0.39$ $\pm$ 0.02 (12) & $ +0.04$ $\pm$ 0.07 ( 1) & $ -0.20$ $\pm$ 0.05 ( 4) \\
210.7513  & $-2.12$ & 5.38 $\pm$ 0.02 ( 83) & 5.42 $\pm$ 0.04 (11) & $ -0.30$ $\pm$ 0.22 & $ +0.42$ $\pm$ 0.04 ( 4) & $ +0.35$ $\pm$ 0.03 (12) & $ -1.10$ $\pm$ 0.09 ( 1) & $ +0.49$ $\pm$ 0.05 ( 4) \\
210.8633  & $-1.95$ & 5.55 $\pm$ 0.03 ( 66) & 5.57 $\pm$ 0.04 ( 7) & $<+0.16$            & $ +0.20$ $\pm$ 0.12 ( 1) & $ +0.07$ $\pm$ 0.06 ( 4) & $ +0.46$ $\pm$ 0.12 ( 1) & $ +0.63$ $\pm$ 0.12 ( 1) \\
211.2766  & $-1.39$ & 6.11 $\pm$ 0.02 ( 88) & 6.06 $\pm$ 0.03 (11) & $ +0.08$ $\pm$ 0.23 & $ +0.64$ $\pm$ 0.04 ( 5) & $ +0.27$ $\pm$ 0.03 (14) & $ +0.67$ $\pm$ 0.10 ( 1) & $ +1.12$ $\pm$ 0.05 ( 4) \\
211.7184  & $-2.42$ & 5.08 $\pm$ 0.02 ( 79) & 5.13 $\pm$ 0.04 (11) & $ -0.18$ $\pm$ 0.23 & $ +0.42$ $\pm$ 0.04 ( 4) & $ +0.29$ $\pm$ 0.04 ( 7) & $ -0.58$ $\pm$ 0.17 ( 1) & $ -0.31$ $\pm$ 0.05 ( 4) \\
212.5834  & $-1.79$ & 5.71 $\pm$ 0.03 ( 47) & 5.65 $\pm$ 0.07 ( 2) & $<+0.59$            & $ +0.06$ $\pm$ 0.07 ( 2) & $ +0.24$ $\pm$ 0.04 ( 8) & $ +0.22$ $\pm$ 0.10 ( 1) & $ +0.02$ $\pm$ 0.10 ( 1) \\
213.7878  & $-2.45$ & 5.05 $\pm$ 0.03 ( 54) & 5.04 $\pm$ 0.04 ( 9) & $ +0.73$ $\pm$ 0.23 & $ +0.71$ $\pm$ 0.06 ( 4) & $ +0.38$ $\pm$ 0.04 ( 6) & $ +0.43$ $\pm$ 0.10 ( 1) & $ -0.69$ $\pm$ 0.10 ( 3) \\

\hline
\end{tabular}
\end{table*}

\setlength{\tabcolsep}{4pt}
\begin{table*}
\contcaption{}
\label{tab:abundances1}

\begin{tabular}{lllllllll}
  
\hline\hline\Tstrut
Star & [Fe/H] & log(\ion{Fe}{I}) $\pm$ $\sigma$ (N) & log(\ion{Fe}{II}) $\pm$ $\sigma$ (N) & [C/Fe] $\pm$ $\sigma$ & [Mg/Fe] $\pm$ $\sigma$ (N) & [Ca/Fe] $\pm$ $\sigma$ (N) & [Sr/Fe] $\pm$ $\sigma$ (N) & [Ba/Fe] $\pm$ $\sigma$ (N) \\
\hline\Tstrut
214.5557  & $-2.14$ & 5.36 $\pm$ 0.04 ( 15) & 5.31 $\pm$ 0.07 ( 3) & $ +2.02$ $\pm$ 0.30 & $ +0.46$ $\pm$ 0.07 ( 2) &            --            &            --            & $ +1.90$ $\pm$ 0.07 ( 4) \\
215.6129  & $-1.92$ & 5.58 $\pm$ 0.01 ( 86) & 5.58 $\pm$ 0.03 (11) & $ +0.15$ $\pm$ 0.21 & $ +0.33$ $\pm$ 0.04 ( 4) & $ +0.34$ $\pm$ 0.02 (20) & $ +0.57$ $\pm$ 0.09 ( 1) & $ -0.05$ $\pm$ 0.05 ( 4) \\
215.6783  & $-0.35$ & 7.15 $\pm$ 0.02 ( 74) & 7.22 $\pm$ 0.02 (19) & $ -0.01$ $\pm$ 0.20 & $ +0.16$ $\pm$ 0.07 ( 2) & $ +0.10$ $\pm$ 0.03 (11) & $ -0.42$ $\pm$ 0.09 ( 1) & $ +0.08$ $\pm$ 0.05 ( 3) \\
216.1245  & $-2.21$ & 5.29 $\pm$ 0.03 ( 64) & 5.24 $\pm$ 0.04 ( 9) & $ +0.82$ $\pm$ 0.23 & $ +0.28$ $\pm$ 0.07 ( 2) & $ +0.34$ $\pm$ 0.07 ( 2) &            --            & $ +0.04$ $\pm$ 0.06 ( 3) \\
217.3862  & $-1.97$ & 5.53 $\pm$ 0.03 ( 65) & 5.58 $\pm$ 0.04 ( 9) & $ +0.51$ $\pm$ 0.23 & $ +0.41$ $\pm$ 0.05 ( 4) & $ +0.41$ $\pm$ 0.03 (11) & $ +0.19$ $\pm$ 0.10 ( 1) & $ -0.51$ $\pm$ 0.06 ( 3) \\
217.5786  & $-2.66$ & 4.84 $\pm$ 0.01 ( 57) & 4.99 $\pm$ 0.04 ( 9) & $ +0.02$ $\pm$ 0.25 & $ +0.47$ $\pm$ 0.06 ( 3) & $ +0.38$ $\pm$ 0.04 ( 4) & $ +1.50$ $\pm$ 0.09 ( 1) & $ +0.35$ $\pm$ 0.06 ( 3) \\
217.6444  & $-1.82$ & 5.68 $\pm$ 0.03 ( 24) &        --            & $<+1.27$            & $ +0.20$ $\pm$ 0.08 ( 3) &            --            &            --            & $ +0.63$ $\pm$ 0.07 ( 1) \\
218.4256  & $-0.60$ & 6.90 $\pm$ 0.01 ( 91) & 6.91 $\pm$ 0.03 (17) & $ +0.02$ $\pm$ 0.19 & $ +0.25$ $\pm$ 0.07 ( 2) & $ +0.41$ $\pm$ 0.03 (11) & $ -0.01$ $\pm$ 0.17 ( 2) & $ +0.01$ $\pm$ 0.06 ( 3) \\
218.4622  & $-2.40$ & 5.10 $\pm$ 0.03 ( 24) & 5.15 $\pm$ 0.09 ( 9) & $<+0.80$            &            --            & $ +0.39$ $\pm$ 0.14 ( 2) & $ +1.00$ $\pm$ 0.12 ( 1) &            --            \\
218.4977  & $-0.16$ & 7.34 $\pm$ 0.01 ( 50) & 7.35 $\pm$ 0.03 (18) & $ +0.12$ $\pm$ 0.19 & $ -0.02$ $\pm$ 0.07 ( 1) & $ +0.05$ $\pm$ 0.03 ( 7) & $ -0.38$ $\pm$ 0.07 ( 1) & $ -0.02$ $\pm$ 0.09 ( 3) \\
223.5283  & $-2.30$ & 5.20 $\pm$ 0.01 ( 89) & 5.32 $\pm$ 0.03 (10) & $ -0.39$ $\pm$ 0.21 & $ +0.58$ $\pm$ 0.04 ( 4) & $ +0.44$ $\pm$ 0.03 (10) & $ +0.74$ $\pm$ 0.09 ( 1) & $ -0.14$ $\pm$ 0.05 ( 4) \\
227.2895  & $-0.35$ & 7.15 $\pm$ 0.02 ( 92) & 7.15 $\pm$ 0.02 (23) & $ +0.13$ $\pm$ 0.23 & $ +0.20$ $\pm$ 0.10 ( 2) & $ +0.29$ $\pm$ 0.03 (11) & $ -0.04$ $\pm$ 0.09 ( 1) & $ +0.21$ $\pm$ 0.05 ( 3) \\
228.4607  & $-2.20$ & 5.30 $\pm$ 0.04 ( 17) & 5.15 $\pm$ 0.09 ( 3) & $<+1.30$            & $ +0.13$ $\pm$ 0.09 ( 3) &            --            & $ +0.34$ $\pm$ 0.14 ( 1) & $<-0.52$  ~~~~~~~   ( 1) \\
228.6558  & $-2.26$ & 5.24 $\pm$ 0.03 ( 17) & 5.18 $\pm$ 0.06 ( 4) & $ +1.84$ $\pm$ 0.25 & $ +0.42$ $\pm$ 0.12 ( 1) & $ +0.28$ $\pm$ 0.07 ( 3) & $ -0.21$ $\pm$ 0.12 ( 1) &            --            \\
228.8159  & $-2.02$ & 5.48 $\pm$ 0.02 ( 36) & 5.55 $\pm$ 0.05 ( 4) & $ +1.39$ $\pm$ 0.25 & $ +0.18$ $\pm$ 0.04 ( 4) & $ +0.37$ $\pm$ 0.04 ( 5) & $ +1.07$ $\pm$ 0.10 ( 1) & $<-0.29$  ~~~~~~~   ( 1) \\
229.0409  & $-0.10$ & 7.40 $\pm$ 0.03 ( 50) & 7.47 $\pm$ 0.04 (15) & $ -0.05$ $\pm$ 0.22 & $ +0.04$ $\pm$ 0.07 ( 2) &            --            & $ -0.54$ $\pm$ 0.10 ( 1) & $ +0.29$ $\pm$ 0.06 ( 3) \\
229.1219  & $-2.25$ & 5.25 $\pm$ 0.03 ( 19) & 5.26 $\pm$ 0.06 ( 6) & $ +1.74$ $\pm$ 0.25 & $ +0.16$ $\pm$ 0.12 ( 1) & $ +0.22$ $\pm$ 0.07 ( 3) & $ +0.92$ $\pm$ 0.12 ( 1) &            --            \\
229.8911  & $+0.10$ & 7.60 $\pm$ 0.02 ( 62) & 7.59 $\pm$ 0.03 (20) & $ +0.11$ $\pm$ 0.22 & $ +0.04$ $\pm$ 0.09 ( 1) & $ +0.03$ $\pm$ 0.03 (10) &            --            & $ -0.01$ $\pm$ 0.05 ( 3) \\
230.4663  & $-1.15$ & 6.35 $\pm$ 0.02 (102) & 6.35 $\pm$ 0.04 (13) & $ +0.13$ $\pm$ 0.23 & $ +0.49$ $\pm$ 0.06 ( 4) & $ +0.38$ $\pm$ 0.03 (12) & $ +0.45$ $\pm$ 0.10 ( 1) & $ +0.13$ $\pm$ 0.06 ( 3) \\
231.0318  & $-0.35$ & 7.15 $\pm$ 0.02 ( 69) & 7.19 $\pm$ 0.03 (22) & $ -0.16$ $\pm$ 0.19 & $ +0.10$ $\pm$ 0.06 ( 2) & $ +0.17$ $\pm$ 0.03 ( 8) & $ -0.39$ $\pm$ 0.09 ( 1) & $ -0.06$ $\pm$ 0.08 ( 3) \\
232.6956  & $-2.22$ & 5.28 $\pm$ 0.03 ( 63) & 5.25 $\pm$ 0.04 ( 8) & $ +0.65$ $\pm$ 0.23 & $ +0.22$ $\pm$ 0.06 ( 3) & $ +0.26$ $\pm$ 0.10 ( 6) & $ -0.11$ $\pm$ 0.13 ( 1) & $ +0.15$ $\pm$ 0.07 ( 2) \\
232.8039  & $-2.26$ & 5.24 $\pm$ 0.03 ( 54) & 5.21 $\pm$ 0.04 ( 7) & $ +0.56$ $\pm$ 0.22 &            --            & $ +0.15$ $\pm$ 0.06 ( 6) & $ +0.15$ $\pm$ 0.10 ( 1) & $ -1.17$ $\pm$ 0.07 ( 2) \\
233.5730  & $-2.74$ & 4.76 $\pm$ 0.02 ( 53) & 4.76 $\pm$ 0.04 ( 7) & $ +0.74$ $\pm$ 0.25 & $ +0.39$ $\pm$ 0.06 ( 2) & $ +0.26$ $\pm$ 0.04 ( 5) & $ +0.31$ $\pm$ 0.09 ( 1) & $ -0.48$ $\pm$ 0.06 ( 2) \\
233.9312  & $-2.20$ & 5.30 $\pm$ 0.02 ( 50) & 5.22 $\pm$ 0.03 ( 7) & $ +0.53$ $\pm$ 0.22 & $ +0.33$ $\pm$ 0.04 ( 3) & $ +0.37$ $\pm$ 0.03 ( 5) &            --            & $ -0.23$ $\pm$ 0.06 ( 3) \\
234.4403  & $-0.40$ & 7.10 $\pm$ 0.02 ( 72) & 7.09 $\pm$ 0.04 (10) & $ +0.25$ $\pm$ 0.22 & $ +0.33$ $\pm$ 0.10 ( 2) & $ +0.32$ $\pm$ 0.04 ( 9) &            --            & $ +0.10$ $\pm$ 0.06 ( 4) \\
235.1448  & $-2.54$ & 4.96 $\pm$ 0.04 ( 22) & 4.72 $\pm$ 0.05 ( 4) & $<+1.29$            & $ +0.55$ $\pm$ 0.07 ( 2) & $ +0.24$ $\pm$ 0.07 ( 4) & $ +0.72$ $\pm$ 0.10 ( 1) & $ +0.01$ $\pm$ 0.10 ( 1) \\
235.7578  & $-1.81$ & 5.69 $\pm$ 0.03 ( 21) & 5.74 $\pm$ 0.04 ( 6) & $ +1.38$ $\pm$ 0.25 & $ +0.12$ $\pm$ 0.07 ( 4) & $ +0.32$ $\pm$ 0.03 ( 8) & $ +0.99$ $\pm$ 0.09 ( 1) & $<-0.42$  ~~~~~~~   ( 2) \\
235.9710  & $-1.83$ & 5.67 $\pm$ 0.03 ( 40) & 5.67 $\pm$ 0.03 ( 9) & $ +1.06$ $\pm$ 0.23 & $ +0.35$ $\pm$ 0.05 ( 4) & $ +0.33$ $\pm$ 0.04 ( 6) & $ +0.56$ $\pm$ 0.10 ( 1) & $ -0.09$ $\pm$ 0.07 ( 2) \\
236.1077  & $-2.55$ & 4.95 $\pm$ 0.01 ( 77) & 4.95 $\pm$ 0.04 ( 9) & $ -0.22$ $\pm$ 0.23 & $ +0.42$ $\pm$ 0.04 ( 4) & $ +0.27$ $\pm$ 0.03 ( 7) & $ -0.07$ $\pm$ 0.10 ( 1) & $ +0.09$ $\pm$ 0.05 ( 3) \\
236.4855  & $-0.27$ & 7.23 $\pm$ 0.01 ( 84) & 7.27 $\pm$ 0.02 (18) & $ -0.07$ $\pm$ 0.22 & $ -0.08$ $\pm$ 0.05 ( 3) & $ +0.07$ $\pm$ 0.03 (12) & $ +0.39$ $\pm$ 0.09 ( 1) & $ +0.09$ $\pm$ 0.07 ( 3) \\
236.7138  & $-0.85$ & 6.65 $\pm$ 0.02 ( 92) & 6.70 $\pm$ 0.03 (19) & $ -0.18$ $\pm$ 0.20 & $ +0.34$ $\pm$ 0.06 ( 2) & $ +0.32$ $\pm$ 0.02 (14) &            --            & $ -0.10$ $\pm$ 0.10 ( 3) \\
237.8246  & $-3.23$ & 4.27 $\pm$ 0.04 ( 16) & 4.16 $\pm$ 0.07 ( 3) & $<+0.58$            & $ +0.66$ $\pm$ 0.07 ( 3) & $ +0.34$ $\pm$ 0.07 ( 3) & $ +0.05$ $\pm$ 0.13 ( 1) & $ +0.37$ $\pm$ 0.07 ( 3) \\
237.8353  & $-2.32$ & 5.18 $\pm$ 0.02 ( 58) & 5.18 $\pm$ 0.04 ( 8) & $ +0.35$ $\pm$ 0.23 & $ +0.35$ $\pm$ 0.05 ( 2) & $ +0.30$ $\pm$ 0.03 (10) & $ -0.09$ $\pm$ 0.12 ( 1) & $ -0.07$ $\pm$ 0.05 ( 3) \\
237.9609  & $-1.90$ & 5.60 $\pm$ 0.03 ( 46) & 5.59 $\pm$ 0.07 ( 8) & $<+0.76$            & $ +0.26$ $\pm$ 0.07 ( 3) & $ +0.26$ $\pm$ 0.07 ( 3) & $ +0.28$ $\pm$ 0.13 ( 1) & $ -0.08$ $\pm$ 0.13 ( 1) \\
238.7217  & $-2.06$ & 5.44 $\pm$ 0.03 ( 19) & 5.15 $\pm$ 0.09 ( 5) & $<+0.59$            & $ -0.21$ $\pm$ 0.07 ( 2) & $ +0.01$ $\pm$ 0.10 ( 1) &            --            & $<-0.32$  ~~~~~~~   ( 1) \\
240.0348  & $-2.30$ & 5.20 $\pm$ 0.02 ( 85) & 5.27 $\pm$ 0.03 (12) & $ +0.12$ $\pm$ 0.22 & $ +0.43$ $\pm$ 0.04 ( 5) & $ +0.39$ $\pm$ 0.03 ( 6) & $ +0.46$ $\pm$ 0.07 ( 1) & $ -0.12$ $\pm$ 0.05 ( 4) \\
240.4216  & $-2.98$ & 4.52 $\pm$ 0.02 ( 27) & 4.53 $\pm$ 0.05 ( 5) & $ +0.74$ $\pm$ 0.25 & $ +0.27$ $\pm$ 0.06 ( 3) & $ +0.30$ $\pm$ 0.04 ( 4) & $ +0.77$ $\pm$ 0.08 ( 1) & $ +0.84$ $\pm$ 0.07 ( 4) \\
241.1186  & $-1.92$ & 5.58 $\pm$ 0.03 ( 32) & 5.59 $\pm$ 0.06 ( 5) & $<+1.14$            & $ +0.13$ $\pm$ 0.12 ( 1) & $ +0.24$ $\pm$ 0.08 ( 2) & $ +0.16$ $\pm$ 0.12 ( 1) & $ -0.48$ $\pm$ 0.12 ( 1) \\
241.7900  & $-2.51$ & 4.99 $\pm$ 0.04 ( 15) & 4.65 $\pm$ 0.20 ( 3) & $<+2.38$            & $ +0.30$ $\pm$ 0.12 ( 2) & $ +0.33$ $\pm$ 0.07 ( 6) & $ +0.12$ $\pm$ 0.18 ( 1) & $<+0.13$  ~~~~~~~   ( 1) \\
242.3556  & $-1.95$ & 5.55 $\pm$ 0.03 ( 22) & 5.43 $\pm$ 0.07 ( 5) & $<+1.13$            & $ +0.01$ $\pm$ 0.15 ( 1) & $<+0.17$  ~~~~~~~   ( 4) & $ -0.08$ $\pm$ 0.15 ( 1) & $<-0.82$  ~~~~~~~   ( 2) \\
243.8390  & $-1.95$ & 5.55 $\pm$ 0.03 ( 52) & 5.49 $\pm$ 0.06 (10) & $<+0.15$            & $ +0.43$ $\pm$ 0.09 ( 3) & $ +0.39$ $\pm$ 0.04 ( 5) & $ +0.72$ $\pm$ 0.09 ( 1) & $ +0.39$ $\pm$ 0.09 ( 1) \\
244.4872  & $-2.10$ & 5.40 $\pm$ 0.07 ( 10) & 5.34 $\pm$ 0.10 ( 5) & $ +0.83$ $\pm$ 0.30 &            --            &            --            &            --            & $<+0.03$  ~~~~~~~   ( 2) \\
245.1096  & $-0.35$ & 7.15 $\pm$ 0.01 ( 69) & 7.14 $\pm$ 0.03 (15) & $ -0.02$ $\pm$ 0.20 & $ +0.06$ $\pm$ 0.06 ( 2) & $ +0.16$ $\pm$ 0.03 (10) & $ +0.21$ $\pm$ 0.07 ( 1) & $ -0.06$ $\pm$ 0.05 ( 4) \\
245.4387  & $-1.67$ & 5.83 $\pm$ 0.03 ( 44) & 5.74 $\pm$ 0.04 ( 9) & $<+0.94$            & $ +0.08$ $\pm$ 0.12 ( 3) & $ +0.26$ $\pm$ 0.03 ( 9) & $ +1.05$ $\pm$ 0.07 ( 1) & $ -0.32$ $\pm$ 0.07 ( 2) \\
245.5747  & $-3.17$ & 4.33 $\pm$ 0.03 ( 10) & 4.17 $\pm$ 0.07 ( 2) & $ +1.26$ $\pm$ 0.25 & $ +0.45$ $\pm$ 0.06 ( 2) &            --            & $ -0.11$ $\pm$ 0.07 ( 1) &            --            \\
245.8364  & $-3.06$ & 4.44 $\pm$ 0.02 ( 26) & 4.49 $\pm$ 0.06 ( 3) & $ +0.63$ $\pm$ 0.23 & $ +0.58$ $\pm$ 0.05 ( 4) & $ +0.44$ $\pm$ 0.05 ( 4) & $ +0.38$ $\pm$ 0.10 ( 1) & $ -0.50$ $\pm$ 0.07 ( 2) \\
246.8588  & $-2.25$ & 5.25 $\pm$ 0.02 ( 78) & 5.27 $\pm$ 0.04 (10) & $ -0.18$ $\pm$ 0.23 & $ +0.44$ $\pm$ 0.06 ( 3) & $ +0.42$ $\pm$ 0.04 ( 6) & $ +0.13$ $\pm$ 0.10 ( 1) & $ -0.03$ $\pm$ 0.05 ( 4) \\
248.4394  & $-1.72$ & 5.78 $\pm$ 0.01 ( 51) & 5.75 $\pm$ 0.04 ( 8) & $ +0.61$ $\pm$ 0.22 & $ +0.23$ $\pm$ 0.11 ( 3) & $ +0.35$ $\pm$ 0.03 ( 7) & $ +0.53$ $\pm$ 0.06 ( 1) & $ +0.11$ $\pm$ 0.05 ( 3) \\
248.4959  & $-2.63$ & 4.87 $\pm$ 0.02 ( 58) & 4.88 $\pm$ 0.06 ( 6) & $ +0.02$ $\pm$ 0.25 & $ +0.33$ $\pm$ 0.05 ( 3) & $ +0.30$ $\pm$ 0.03 ( 7) & $ -0.31$ $\pm$ 0.09 ( 1) & $ -0.31$ $\pm$ 0.05 ( 3) \\
248.5263  & $-2.07$ & 5.43 $\pm$ 0.03 ( 49) & 5.36 $\pm$ 0.04 ( 4) & $ +0.25$ $\pm$ 0.30 & $ +0.47$ $\pm$ 0.05 ( 3) & $ +0.30$ $\pm$ 0.04 ( 4) & $ +0.24$ $\pm$ 0.09 ( 1) & $<-0.09$  ~~~~~~~   ( 4) \\
250.6971  & $-2.66$ & 4.84 $\pm$ 0.03 ( 58) & 4.99 $\pm$ 0.04 ( 8) & $<-0.01$            & $ +0.36$ $\pm$ 0.09 ( 3) & $ +0.23$ $\pm$ 0.06 ( 2) & $ +0.84$ $\pm$ 0.09 ( 1) & $ -0.25$ $\pm$ 0.05 ( 3) \\
250.8797  & $+0.05$ & 7.55 $\pm$ 0.01 ( 65) & 7.55 $\pm$ 0.02 (18) & $ -0.06$ $\pm$ 0.20 & $ -0.08$ $\pm$ 0.06 ( 2) & $ +0.05$ $\pm$ 0.03 ( 8) & $ -0.19$ $\pm$ 0.09 ( 1) & $ -0.09$ $\pm$ 0.05 ( 3) \\
251.4082  & $-3.22$ & 4.28 $\pm$ 0.02 ( 31) & 4.20 $\pm$ 0.05 ( 4) & $ +0.58$ $\pm$ 0.25 & $ +0.14$ $\pm$ 0.07 ( 2) & $ +0.10$ $\pm$ 0.10 ( 1) & $<+1.20$  ~~~~~~~   ( 1) & $ -0.68$ $\pm$ 0.06 ( 3) \\
252.1648  & $-2.43$ & 5.07 $\pm$ 0.01 ( 69) & 5.18 $\pm$ 0.05 ( 3) & $ -0.06$ $\pm$ 0.23 & $ +0.31$ $\pm$ 0.04 ( 4) & $ +0.29$ $\pm$ 0.04 ( 7) & $ -0.73$ $\pm$ 0.09 ( 1) & $ -0.72$ $\pm$ 0.05 ( 4) \\
252.4208  & $-1.30$ & 6.20 $\pm$ 0.01 ( 96) & 6.24 $\pm$ 0.03 (16) & $ +0.02$ $\pm$ 0.22 & $ +0.56$ $\pm$ 0.06 ( 4) & $ +0.35$ $\pm$ 0.03 (13) & $ +0.03$ $\pm$ 0.09 ( 1) & $ +0.05$ $\pm$ 0.05 ( 4) \\
252.4917  & $-0.60$ & 6.90 $\pm$ 0.01 ( 86) & 6.99 $\pm$ 0.02 (14) & $ -0.07$ $\pm$ 0.24 & $ +0.22$ $\pm$ 0.04 ( 4) & $ +0.10$ $\pm$ 0.02 (14) & $ +0.17$ $\pm$ 0.07 ( 1) & $ -0.20$ $\pm$ 0.05 ( 4) \\
252.6179  & $-2.52$ & 4.98 $\pm$ 0.01 ( 72) & 5.07 $\pm$ 0.03 ( 8) & $ +0.34$ $\pm$ 0.22 & $ +0.57$ $\pm$ 0.04 ( 4) & $ +0.47$ $\pm$ 0.03 ( 9) & $ +0.69$ $\pm$ 0.07 ( 1) & $ -0.09$ $\pm$ 0.05 ( 4) \\
253.8582  & $-2.58$ & 4.92 $\pm$ 0.03 ( 44) & 4.94 $\pm$ 0.04 ( 5) & $<+0.55$            & $ +0.47$ $\pm$ 0.06 ( 3) & $ +0.17$ $\pm$ 0.04 ( 5) & $<-1.16$  ~~~~~~~   ( 1) & $ -0.31$ $\pm$ 0.06 ( 3) \\
254.0662  & $-0.56$ & 6.94 $\pm$ 0.02 ( 92) & 7.01 $\pm$ 0.02 (21) & $ +0.03$ $\pm$ 0.19 & $ +0.28$ $\pm$ 0.04 ( 4) & $ +0.27$ $\pm$ 0.02 (13) & $ -0.02$ $\pm$ 0.07 ( 1) & $ +0.02$ $\pm$ 0.05 ( 4) \\
254.5215  & $-0.09$ & 7.41 $\pm$ 0.01 ( 66) & 7.41 $\pm$ 0.01 (19) & $ +0.07$ $\pm$ 0.18 & $ -0.07$ $\pm$ 0.05 ( 2) & $ +0.09$ $\pm$ 0.02 (10) & $ -0.36$ $\pm$ 0.07 ( 1) & $ +0.12$ $\pm$ 0.06 ( 4) \\
254.5478  & $-2.15$ & 5.35 $\pm$ 0.02 ( 64) & 5.33 $\pm$ 0.04 ( 9) & $ +0.22$ $\pm$ 0.21 & $ +0.23$ $\pm$ 0.04 ( 4) & $ +0.26$ $\pm$ 0.04 ( 6) & $ +0.12$ $\pm$ 0.09 ( 1) & $ +0.23$ $\pm$ 0.05 ( 4) \\
254.7768  & $-0.35$ & 7.15 $\pm$ 0.01 ( 75) & 7.12 $\pm$ 0.01 (20) & $ +0.14$ $\pm$ 0.20 & $ +0.07$ $\pm$ 0.05 ( 2) & $ +0.18$ $\pm$ 0.02 (11) & $ -0.09$ $\pm$ 0.07 ( 1) & $ -0.13$ $\pm$ 0.05 ( 3) \\
255.2679  & $-2.09$ & 5.41 $\pm$ 0.03 ( 33) & 5.34 $\pm$ 0.07 ( 7) & $<+1.42$            & $ +0.11$ $\pm$ 0.08 ( 3) & $ +0.17$ $\pm$ 0.08 ( 3) &            --            & $<-0.49$  ~~~~~~~   ( 1) \\
255.5564  & $-2.55$ & 4.95 $\pm$ 0.03 ( 33) & 5.12 $\pm$ 0.06 ( 4) & $ +0.56$ $\pm$ 0.25 & $ +0.61$ $\pm$ 0.07 ( 1) &            --            & $ +0.50$ $\pm$ 0.07 ( 1) & $ +0.59$ $\pm$ 0.06 ( 2) \\
255.8043  & $-2.99$ & 4.51 $\pm$ 0.05 (  9) & 4.58 $\pm$ 0.09 ( 3) & $ +1.23$ $\pm$ 0.30 & $ +0.63$ $\pm$ 0.09 ( 3) & $ +0.74$ $\pm$ 0.09 ( 3) & $ -0.54$ $\pm$ 0.14 ( 1) & $<-0.45$  ~~~~~~~   ( 2) \\

\hline

\end{tabular}
\end{table*}


\subsection{Specific comments on individual abundances}  \label{Sec:abundances}

\begin{figure}
    \centering
    \includegraphics[width=1\columnwidth]{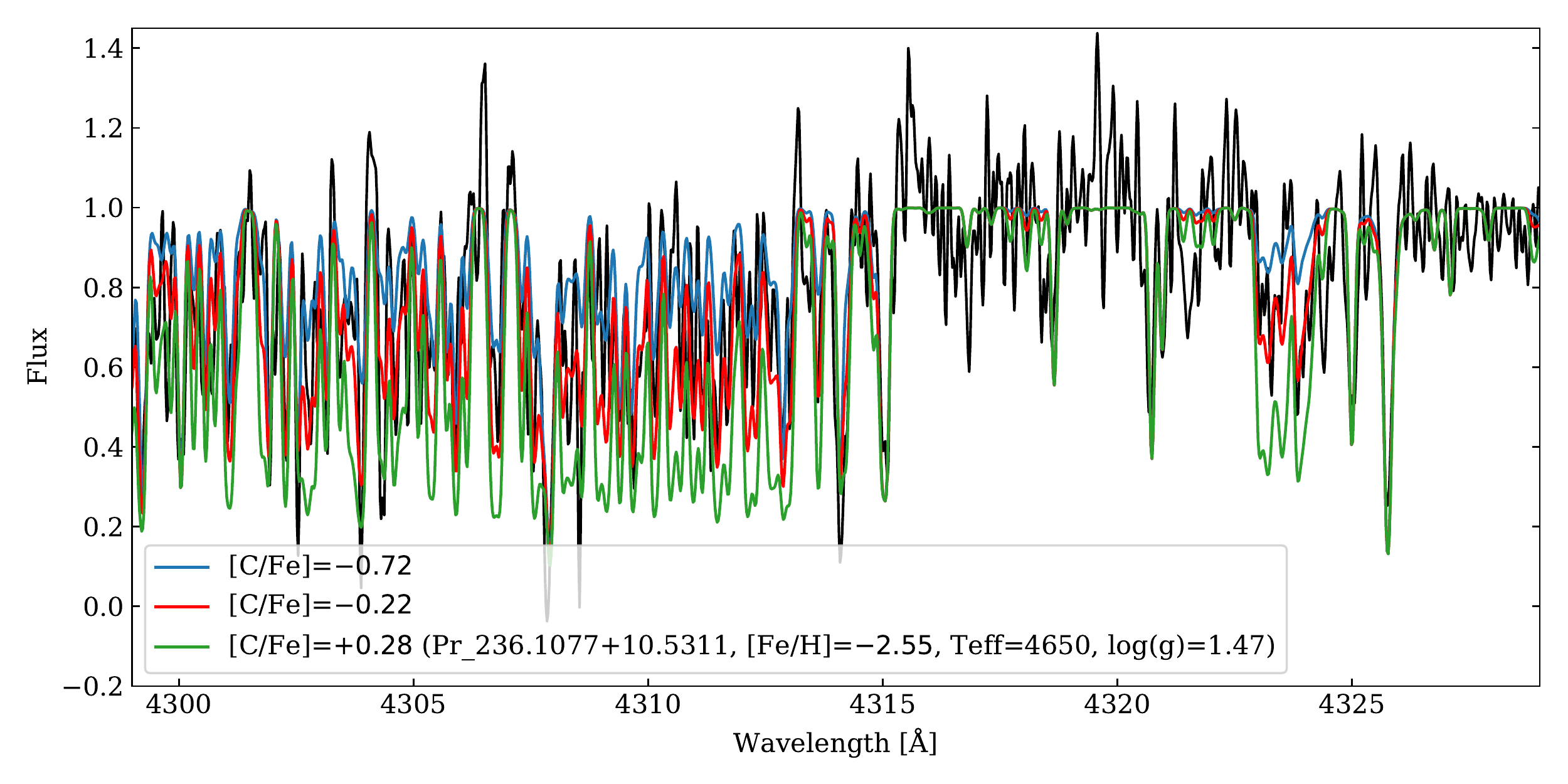}
    \includegraphics[width=1\columnwidth]{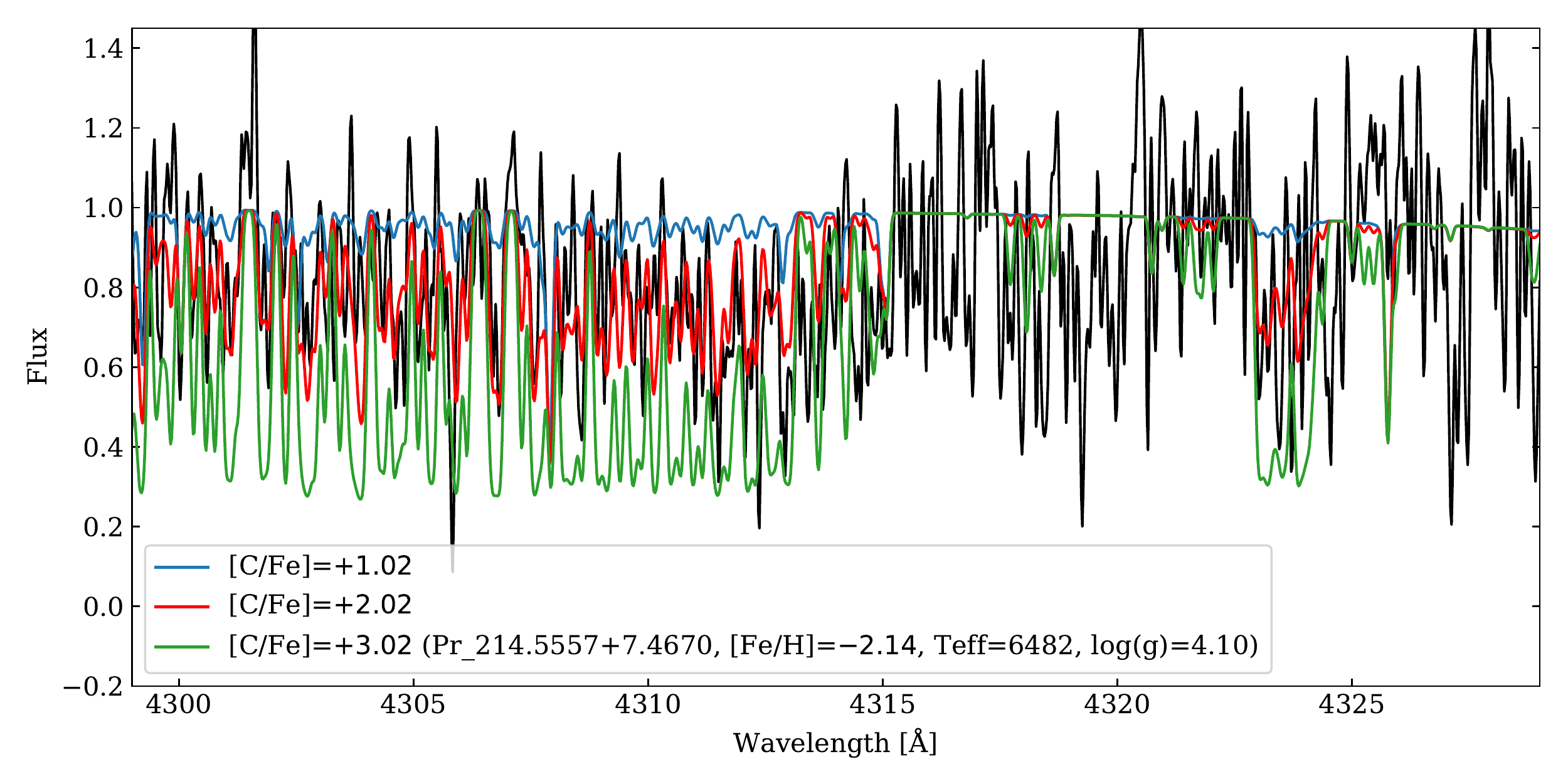}
    \includegraphics[width=1\columnwidth]{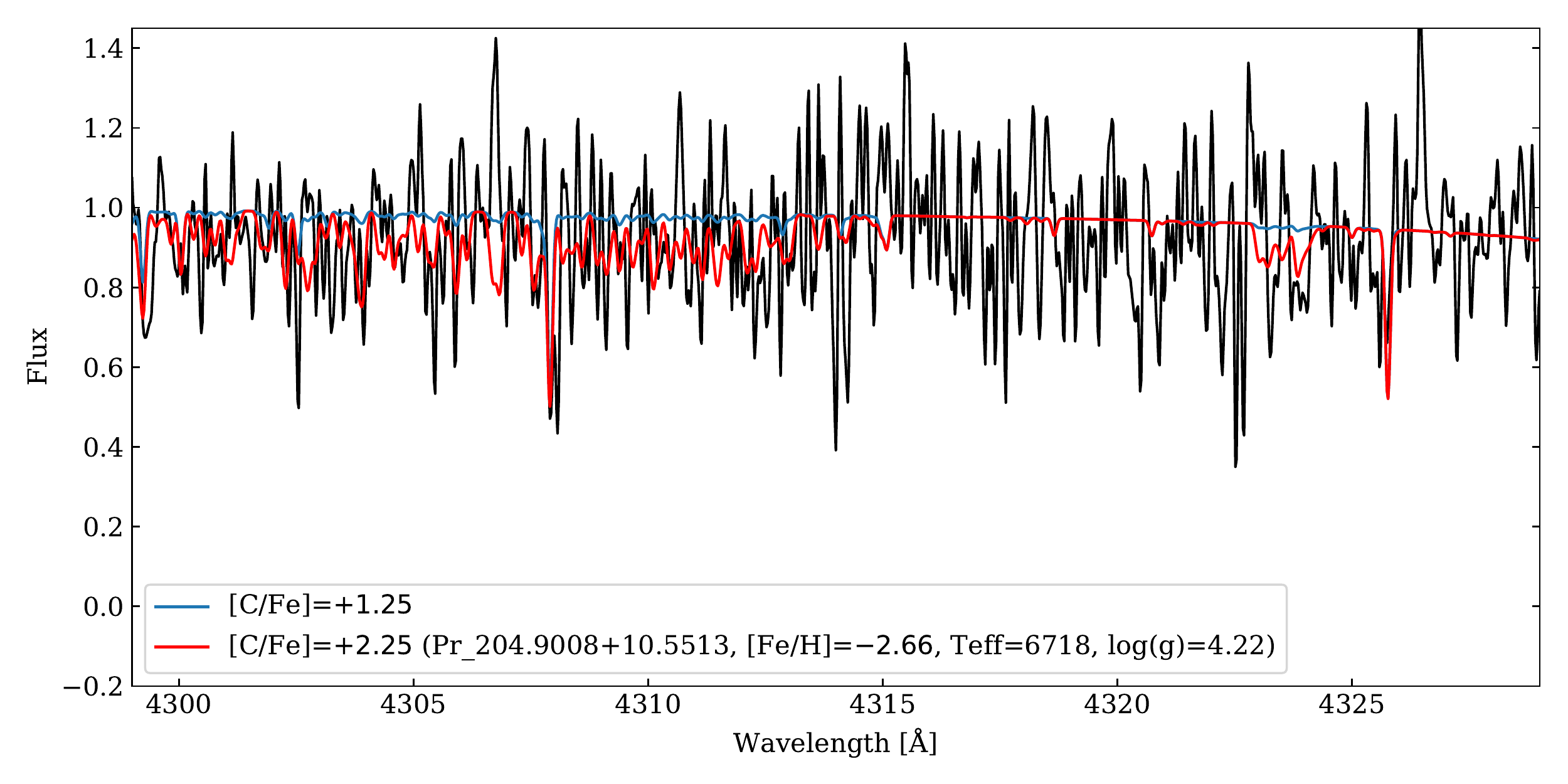}
    \caption{Examples of observed and synthetic spectra in the region of the CH molecular absorption band {\bf Top panel:} The cool (4650~K) RGB star Pr\_236.1077+10.5311 is presented. The red line shows the best synthetic spectrum at [C/Fe]~=~$-$0.22. The blue and green lines correspond to the synthetic spectrum when the carbon abundance is decreased and increased by 0.5~dex ([C/Fe]~=~$-0.72$ and +0.28) respectively. The difference is clearly seen, especially in the $\lambda$4323~\AA\ region, illustrating that the determination of the carbon abundance is robust in cool giants. \\
    {\bf Middle panel:} The warm ($\sim$6500~K) turn-off star Pr\_214.5557+7.4670 is presented. At these very high T$_{\mathrm{eff}}$, the CH feature is very weak, only high carbon-enhancement can be measured. The red line represents the best synthetic spectrum with [C/Fe]~=~+2.02, the blue and green color shows how the CH feature change when the C abundance is decreased and increased by 1.0~dex ([C/Fe]~=~+1.02 and +3.02) respectively.
    {\bf Bottom panel:} The warm ($\sim$6700~K) turn-off star Pr\_204.9008+10.5513 is presented. 
    The carbon upper limit is represented by the red synthetic spectrum with [C/Fe]~=~+2.25. 
    The blue line corresponds to the synthetic spectrum when the carbon abundance is decreased by 1.0~dex ([C/Fe]~=~+1.25). Both synthetic spectra are at the level of the noise, illustrating that only upper limits at a high [C/Fe] level can be placed for the warmest and most metal-poor stars of the sample.}
    \label{Fig:CH-band}
\end{figure}

The abundance of carbon was derived by spectral synthesis of the CH
absorption at 4300~\AA, assuming [O/Fe]~=~[Mg/Fe] and [N/Fe]~=~0 to take into account that some of C can be locked into the CO and CN molecules. While the $\chi^2$ minimization was performed in the 4309-4315 \AA\ window, the continuum was estimated from a wider 60~\AA\  region around the molecular band. The resulting C abundance was finally checked against the 4323~\AA\ absorption band and only concordant band strengths were considered as robust measurements. Fig. \ref{Fig:CH-band} provides three examples of our synthesis, corresponding to actual measurements for a cool giant and a hot dwarf stars, and finally the case of an upper limit for a hot star.

\begin{figure*}
    \centering
    \includegraphics[width=0.98\textwidth]{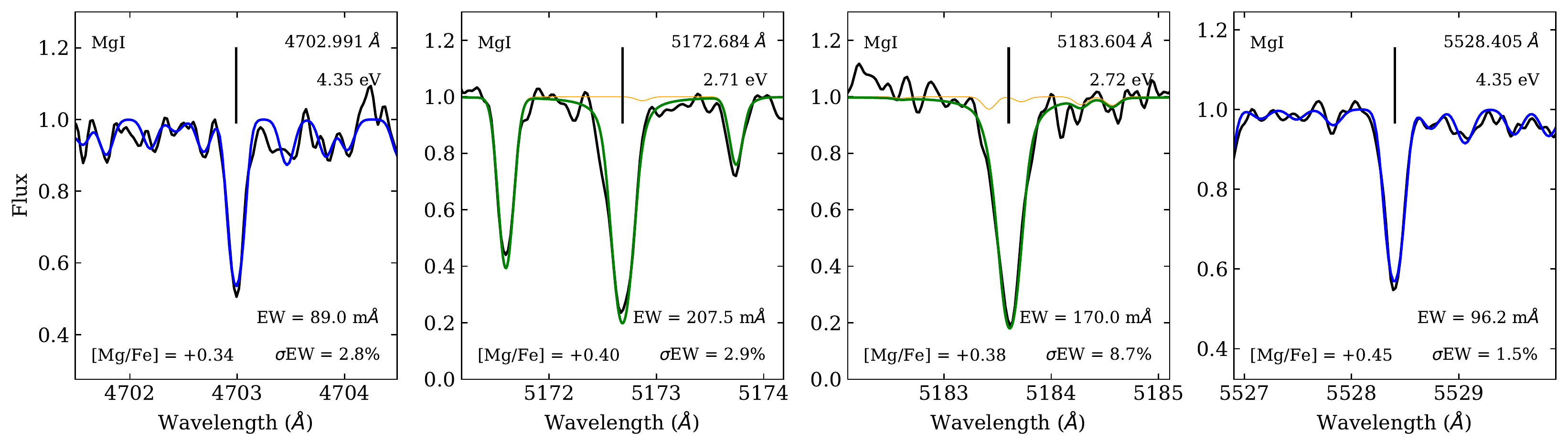}
    \caption{Comparison for star Pr\_211.7184+15.5516 of the derived abundances from the 4 Mg lines available in the spectrum. Abundances from Mg~4702.991~\AA\ and 5528.405~\AA\ were obtained through their EWs (blue colour), abundances of the stronger Mg~5172.684~\AA\ and 5183.604~\AA\ lines were obtained by spectral synthesis (green colour). EW measurements of the latest two are given as indication only and not used for the abundance determination. Abundances are obtained by spectral synthesis, they }are very consistent, and they well agree with the EW results with a standard deviation of $\sigma$~=~0.05~dex. Orange lines are synthetic spectra without Mg, allowing the identification of blends.
    \label{Fig:Mg_comp}
\end{figure*}

The abundance of magnesium was obtained from 2 to 5 \ion{Mg}{I} lines
at 4571.096~\AA, 4702.991~\AA, 5172.684~\AA, 5183.604~\AA, and
5711.088~\AA. Spectral synthesis was performed on the strong \ion{Mg}{I} lines at 5172.684~\AA\ and 5183.604~\AA\ with
equivalent widths larger than 110~m\AA.  Figure~\ref{Fig:Mg_comp}
illustrates the case of Pr\_211.7184+15.5516, for which both strong and weaker
lines are available. The results from the four Mg lines are consistent, with a
mean [Mg/Fe]~=~+0.39 and a standard deviation of $\sigma$~=~0.05.

The abundance of \ion{Ca}{I} was obtained from 2 to 20 lines, with equivalent
widths between 25 and 100~m\AA. All lines were carefully inspected to retain
only the best fitted ones. This was usually the case in the red part of the
spectrum between 5300~\AA\ and 6700~\AA\, which has a higher SNR.
    
The abundance of strontium was determined from the single
\ion{Sr}{II}~4215.519~\AA\ line, since both \ion{Sr}{II}~4161.792~\AA\ and
4077.709~\AA\ are too weak and noisy to be measured. Strontium abundances were derived primarily from EW analysis. 
Additionally, spectral synthesis were performed in case of weak and noisy features. 
A careful comparison was carried out between results from different methods in order to keep only lines with a good fit or a clear $\chi^2$ convergence.

Barium was measured from 1 to 4 \ion{Ba}{II} lines at 4934.076, 5853.668,
6141.713, and 6496.897~\AA\ by spectral synthesis to take into account the
hyperfine structure (HFS) of the lines and some small blends with iron
lines. The HFS data in the line list are from \citet{Prochaska2000} and \citet{Arlandini1999}.

\subsection{Uncertainties}

\begin{enumerate}

\item {\em \textup{Uncertainties due to the atmospheric parameters}.}
  
 To estimate the sensitivity of the  abundances to the adopted
 atmospheric parameters, we chose three stars representative of three regions
 in the T$_{\mathrm{eff}}$ vs. $\log g$ diagram (Fig~\ref{Fig:CMD}).  We
 repeated the abundance analysis and varied only one stellar atmospheric
 parameter at a time by its corresponding uncertainty, keeping the others
 fixed. The estimated internal errors are $\pm$150~K in T$_{\mathrm{eff}}$,
 $\pm$0.15~dex in $\log$ (g), and $\pm$0.15~km s$^{-1}$ in
 $v_\mathrm{t}$ for cool stars (T$_{\mathrm{eff}}<$6000~K), and $\pm$200~K in T$_{\mathrm{eff}}$,
 $\pm$0.20~dex in $\log$ (g), and $\pm$0.20~km s$^{-1}$ in
 $v_\mathrm{t}$ for the warmest stars of the sample (T$_{\mathrm{eff}}>$6000~K). These errors are presented in Table \ref{uncertainties}.\\

\item {\em \textup{Uncertainties due to EWs measurement}. }
The uncertainties on the individual EW measurements $\delta_{EWi}$ 
are provided by {\tt DAOSPEC} and computed according to the following formula \citep{Stetson2008} :

\begin{equation}
\delta_{EWi} = \sqrt{\sum_{p}^{}\left(\delta I_p\right)^2 \left(\frac{\partial EW}{\partial I_p}\right)^2+ \sum_{p}^{} \left(\delta I_{C_p}\right)^2 \left(\frac{\partial EW}{\partial I_{C_p}}\right)^2}
\end{equation}

where $I_p$ and $\delta I_p$ are the intensity of the observed line profile at pixel $p$ and its uncertainty, and $I_{C_p}$ and $\delta I_{C_p}$ are the intensity and uncertainty of the corresponding continuum. The uncertainties on the intensities are estimated from the scatter of the residuals that remain after subtraction of the fitted line. The corresponding uncertainties $\sigma_{EWi}$ on individual line abundances are propagated by {\tt Turbospectrum}.

\end{enumerate}

The final errors listed in Table~\ref{tab:abundances} were computed following the recipes outlined in  \citet{Jablonka2015}, \citet{Hill2019}, and \citet{Lucchesi2020}. 

The dispersion $\sigma_X$ around the weighted mean abundance of an element X measured from several lines is computed as
\begin{equation}
\sigma_X = \sqrt{\dfrac{\sum_i(\epsilon_i - \overline{\epsilon})^2}{N_X-1}}
\end{equation}
where $N_X$ represents the number of lines measured for element X, and  $\epsilon$ stands for the logarithmic abundance.\\

The final error on the elemental abundances is defined as 
\begin{equation}
\sigma_{fin} = \max (\dfrac{\sigma_X }{\sqrt{N_X}}, \dfrac{\sigma_{Fe}}{\sqrt{N_X}})
\end{equation}
As a consequence, no element X can have an estimated dispersion $\sigma_X$~<~$\sigma_{Fe}$; this is particularly important for species with very few lines.

\begin{table}
\centering
\caption{Changes in the mean abundances $\delta\log\epsilon$(X) caused by a change of $\pm$150~K on T$_\mathrm{eff}$, $\pm$~0.15 on $\log$g and $\pm$~0.15 on v$_\mathrm{t}$ for cool stars ($<$6000~K) and $\pm$200~K on T$_\mathrm{eff}$, $\pm$~0.20 on $\log$g and $\pm$~0.20 on v$_\mathrm{t}$ for warm stars ($>$6000~K), corresponding to the typical uncertainties on the stellar parameters. We provide T$_\mathrm{eff}$, $\log$g, v$_\mathrm{t}$ and [Fe/H] for each of the three representative stars.} 
\resizebox{\linewidth}{!}{
\begin{tabular}{c|cc|cc|cc}
\hline\hline\Tstrut
El. & $+\Delta$T$_\mathrm{eff}$ & $-\Delta$T$_\mathrm{eff}$ & $+\Delta\log$g & $-\Delta\log$g & $+\Delta$v$_\mathrm{t}$ & $-\Delta$v$_\mathrm{t}$ \\

\hline\Tstrut
 & \multicolumn{6}{c}{ Pr\_203-2831+13-6326 ( 5008 1.95 1.45 -2.7 )} \\
\hline\Tstrut
$\ion{Fe}{I} $ & $+0.12$ & $-0.15$ & $+0.00$ & $+0.01$ & $-0.02$ & $+0.03$ \\
$\ion{Fe}{II}$ & $+0.01$ & $-0.01$ & $+0.04$ & $-0.06$ & $-0.03$ & $+0.03$ \\
$\ion{C}{I}  $ & $+0.35$ & $-0.38$ & $-0.05$ & $+0.05$ & $+0.00$ & $+0.01$ \\
$\ion{Mg}{I} $ & $+0.09$ & $-0.08$ & $+0.00$ & $+0.01$ & $+0.00$ & $+0.01$ \\
$\ion{Ca}{I} $ & $+0.10$ & $-0.12$ & $-0.01$ & $+0.00$ & $-0.02$ & $+0.02$ \\
$\ion{Sr}{II}$ & $+0.21$ & $-0.24$ & $-0.03$ & $+0.03$ & $+0.00$ & $+0.00$ \\
$\ion{Ba}{II}$ & $+0.11$ & $-0.11$ & $+0.05$ & $-0.05$ & $-0.02$ & $+0.02$ \\
\hline\Tstrut
 & \multicolumn{6}{c}{ Pr\_196-6013+15-6768 ( 5600 4.6 1.08 -0.69 )} \\
\hline\Tstrut
$\ion{Fe}{I} $ & $+0.06$ & $-0.10$ & $-0.02$ & $+0.01$ & $-0.02$ & $+0.02$ \\
$\ion{Fe}{II}$ & $-0.04$ & $+0.05$ & $+0.05$ & $-0.06$ & $-0.04$ & $+0.04$ \\
$\ion{C}{I}  $ & $+0.20$ & $-0.21$ & $-0.03$ & $+0.02$ & $-0.02$ & $+0.02$ \\
$\ion{Mg}{I} $ & $+0.09$ & $-0.09$ & $-0.04$ & $+0.03$ & $-0.01$ & $+0.01$ \\
$\ion{Ca}{I} $ & $+0.10$ & $-0.11$ & $-0.04$ & $+0.03$ & $-0.03$ & $+0.02$ \\
$\ion{Sr}{II}$ & $+0.16$ & $-0.22$ & $-0.05$ & $+0.01$ & $-0.05$ & $+0.01$ \\
$\ion{Ba}{II}$ & $+0.07$ & $-0.07$ & $+0.03$ & $-0.03$ & $-0.04$ & $+0.05$ \\
\hline\Tstrut
 & \multicolumn{6}{c}{ Pr\_206-3487+9-3099 ( 6522 3.94 1.21 -1.8 )} \\
\hline\Tstrut
$\ion{Fe}{I} $ & $+0.15$ & $-0.14$ & $-0.02$ & $+0.02$ & $-0.04$ & $+0.06$ \\
$\ion{Fe}{II}$ & $+0.03$ & $-0.03$ & $+0.10$ & $-0.10$ & $-0.10$ & $+0.10$ \\
$\ion{C}{I}  $ & $+0.29$ & $-0.28$ & $-0.08$ & $+0.09$ & $+0.00$ & $+0.00$ \\
$\ion{Mg}{I} $ & $+0.13$ & $-0.12$ & $-0.05$ & $+0.04$ & $+0.0$ & $+0.02$ \\
$\ion{Ca}{I} $ & $+0.12$ & $-0.10$ & $-0.03$ & $+0.02$ & $-0.05$ & $+0.03$ \\
$\ion{Sr}{II}$ & $+0.15$ & $-0.14$ & $+0.09$ & $-0.07$ & $-0.17$ & $+0.19$ \\
$\ion{Ba}{II}$ & $+0.14$ & $-0.15$ & $+0.07$ & $-0.09$ & $-0.05$ & $+0.04$ 

\label{uncertainties}
\end{tabular}}
\end{table}

\section{Results} \label{Results}

In the following, we compare results from our spectroscopic analysis to literature and discuss the derived elemental abundances in the broader context of evolution of low-mass stars and Galactic chemical evolution.

\subsection{Comparison with previous work}

To verify the reliability of the derived chemical abundances, we compare our abundance estimates to literature values.

\begin{figure}
\centering
\includegraphics[width=0.98\columnwidth]{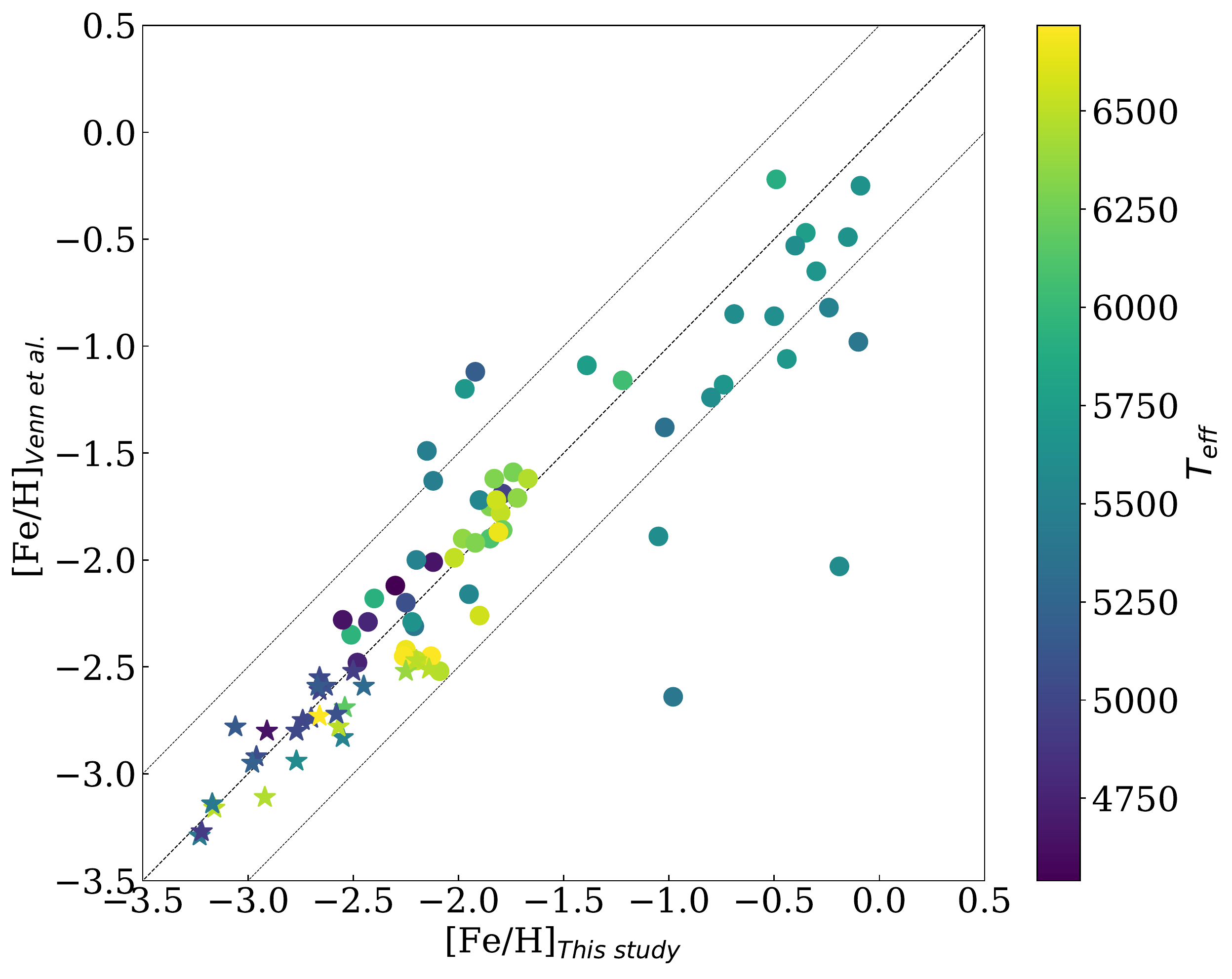}
\caption{Comparison with the metallicities of \citet{Venn2020} for the 86 stars in common. Colours code the stellar effective temperature. Circles  represent the stars whose metallicities were previously only based on the Q6 method of \citet{Venn2020}. Star symbols identify stars for which  \citet{Venn2020} provided full chemical characterisation.  For convenience, the one-to-one line and the $\pm$0.5~dex lines are also shown.}
\label{Fig:Q6comparison}
\end{figure}

We show in Figure~\ref{Fig:Q6comparison} a comparison between our spectroscopic metallicities and the results derived by \citet{Venn2020} for the stars in common (see \S~\ref{sample}).
The two investigations agree very well for the 28 stars with  [Fe/H] $\leq$ --2.5 for which \citet{Venn2020} provided full spectroscopic analysis
 (star symbols in Fig~\ref{Fig:Q6comparison}). For these stars, we compute an average difference in metallicity as small as $\Delta$[Fe/H]$_{\rm (Venn~et ~al.-this~work)}$ ~=~$-0.06$~dex ($\sigma$~=~0.15~dex). Differences in the adopted atmospheric parameters are also small 
--e.g., $\Delta$T$_{\mathrm{eff}}$$_{\rm (Venn~et ~al.-this~work)}$=~20~K
($\sigma$~=~40~K) and $\Delta\log$~g$_{\rm (Venn~et ~al.-this~work)}$=~$-0.09$~dex ($\sigma$~=~0.17~dex).

For the  stars with previous Q6 analysis only (circles in
Fig~\ref{Fig:Q6comparison}), the agreement is good
($\Delta$[Fe/H]$_{\rm (Venn~et ~al.-this~work)}$=~$-0.15$~dex), however we also note a significant dispersion
($\sigma$~=~0.5~dex) towards the metal-rich tail of the [Fe/H] distribution. 
In this case, part of the observed differences likely results from the adoption of different atmospheric parameters in the two studies. 
In this work, we derive parameters using the information encoded in the spectra (see Section~\ref{analysis}). In contrast, \citet{Venn2020} adopt the ``Bayesian inference method'' \citep{Sestito2019} which makes
combined use of SDSS and Gaia DR2 photometry, and adopts {\em a priori} photometric metallicities from {\em Pristine} to select the appropriate 
MESA/MIST isochrone with solar-scaled composition \citep{Paxton2011,Dotter2016} to infer parameters. 
Hence, the assumption of an incorrect metallicity affects the derivation of atmospheric parameters and the estimation of the final [Fe/H]. 
This is particularly true for stars in the metal-rich regime ([Fe/H]~$\geq$~--1.5) in Figure~\ref{Fig:Q6comparison}. 
For stars with [Fe/H]$\geq$--1.5 we compute an average difference of 
$\Delta$T$_{\mathrm{eff (Venn~et ~al.-this~work)}}$=~175~K ($\sigma$~=~300~K), 
and $\Delta\log$~g$_{\rm(Venn~et ~al.-this~work)}$=~--0.20 dex ($\sigma$~=~0.60~dex) 
which is significantly larger than the one derived for stars with [Fe/H]~$\leq$~--1.5 
$\Delta$T$_{\mathrm{eff (Venn~et ~al.-this~work)}}$=~120~K ($\sigma$~=~200~K), 
and $\Delta\log$~g$_{\rm(Venn~et ~al.-this~work)}$=~0.05~dex ($\sigma$~=~0.25~dex).

As \citet{Venn2020} adopted the Pristine photometric metallicity estimates {\em a priori}, metal-rich stars turn out to be not well calibrated (e.g. T$_{{\rm eff}}$ and [Fe/H] are degenerate; see~\S~4.3 in \citealp{Venn2020}).
However, we note that the spectra presented in \citet{Venn2020} were collected between 2016 and 2018 and the {\em Pristine} metallicity calibrations have improved over the course of these spectroscopic follow-up observations \citep[see discussion in][]{Venn2020}.

Different line list, codes and minimizing procedures can also
play a role. A careful investigation of the observed discrepancy is not the
main goal of this study.

\subsection{Newly detected very metal-poor stars}

We identify 31 new very metal-poor stars in total that were whether missed by the quick analysis in \citet{Venn2020}, or are presented in this work for the first time (see ~\S~\ref{sample}).

Among the sample of 20 stars presented in this study for the first time, eight VMP stars (including five stars with [Fe/H]~<~$-2.5$ and one EMP star at [Fe/H]~=~$-3$, Pr\_255.8043+10.8443) have been identified. Interestingly, two of these VMP stars were removed from the more recent versions of the \Pristine\ catalog.

Among the 57 stars from \citet{Venn2020} with a Q6 estimation only, no star previously marked at
[Fe/H]~<~$-2.5$ was missed by the quick analysis. However an additional set of 19 
stars are now identified as very metal-poor, with [Fe/H]~<~$-2.0$. From those three have
[Fe/H]~$\sim$~$-2.5$.

Among the 26 stars from \citet{Venn2020} that were rejected by the
\Pristine\ photometric selection before any analysis, 4 turn out to be
actually very metal-poor with $-2.5$~<~[Fe/H]~<~$-2.0$.

The 6 very metal-poor stars ([Fe/H]~$\le$~=~$-2$) that are erroneously rejected from
the more recent version \Pristine\ catalogue, are highlighted in red in
Fig~\ref{Fig:CMD}. Because we are working at the bright end of the catalogue,
these stars have saturated flags in one of the ugri magnitudes of SDSS.
Since \Pristine\ aims at maximizing the rejection of contaminants, the later versions of the catalogue conservatively rejects any star that could {\it potentially} be affected by saturation. In some cases, the g and i magnitudes that are mainly used to infer the \Pristine\ metallicities may in fact not be affected and the inferred metallicity can be accurate, as shown with some of the rejected stars studied here that we determine to be very metal-poor.\\

\subsection{Abundances for the \texorpdfstring{$\alpha$}{a}-elements: Mg and Ca}

\begin{figure}
    \centering
    \includegraphics[width=1.0\columnwidth]{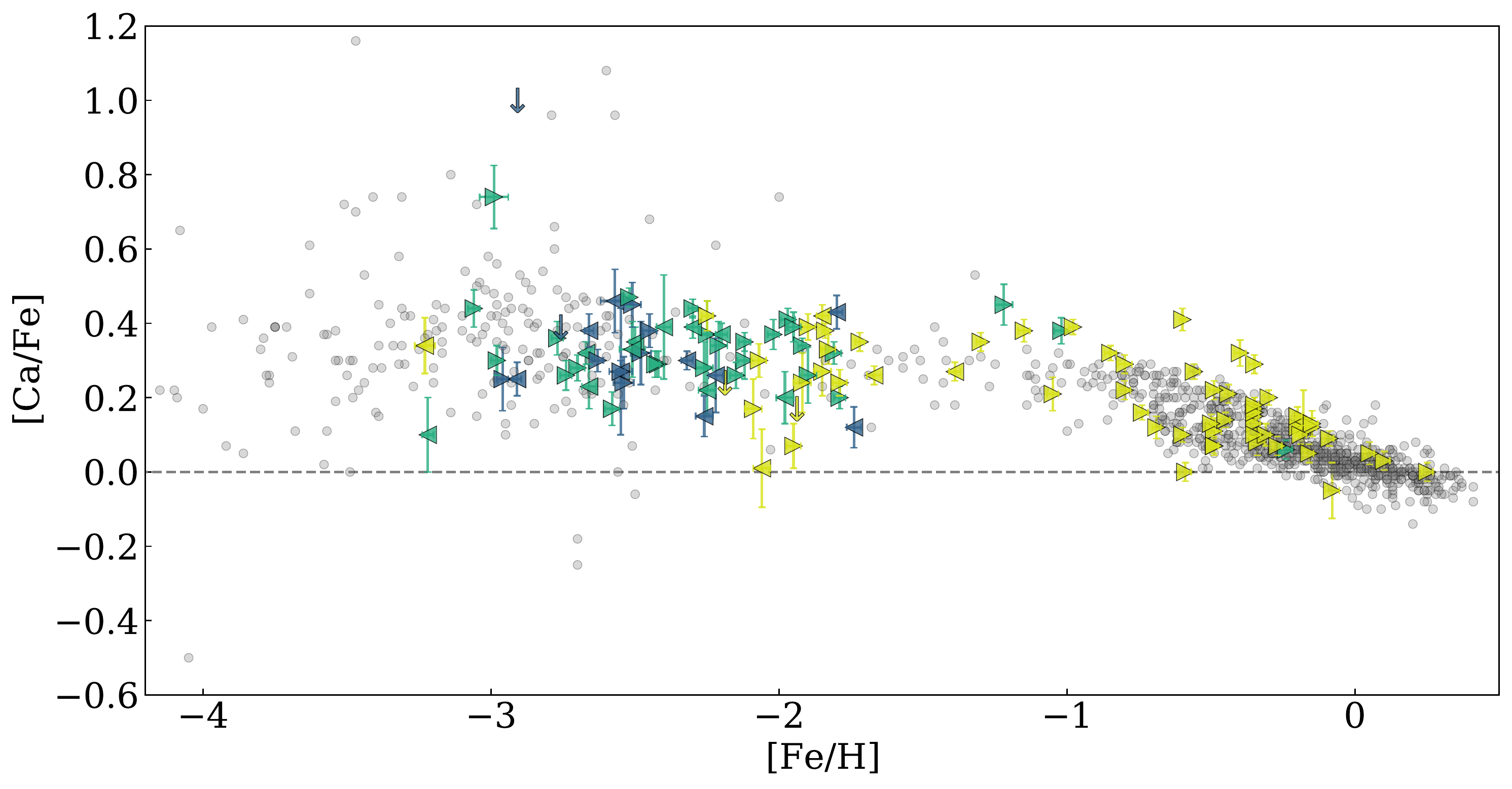}
    \includegraphics[width=1.0\columnwidth]{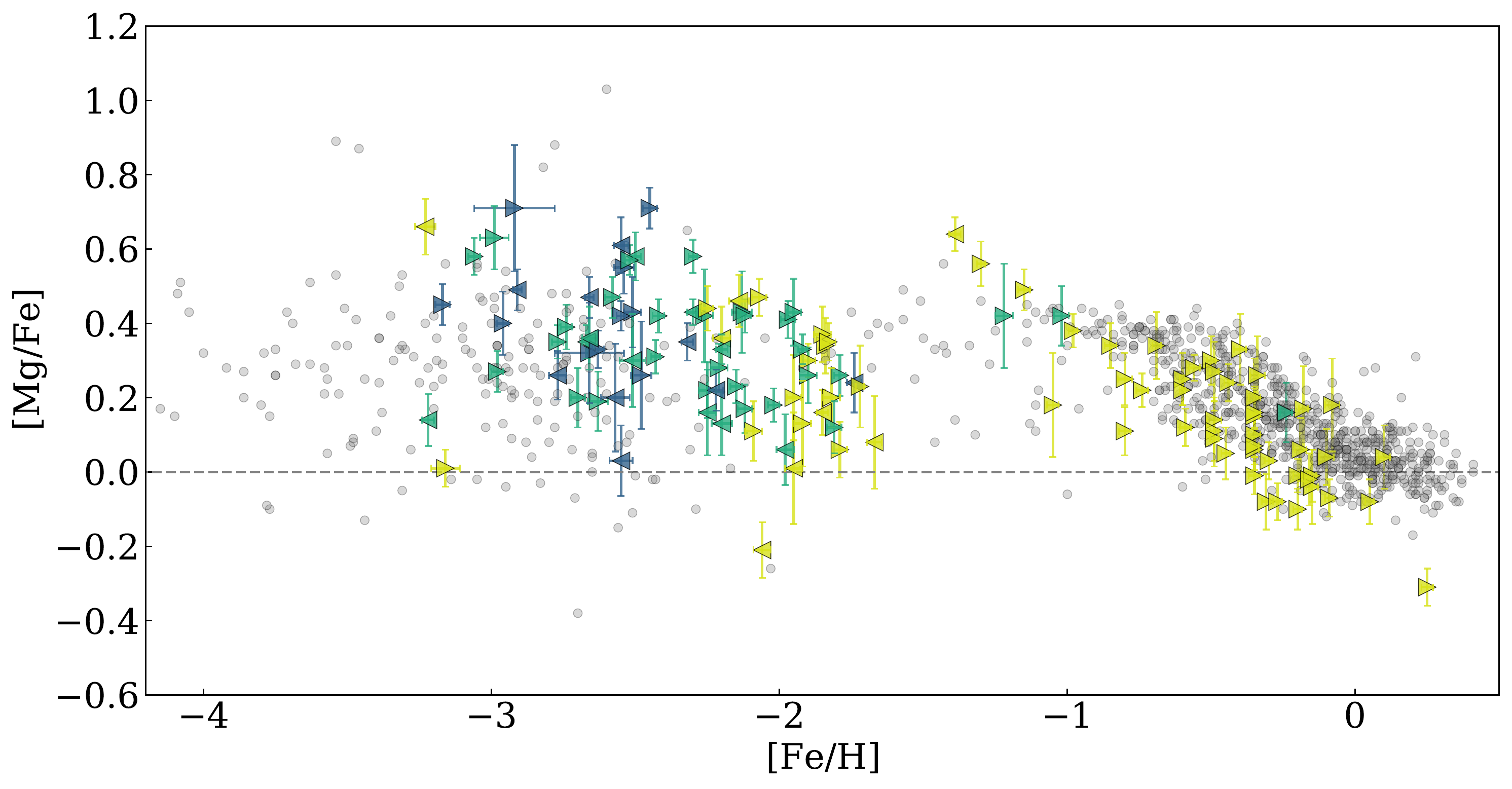}
    \caption{{\em $\alpha$-elements :} Calcium (top panel) and magnesium (bottom panel) abundances and upper limits are plotted against metallicity for the full sample. Left and right arrows stand for stars with prograde and retrograde motions, respectively.  Yellow, green, and dark blue identify stars in the MW disc, inner halo, outer halo, respectively (see Sect. \ref{orbits}). Comparison galactic stars (in grey) are from \citet{Yong2013,Bensby2014}.}
    \label{Fig:alpha}
\end{figure}

In order to gauge the general composition of our sample, Fig. ~\ref{Fig:alpha}
presents the trend of the $\alpha$-elements, calcium and magnesium, with [Fe/H].
While Mg is produced in hydrostatic H and He nuclear burning in massive
core-collapse supernovae (SN) progenitors, Ca originates mainly from the
pre-supernova explosion, and can also be produced later by type Ia supernovae
\citep{Woosley1986}.\\

Our sample follows closely the MW halo distribution in Ca and Mg
\citep[e.g.,][]{Yong2013}.  The large majority of the stars in the metal-poor regime ([Fe/H]<--1) are enhanced in $\alpha$-elements.  There is a well-defined
plateau at [Ca/Fe]~=~+0.3~dex with a small dispersion of
$\sigma$~$\sim$~0.1~dex. We do not find any sub-solar [Ca/Fe] stars in our
sample as it had been the case for \cite{Caffau2020}. Metal-poor stars are also enhanced in Mg, with a larger dispersion
($\sigma$~$\sim$~0.3~dex) though, arising from the larger uncertainties induced
by the smaller number of lines for Mg than for Ca.

Two stars have  [Mg/Fe] significantly lower than the rest of our sample at similar [Fe/H]. Pr\_192.2121+7.4778 at [Fe/H]=+0.25 (T$_\mathrm{eff}$ =5600K, $\log g$=4.3) and Pr\_238.7217+6.1945 at [Fe/H]=$-2.06$ (T$_\mathrm{eff}$ =6551, $\log g$=4.2). The calcium abundance places Pr\_238.7217+6.1945 at the low edge of the distribution, at solar [Ca/Fe], as well. Considering that it is among the hottest stars of our sample, very few lines are accessible to the analysis. It looks also depleted in Mg with respect to stars with the same metallicity. However, its [Mg/Fe] abundance ratio is compatible with the main body of the ESPaDOnS sample when errors due to atmospheric parameters (see Table~\ref{uncertainties}) are taken into account.

Pr\_192.2121+7.4778 is the most metal-rich star of our sample. At this metallicity, with [Mg/Fe]=$-0.31 \pm 0.05$, it could resemble the Mg-poor Milky Way field stars with globular cluster chemical patterns that \citet{Fernandez2017} have analysed. However Pr\_192.2121+7.4778 is not particularly enriched in Al. Further specific investigation for this star shows that it is not particularly abundant in supernovae type Ia products such as Cr or Co either. Moreover [Si/Fe]$\sim$0, just as [Ca/Fe], typical of the Milky Way disc stars.  \citet{Mackereth2019} have shown from the analysis of APOGEE DR14 the correspondence between the lowest [Mg/Fe] and lowest eccentricity even in the Milky Way disc. At this stage, one can only say that, with e=0.175  this star is indeed in the low quartile of the e-distribution in our sample. 

Pr\_251.4082+12.3657 at [Fe/H]~=~$-3.22$ has relatively low $\alpha$ ratios, [Ca/Fe]~=~+0.1 and [Mg/Fe]~=~+0.14, but we do not confirm the sub-solar [Mg/Fe] value found by \citet{Venn2020}.  However this star is one of the most metal-poor one of the sample and the single Ca line measurable is weak (<~20~m\AA), making its abundance difficult to ascertain.

At the other end of the abundance ratio distribution, Pr\_255.8043+10.8443, at
[Fe/H]~=~$-$3, the EMP from the new sample, is very enriched
in both Ca and Mg, at $+0.63$ and $+0.74$, respectively, however with no other
outstanding chemical feature. In particular, only an upper limit to its C
abundance could be estimated. As found by \citet{Venn2020} Pr\_181.2243+07.4160, has high [Mg/Fe]~=~+0.7~dex and high calcium abundance,
however this is a warm turn-off star (6450~K) with small Ca features, hence only
an upper limit can be placed.

\subsection{Carbon abundances and internal mixing}

Our sample allows for the investigation of the carbon abundance of Milky Way stars over a wide range of metallicities and to simultaneously explore the impact of
internal mixing.

Indeed, during the evolution on the red-giant branch (RGB), carbon is converted into nitrogen due to the CN
cycle, then mixed to the surface of the star.
This mechanisms occurs when low mass stars ($\leq$2.5 M$_{\odot}$) evolve off the main sequence.  Their outer convective envelope starts to move inward, dredging up
material that has been processed through the CN-cycle in the inner regions
\citep[First Dredge-up,][]{Iben1964}.  In a more advanced stage of evolution along
the RGB, these stars experience an additional mixing episode just after the RGB
bump, when the molecular weight barrier (the $\mu$-barrier) left by the
convective envelope at the point of deepest inward progress is canceled out by
the outward expansion of the H-burning shell. This extra-mixing episode
\citep{Sweigart1979,Charbonnel1995,Angelou2012} produces a decline in the surface
abundance of carbon ([C/Fe]) and $^{12}$C/$^{13}$C \citep{Briley1990,Gratton2000,Martell2008,Gerber2019}
and lithium \citep{Lind2009}, and an increase in the nitrogen abundance \citep{Gratton2000}.

\begin{figure}
    \includegraphics[width=0.5\textwidth]{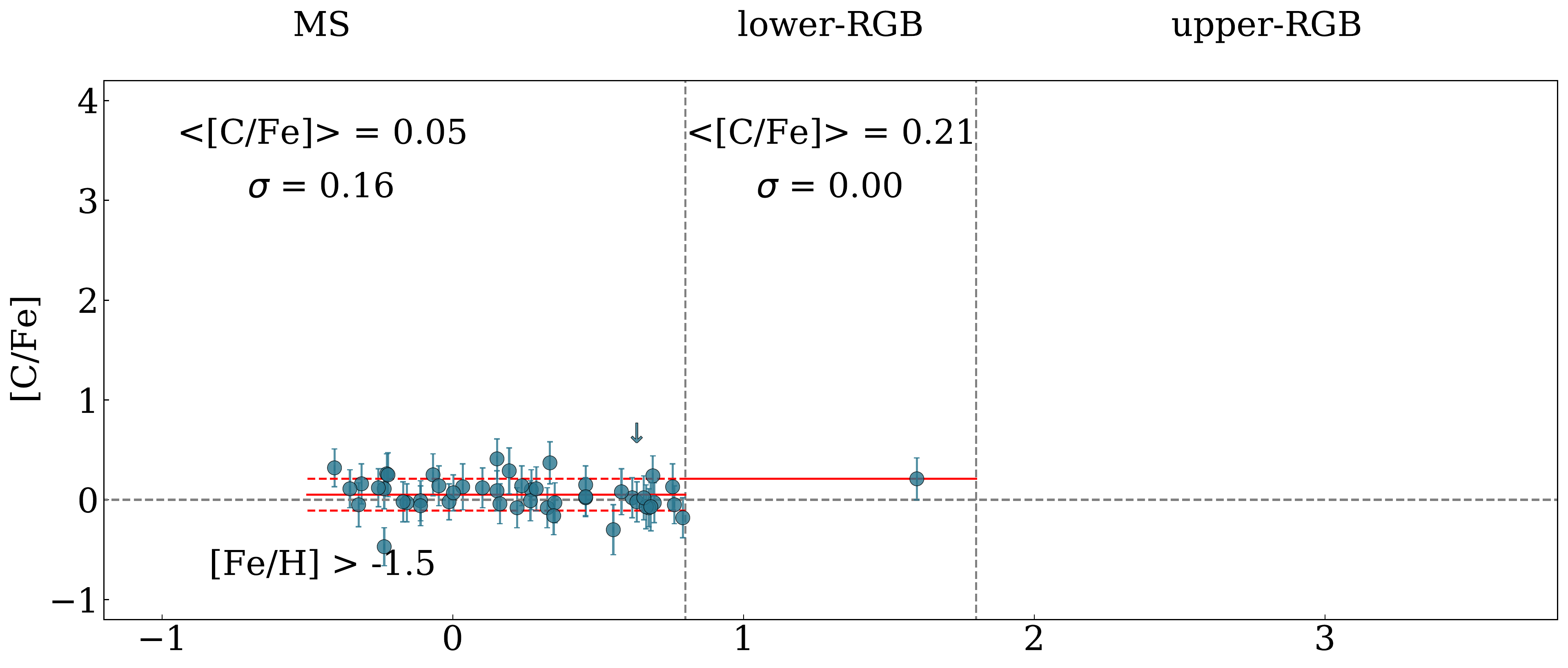}
    \includegraphics[width=0.5\textwidth]{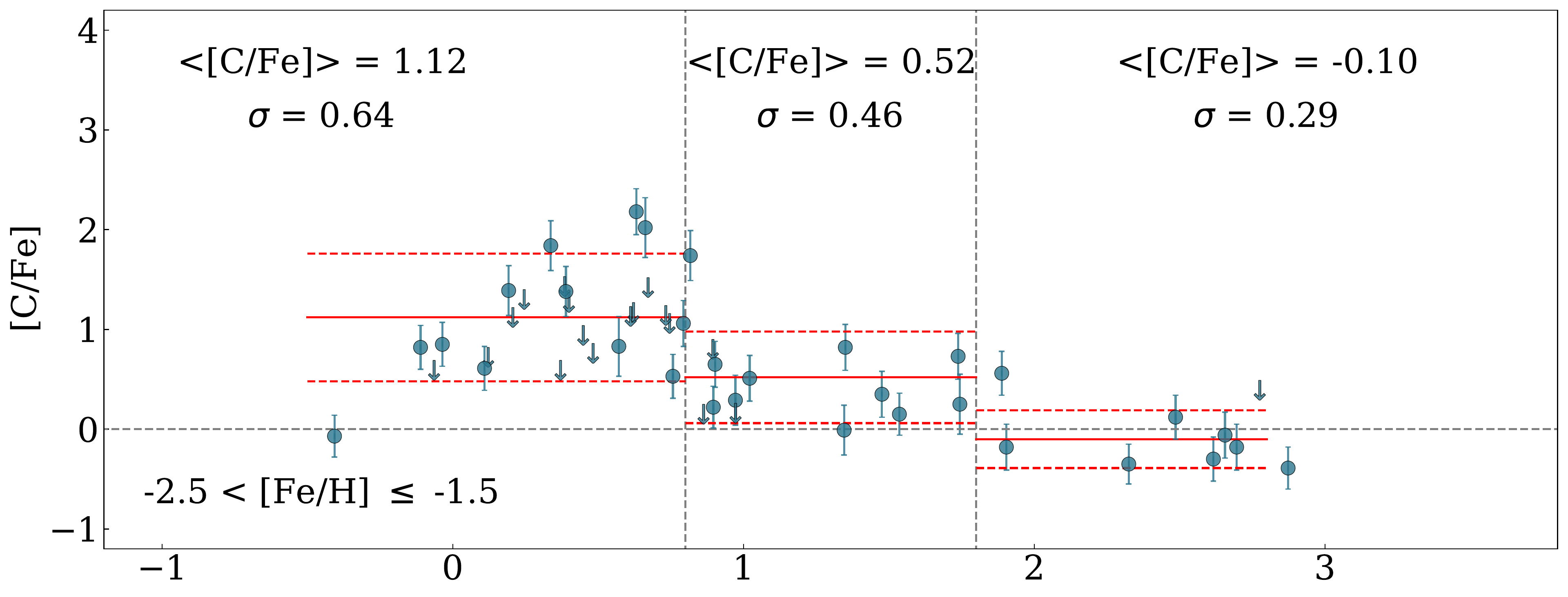}
    \includegraphics[width=0.5\textwidth]{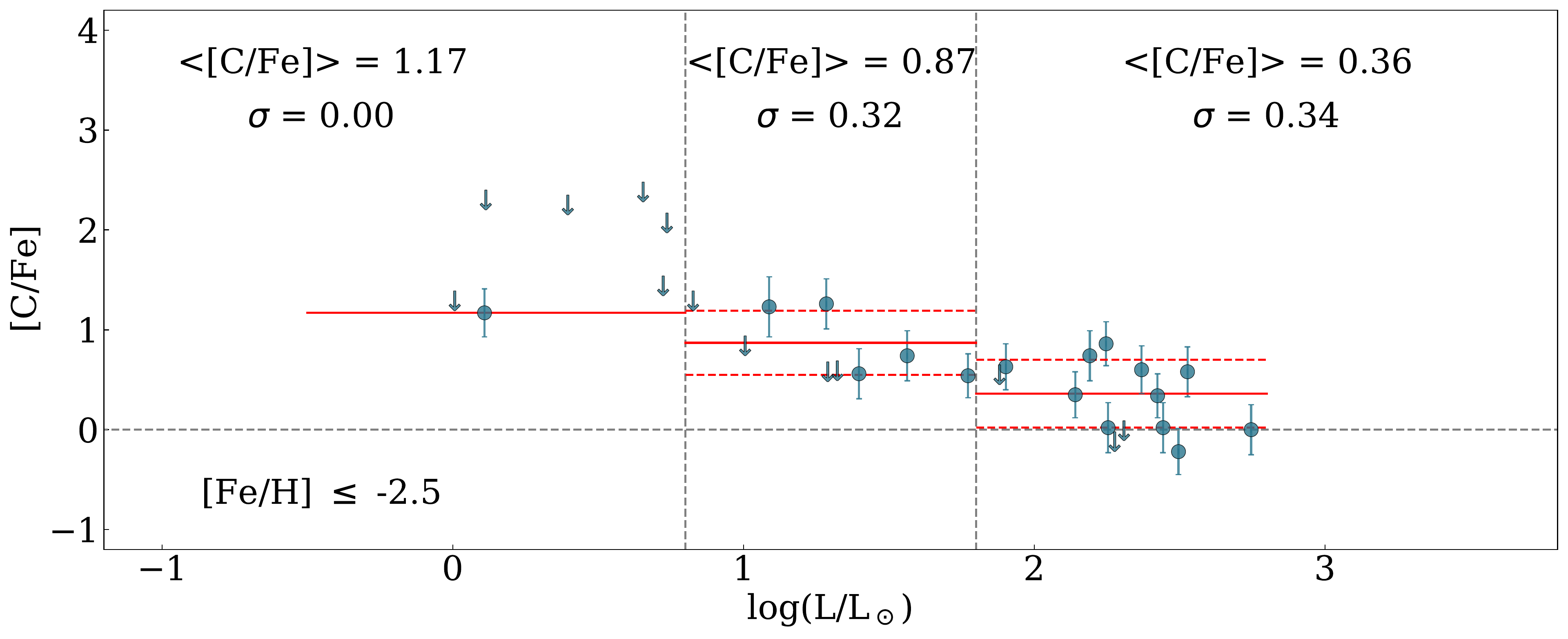}
    \caption{Run of the abundance ratios of [C/Fe] with luminosity for stars with different metallicities ({\em from top to bottom:} [Fe/H]~>~$-1.5$; $-2.5$~<~[Fe/H]~$\leq$~$-1.5$, and [Fe/H]~$\leq$~$-2.5$). The dotted vertical lines separate
different evolutionary phases --main sequence (MS) and turn-off (TO) stars;
lower-RGB and upper-RGB stars -- following the \citet{Gratton2000} classification.
The red lines represent the mean carbon abundances along with their standard
deviation in the various different evolutionary stages (solid and dashed lines;
respectively).}
    \label{Fig:C_logg}
\end{figure}

The degree of carbon depletion is a function of both metallicity and the initial stellar carbon and nitrogen abundances. This is already discussed extensively in the literature \citep{Spite2005,Spite2006,Aoki2007,Placco2014,Shetrone2019}.
Along this line, Figure~\ref{Fig:C_logg}
presents the [C/Fe] abundance ratio of our sample stars as a function of their
luminosity in 3 different metallicity bins. The different evolutionary phases,
main sequence (MS) and turn-off (TO) stars, lower-RGB and upper-RGB stars, are
identified following the \citet{Gratton2000} classification.  In unevolved stars
(log L/L$_{\sun}$ < 0.8), the average C abundance in stars with [Fe/H]~>~$-$1.5
is [C/Fe]~=~0.05~$\pm$~0.16 (48 stars). 
The C abundance appears to increase at first dredge up with a value of [C/Fe]=0.21, however this apparent increase is only based on a single stars on the lower-RGB and thus not significant.  
In the lower metallicity bins, the carbon depletion
is more severe. At the first dredge up, C abundances decrease from
[C/Fe]~=~+1.12 to [C/Fe]~=~+0.52 for the stars in the metallicity range
$-2.5$~<~[Fe/H]~$\le$~$-1.5$, and from [C/Fe]~=~+1.17 to [C/Fe]~=~+0.87 for the
lowest metallicity bin [Fe/H]~$\le$~$-2.5$.

\subsection{Frequency of C-enhanced stars}

As discussed by \citet{Norris2019}, the 3D-NLTE treatment of \ion{Fe}{I} and
CH-based carbon abundances could change our view of the genuine fraction of
carbon enhanced metal-poor stars in the future. Until these calculations are fully
accessible, 1D-LTE studies, such as this one, are important.

CEMP stars are commonly separated into two broad categories, according to their
chemical composition: carbon enriched stars that display an overabundance of
heavy elements formed in slow (s), intermediate (i), or rapid (r) neutron capture processes
(CEMP-s, CEMP-i, CEMP-r, and CEMP-r/s); and CEMP-no, stars that display no such excess
of neutron-capture elements \citep{Spite2013}. The observed chemical pattern of CEMP-s and CEMP-
r/s stars is thought to be the result of mass transfer in a binary
system (e.g.,~\citealp{Masseron2010} but see also~\citealp{Hansen2016}).

\begin{figure}
    \centering
    \includegraphics[width=1.0\columnwidth]{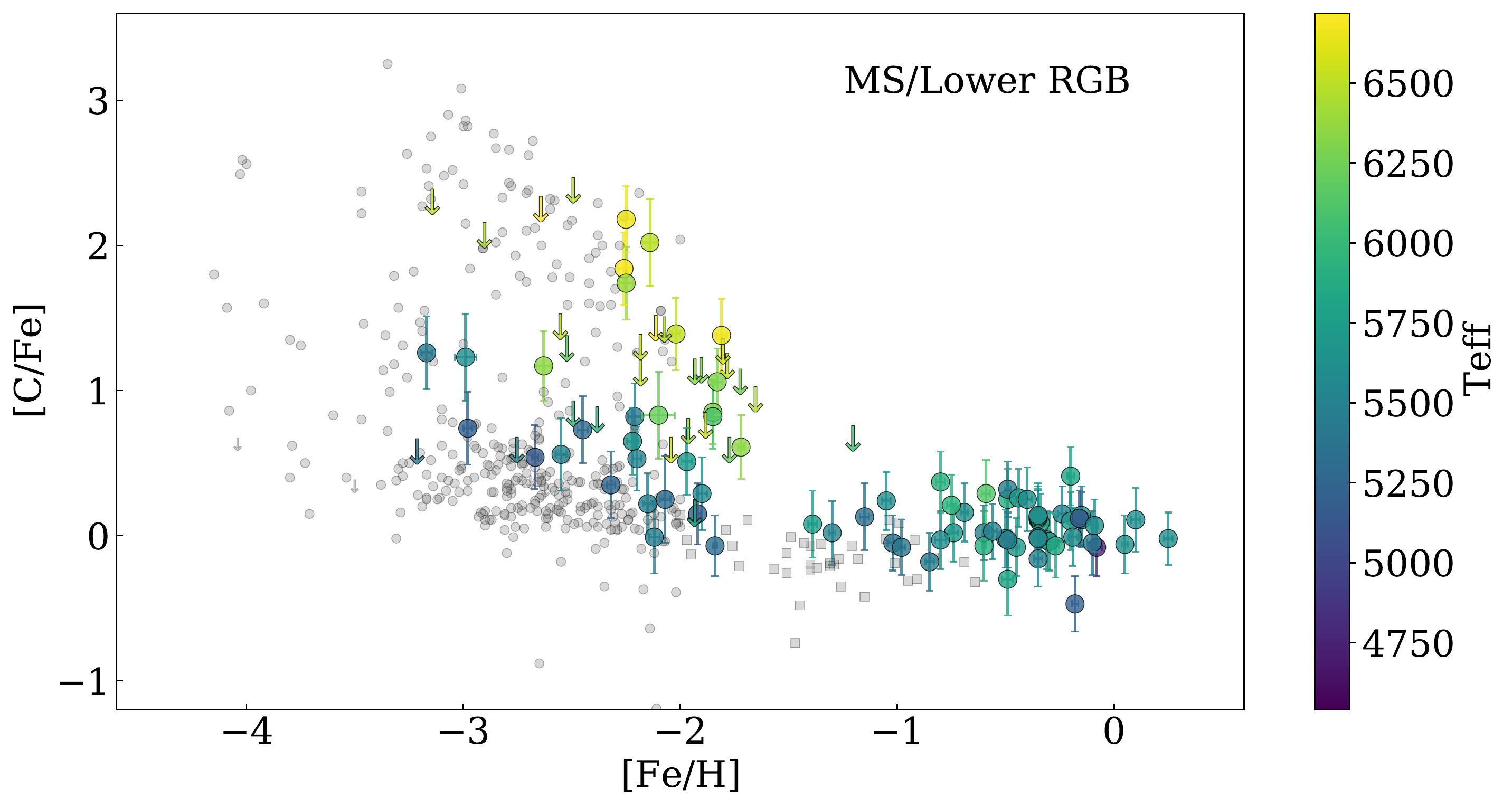}
    \includegraphics[width=1.0\columnwidth]{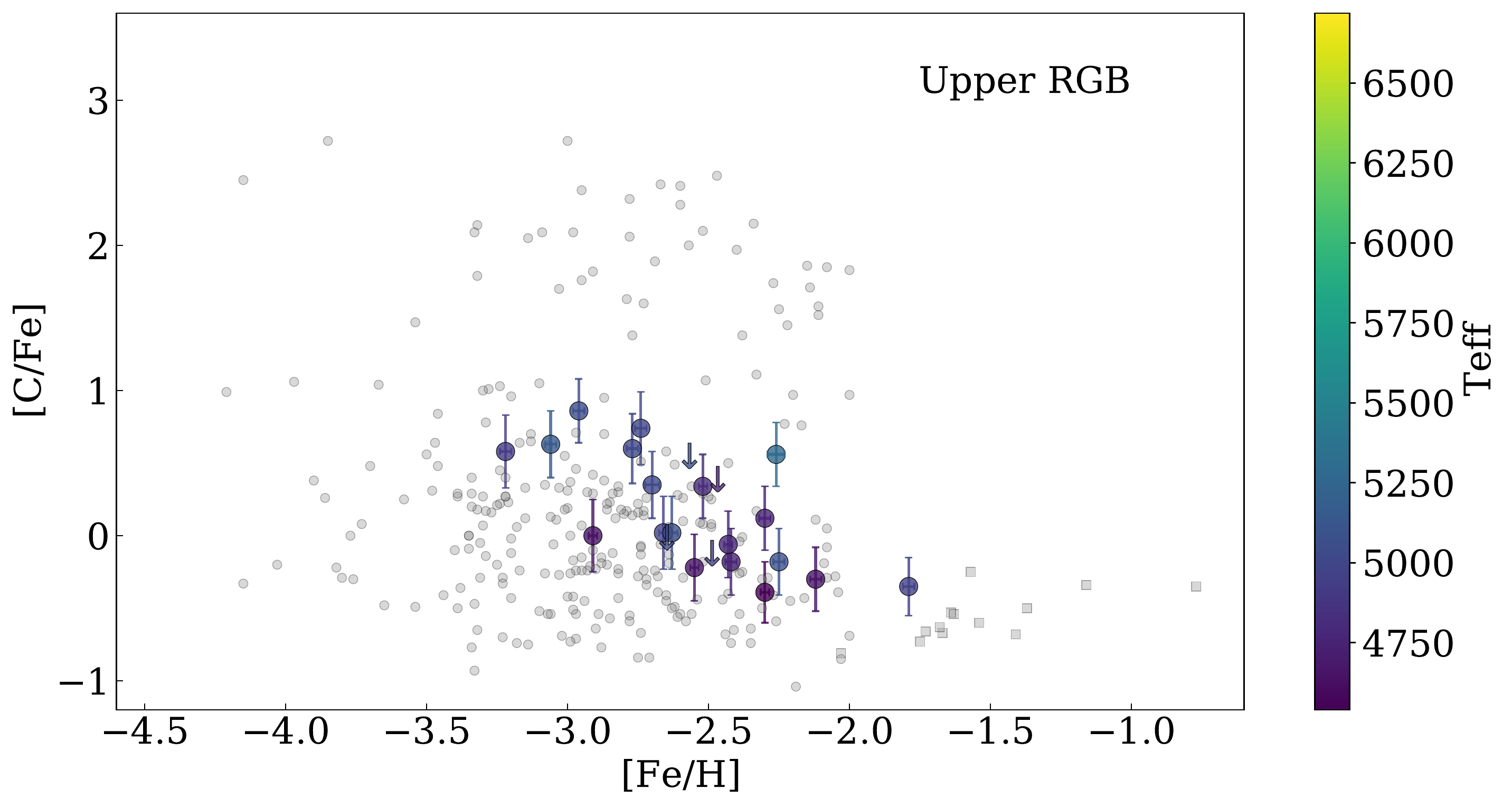}
    \caption{ {\em Top panel~:} [C/Fe] abundance ratios are plotted against stellar
      metallicities [Fe/H] for unevolved stars (e.g. stars with
      $\log$(L/L$_{\sun}$)~<~1.8) and colour-coded by T$_{\mathrm{eff}}$. {\em
        Bottom panel~:} [C/Fe] abundance ratios are plotted against stellar
      metallicities [Fe/H] for evolved stars (e.g. stars with
      $\log$(L/L$_{\sun}$)~>~1.8). Upper limits for carbon are plotted as downward
      arrows. The evolutionary phases are based on the work of
      \citet{Gratton2000}. Grey circles are MW halo stars from \citet{Placco2014}, whereas grey squares are from \citet{Gratton2000}.}
    \label{Fig:Carbon}
\end{figure}

Our sample encompasses 38 stars with [Fe/H]~$\leq$~$-$2.0 and measured C abundance (Table \ref{tab:abundances}). Fourteen  of them 
have [C/Fe]$\ge$0.7 (criterion for CEMP stars of \citet{Aoki2007}) and 8 of them have [C/Fe]$\ge$1 (criterion for CEMP stars of \citet{Beers2005}).
This translates into a frequency of CEMP stars of 37\% and 21\% for the two criteria,
respectively.  This is consistent with the results of \citet{Aguado2019} which
found 41\% and 23\% respectively of CEMP stars in the large medium-resolution
survey of \Pristine. 
However, we note  the number of CEMP stars in \citet{Aguado2019}
 is likely over-estimated due to systematic biases in their derived $\log$g
values and a strong dependence of [C/Fe] on $\log$g in their analysis. This issue
will be further discussed in Arentsen et al. (in preparation).

Figure~\ref{Fig:Carbon} presents the carbon abundances of our \Espadons\ sample
stars depending on their evolutionary stage. The top panel shows the run of
[C/Fe] abundance ratios for unevolved stars in the sample
($\log$(L/L$_\odot$)< 1.8; \citealp{Gratton2000}), while the bottom panel
displays the same trend for the  evolved stars (
$\log$(L/L$_\odot$)~$\ge$~1.8) we have analysed.

\begin{table}
\centering
\caption{Fraction of CEMP stars in our sample of stars compared to the work of \citet{Placco2014}, we consider only the stars with carbon measurements (not the upper limits)} and compare the results with the criterion [C/Fe]~>~+0.7 of \citet{Aoki2007}, and [C/Fe]~>~+1.0 of \citet{Beers2005}
\resizebox{\linewidth}{!}{
\begin{tabular}{lc|cc|cc}
\hline\hline\Tstrut
Study & VMP Dwarfs & [C/Fe]~$\ge$~+0.7 & Fraction & [C/Fe]~$\ge$~+1.0 & Fraction  \\
\hline\Tstrut
Placco 2014   & 348 (56\%) & 123 & 35\% & 101 & 29\%    \\
This work     & 20  (52.5\%) & 12  & 60\% & 8  & 40\%  \\
\hline\hline\Tstrut
 & VMP Giants & [C/Fe]~$\ge$~+0.7 & Fraction & [C/Fe]~$\ge$~+1.0 & Fraction  \\
\hline\Tstrut
Placco 2014   & 268 (44\%) & 60 & 22.4\% & 48 & 18\%    \\
This work     & 18  (47.5\%) & 2  & 11\% & 0  & 0\%     \\
\hline
\end{tabular}}
\label{Tab:carbon_freq}
\end{table}

As summarized in Tab.~\ref{Tab:carbon_freq}, the dwarf sub-sample of
\citet{Placco2014} has a fraction of 35\% and 29\% of carbon-rich stars depending on the criterion adopted ([C/Fe]~>~+0.7 vs. [C/Fe]~>~+1.0), while our sub-sample of very metal-poor dwarfs is composed of 60\% and 40\% of C-rich stars considering the same criteria. 
However, we note that large uncertainties are associated with the derived CEMP fractions  (e.g., of the order of $\sim$15\%) because of the small size of the observed sample. Also, for warm stars, carbon abundances could be measured only for object with a relatively high level of C overabundance. Both factors must be taken into account when the fractions in the table~\ref{Tab:carbon_freq} are compared to literature and and larger (an un-biased) samples of EMP stars are required.

The trend with temperature in the dwarf sub-sample is clearly seen in Fig.~\ref{Fig:Carbon} 
-- the C-rich stars having the highest T$_{\mathrm{eff}}$ (>~5800~K). This correlation is very much driven by the fact that at the low signal-to-noise (7-25) in the blue part of the spectra, normal carbon abundances were out of reach for the hot stars. Only the very strong absorption bands of the CEMP stars were measurable (see Fig. \ref{Fig:CH-band}). In that case, we were facing observational limits rather than a bias in the \Pristine\ selection.

The comparison between the upper and lower panels of Fig.~\ref{Fig:Carbon} illustrates the clear dichotomy between the giants and the dwarfs in our sample. We almost totally lack C-rich giant stars, with  only two stars with [C/Fe]~>~+0.7 (and none with [C/Fe]~>~+1.0) among our 18
very metal-poor giants, i.e. 11\% compared to 22\% in the sample of
\citet{Placco2014}. However, given the small sample size, the errors associated to such fraction is as large as the estimated fraction itself (of the order of $\pm$10\%). Thus we cannot draw firm conclusions on the CEMP fraction for evolved giants from such a comparison.

\citet{Placco2014} developed a procedure to compute corrections for the evolutionary depletion of carbon. The corrections tend to increase
the C abundances and they depend on the surface gravity, the metallicity of the
star and the observed carbon abundance. The corrections reach up to +0.70~dex at
[Fe/H]~=~$-3$, $\log$~g~=~1.0. They would increase the number of giants that can
be considered as C-rich in our sample, however Fig~\ref{Fig:Carbon} and
Tab~\ref{Tab:carbon_freq} compare non-corrected abundances only, thus the low CEMP fractions in our subsample of very metal-poor giant is real and probably results from a bias in the \Pristine\ photometric selection process of the very metal-poor
candidates. This selection appears to favor the warm C-rich turn-off stars, but to be biased against the cooler evolved CEMP stars. Most probably this is also the reason for the very low fraction of C-enhanced stars in \citet{Caffau2020}, in which nearly all very metal-poor stars are cool giants. The origin of this \Pristine\ selection bias will be further discussed in a forthcoming study (Arentsen et al., in prep.).

\subsection{Abundances for the heavy-elements Sr and Ba}

Heavy elements (e.g. elements with atomic number greater than 30; Z > Z$_{Zn}$) are
produced through the slow (s) and rapid (r) neutron-capture processes. The main
sources of s-process elements are asymptotic giant branch stars \citep{Busso1999,Bisterzo2012},  while r-process occurs instead in different types of core-collapse supernovae \citep{Hillebrandt1976, Woosley1994, Wanajo2001,  Nishimura2006,Kratz2014} and neutron star
merger \citep{Lattimer1974, Freiburghaus1999, Rosswog2000, Wanajo2013, Thielemann2017}

The wavelength range of the \Espadons\ spectra includes the spectral features of
two neutron-capture elements, Sr and Ba. Because europium is mostly produced by
the r-process, [Ba/Eu] ratios are commonly used to identify the origin of the
heavy elements. Unfortunately, no Eu lines were measurable in our spectra.

\begin{figure}
    \centering
    \includegraphics[width=1.0\columnwidth]{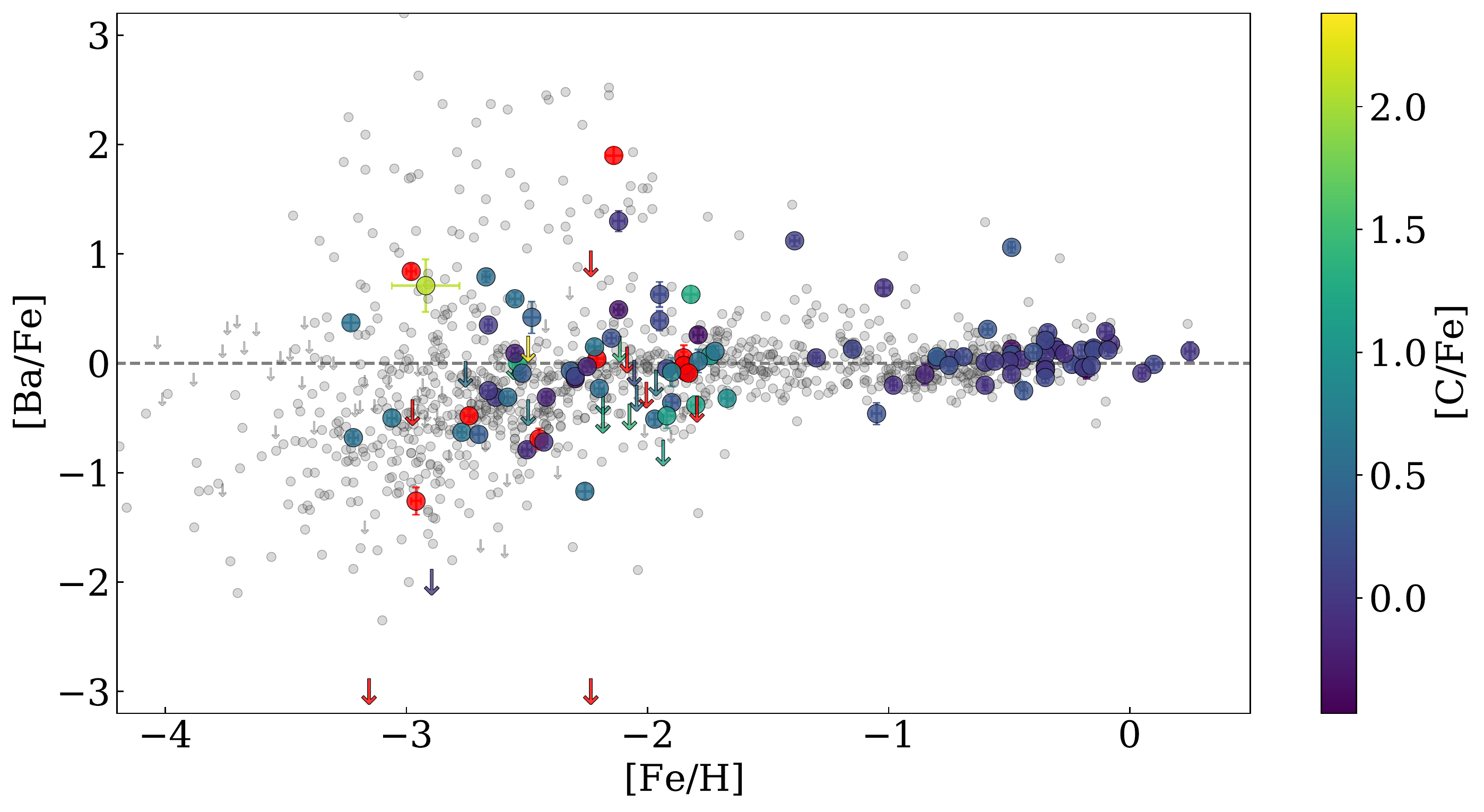}
    \includegraphics[width=1.0\columnwidth]{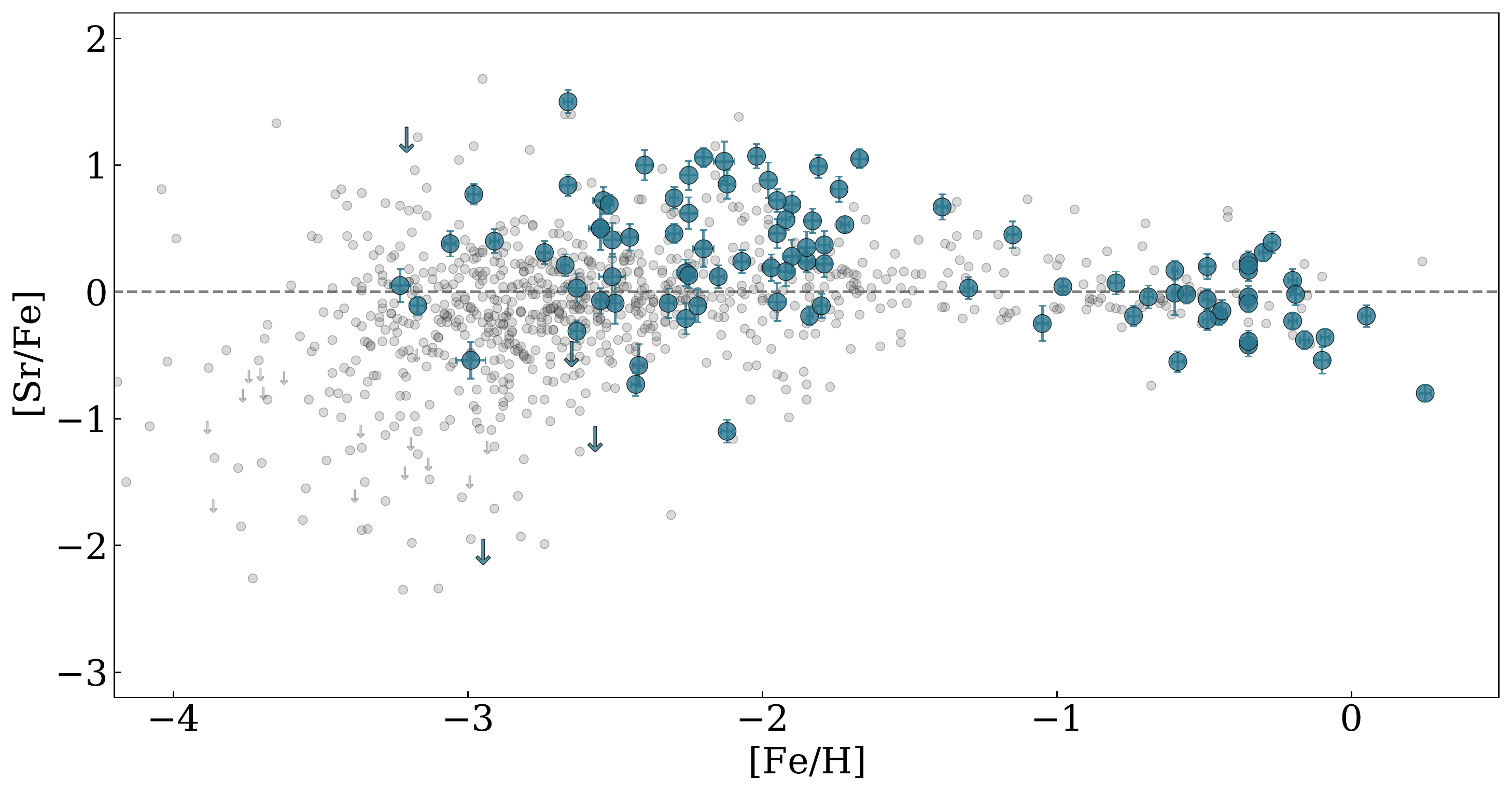}
    \caption{Neutron-capture elements : Barium-to-iron ratio as a function of metallicity at the top, and strontium-to-iron ratio at the bottom. Pointing down arrows are upper limits. Grey dots are galactic comparison stars compiled by \citet{Roederer2013}.\\
    The [Ba/Fe] ratios are colour-coded by their carbon content, red points are carbon-rich stars defined by the criterion of \citet{Aoki2007}.}
    \label{Fig:neutron-capture}
\end{figure}

Figure~\ref{Fig:neutron-capture} presents the [Ba/Fe] and [Sr/Fe] measurements
of the ESPaDOnS data-set compared to  MW halo population
\citep{Roederer2013}. Barium is generally sub-solar at
[Fe/H]~<~$-2$ with nevertheless a large scatter with some stars highly enhanced or depleted in barium. Above [Fe/H]$\sim -2$, [Ba/Fe] converges to the solar value and the scatter is almost completely removed at [Fe/H]~>~$-1$.

The comparison between the C and Ba abundances allows one to identify CEMP-no
and CEMP-s stars \citep{Beers2005}.  In the absence of Eu, we base our
classification on the work of \citet{Matsuno2017}, [Ba/Fe] $>1$ for CEMP-s,
adding [Sr/Ba] when possible following \citet{Hansen2019} ([Ba/Fe]$>0$, [Sr/Ba]$<-1.5$ for CEMP-r, $-1.5<$[Sr/Ba]$<-0.5$ for  CEMP-r/s, and
$-0.5<$[Sr/Ba]$<0.75$ for CEMP-s). While it would not be sufficient to discriminate
between CEMP-s and CEMP-r/s \citep{Goswami2021}, it is probably good enough to
distinguish between CEMP-r and the other categories.

We color code in red in Fig.~\ref{Fig:neutron-capture} the stars that are carbon
enhanced ([C/Fe]~>~+0.7). The arrows indicate the CEMP stars for which only
upper limits in Ba could be derived; most of these stars are hot
(T$_\mathrm{eff}$~>~6200~K).

A few stars stand out from our sample:
Pr\_245.5747+6.8844 ([Fe/H]~=~$-3.17$) is a CEMP-no star ([C/Fe]~=~+1.26)
without any detectable barium line, while at T$_\mathrm{eff}$=5424K it could be
measurable, and its strontium abundance is normal ([Sr/Fe]~=~$-0.1$).  It is
identified in Fig.~\ref{Fig:neutron-capture} by its upper limit arbitrarily put
at [Ba/Fe]~=~$-3$. Again, following the classification of \citet{Hansen2019},
Pr\_180.2206+9.5683 is another possible CEMP-no star ([Fe/H]~=~$-$2.96,
[C/Fe]~=~+0.86) with both low Ba and Sr ([Ba/Fe]~=~$-1.26$, [Sr/Fe]~<~$-2.05$).
Barium was not detectable in two other CEMP stars
(Pr\_134.3232+17.6970, and Pr\_228.6558+9.0914) with [Fe/H]~<~$-2.0$. They are all carbon-rich, but they
are turn-off stars with effective temperatures~$\ge$~6350K. At these temperatures the barium spectral features are very weak and require a much higher SNR.

We confirm the finding of \citet{Venn2020} that Pr\_214.5557+7.4670
([Fe/H]~=~$-2.14$) is enriched in Ba. This is most probably a CEMP-s candidate
with [Ba/Fe]~=~+1.90 and [C/Fe]~=~+2.22 \citep[Fig.7 of][]{Matsuno2017}. Unfortunately, its strontium lines are buried in the noise,
and the Y lines very much so as well.

The bottom panel of Fig.~\ref{Fig:neutron-capture} presents [Sr/Fe] as a
function of [Fe/H]. Although our sample is relatively devoid of low abundance
ratios, it matches the distribution of the MW stars known so far. One star,
Pr\_210.7513+12.7744 at [Fe/H]~=~$-2.12$, stands out of the general
distribution with a significantly lower strontium content level
[Sr/Fe]~=~$-$1.10  for its metallicity, while its Ba
content is normal ([Ba/Fe]~=~$+0.49$). This depletion in Sr has essentially been
observed in (most of) the ultra-faint dwarfs (UFDs) and is so far understood
as the evidence for the second channel of Sr production to be missing in these
faint systems, possibly by undersampling of the initial mass function
\citep[e.g.,][]{Tafelmeyer2010,Jablonka2015,Mashonkina2017,Ji2019,Sitnova2021}.
Interestingly the kinematic analysis of the orbit of
Pr\_210.7513+12.7744 confirms that it is a halo member, with an orbit almost
perpendicular to the plane of the Milky Way (see Section~\ref{orbits}), with an apocenter of R$_{\rm{apo}}$=10.7 $^{+1.8}_{-1.3}$kpc.

\section{Orbits}\label{orbits}

Thanks to Gaia EDR3 \citep{GaiaCollaboration2016,GaiaCollaboration2021}, we can now measure the distances and the orbital parameters of our sample with increased accuracy. The first step for determining the kinematic properties of our stars is to measure their distances. Since it is ill advised to invert the parallax \citep{Bailer2015}, we infer the distances using a Bayesian inference method. The posterior probability on the distance is composed by two factors, a Gaussian likelihood on the parallax and a prior on the stellar density distribution as in Eq. 8 from \citet{Sestito2019}. We choose a method that does not depend on theoretical isochrones, thereby differing from previous \Pristine\ papers \citep[e.g.][]{Sestito2020,Venn2020}. 
For the zero-point on the Gaia EDR3 parallax, we use the python code \texttt{gaiadr3$\_$zeropoint}\footnote{https://gitlab.com/icc-ub/public/gaiadr3\_zeropoint} as described in \citet{Lindegren2020}. Then, we use \texttt{galpy} package \citep{Bovy2015} to determine the orbital parameters. For this analysis, we modify their \texttt{MWPotential14} assuming a more massive halo of $M$ = $1.2 \times 10^{12} M_{\odot}$ compatible with the value from \citet{BlandHawthorn2016}, as fully described in \citet[][and references therein]{Sestito2019}.  We run the orbital inference also for the  112 stars from \citet{Venn2020}, since Gaia EDR3 provides a better measurement of the astrometric solution than Gaia DR2 \citep{GaiaCollaboration2018}, and therefore on the distance and the orbits.

\begin{figure}
    \centering
    \includegraphics[width=1.0\columnwidth]{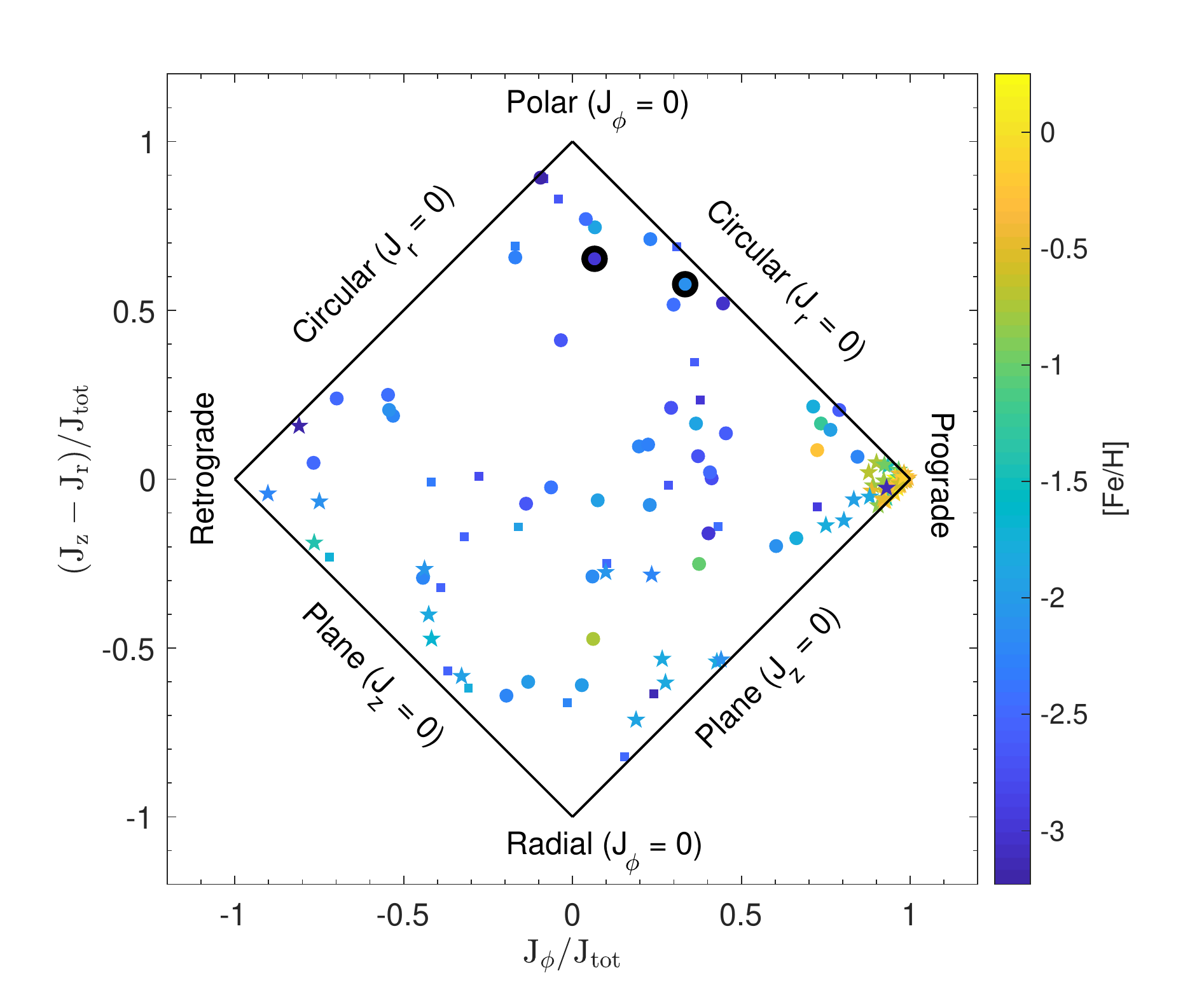}
    \caption{Action plot of the stellar sample colour-coded by metallicity. The x-axis is the azimuthal component of the action vector, proxy for rotation. The vertical axis is the difference between the vertical and the radial component of the action vector. Both the axis are normalised by $J_{tot}$ = $J_r$ + $J_z$ + $|J_{\phi}|$. Squares denote outer halo stars ($R_{apo}\geq 15\kpc$ and $Z_{max}> 3.5\kpc$), circles marked the inner halo stars ($R_{apo}< 15\kpc$ and $Z_{max}> 3.5\kpc$), while the star symbol denotes the stars confined to the MW plane ($Z_{max}\leq 3.5\kpc$). The two bigger markers with edge colour in black represent Pr\_210.7513+12.7744  and Pr\_255.8043+10.8443.}
    \label{Fig:Action}
\end{figure}

Figure \ref{Fig:Action} show the azimuthal component of the action vector vs. the difference between the vertical and the radial components of the action, both axis normalised by the sum of the action components.  In this space, stars with different kinematics occupy  different portions of this diagram. We divide the total sample of 132 stars into three groups: outer halo ($R_{apo}\geq 15\kpc$ and $Z_{max}> 3.5\kpc$), inner halo ($R_{apo}< 15\kpc$ and $Z_{max}> 3.5\kpc$), and confined to the disc ($Z_{max}\leq 3.5\kpc$). The limits on the apocentre distance $R_{apo}$ and the maximum height $Z_{max}$ from the MW plane are arbitrarily chosen and follow \citet{Sestito2019,Sestito2020}. The chemical distribution of these populations is  illustrated in Fig.\ref{Fig:alpha}.

This \Pristine\--\Espadons\ sample is composed of 65 halo stars. The 22 outer halo stars have [Fe/H] $\le -1.75$. The 43  inner halo stars share
the same metallicity distribution at the exception of 4 stars, 3 prograde  
and one J$_\phi$=0, at [Fe/H]$> - 1.5$ that overlay on the region covers by the MW disc stars.  These 4 stars have a $Z_{max}$ between 3.9 to 7 kpc and the apocentre is between 8 to 9.6 kpc. Two of them have a high eccentricity, 0.77 and 0.9. Their maximum height place them clearly above the plane. They might be born in the disc and then heated up afterwards. In total 67 stars, mostly prograde, are confined in the MW disc. This planar subsample contains 10 high eccentricity stars in the range $-2.25 \le $[Fe/H]$\le -1.56 $, 7 of which are prograde, and 4 retrograde. 
There are also two EMP stars on a disk orbit, one is prograde and the other is retrograde. Finally, all but one planar stars at [Fe/H]$\ge -1.5$ are prograde. 
A full analysis of this sample, in particular regarding the possible association with known stellar structures, is beyond the scope of this paper. However, we focus on two objects, Pr\_210.7513+12.7744  and Pr\_255.8043+10.8443, whose chemistry stands out from the rest of the sample. The former has a low content of Sr, while the latter is a new EMP star.
Both the stars are located in the top region of the action plot in Figure \ref{Fig:Action} (see the larger markers), and they display a prograde polar orbit. This region has been shown to be occupied by the recently discovered LMS-1 stream \citep{Yuan2020,Malhan2021}, a 60 degree long structure wrapping around the inner region of the Milky Way. The proper motion and position on the sky of Pr\_255.8043+10.8443 do not match the ones of LMS-1 \citep[see figure 4 of][]{Malhan2021}, hence we can exclude a possible association with this stream. As to Pr\_210.7513+12.7744, we find differences in RV ($RV\sim40$ km s$^{-1}$) and right ascension ($\sim8$ degree) between this star and the best fit orbit of LMS-1 \citep[see figure 4 of][]{Malhan2021}. The comparison of the orbital parameters of Pr\_210.7513+12.7744 with the ones of the simulated stars in this stream indicates that an association with the leading trail can be excluded. However, this star might belong to the older wraps of LMS-1, which display a much larger dispersion on their orbital parameters than the leading trail. Would Pr\_210.7513+12.7744 be confirmed as a member in the future, it would undoubtedly open new insight into the star formation history of the parent galaxy of LMS-1.

\section{Summary and Conclusions}\label{conclusion}

We have presented the homogeneous 1D, LTE analysis of 132 stars observed at
high-resolution with ESPaDOnS, so far the largest sample at high resolution
(R$\sim$40,000) from the \Pristine\ survey. This study expands on the earlier work
of \citet{Venn2020}, in which only 28 very metal-poor stars were fully chemically
characterized. Because this sample is based on the first version of the
\Pristine\ catalog, the success rate of identification of genuine EMP stars is
not as high as in the later versions. As a consequence, the range of metallicity
of our sample extends much beyond $-2$, reaching [Fe/H]=$+0.25$.  Nevertheless,
near half of our sample (58 stars) is composed of very metal-poor stars
([Fe/H]$\le -2$). The more metal-rich stars offer us the
opportunity of a new and detailed study of the Milky Way halo stellar
population. Because it encompasses both dwarf and giant stars, it also enables the analysis of any potential biases induced by the \Pristine\ selection process.

Based on Gaia EDR3, the orbital analysis of this \Pristine\--\Espadons\ sample showed that it is composed of 65 halo stars and 67 disc stars. After a general assessment of the sample chemical properties with the
$\alpha$-elements Mg and Ca, we focused on the abundance of carbon and the neutron capture elements, Ba and Sr. Our results can be summarized as follows.

\begin{itemize}

\item We presented a chemical analysis of 31 newly identified very metal-poor stars, out of which 23 were already presented in \citet{Venn2020} but they were not identified as VMP stars in their Q6 analysis. Eight VMP stars (including 5 stars with [Fe/H]~<~$-2.5$ and one EMP star at
  [Fe/H]~=~$-3$) were identified in the subset of 20 stars analysed in this study for the first time.
  
\item Comparing the earliest and latest version of the \Pristine\ catalogs, it
  appears that some very metal-poor stars (6) are missed because their SDSS
  magnitudes, at the bright end of the selection, are saturated or polluted by
  instrumental failure. The latest \Pristine\ catalogues conservatively reject these objects, even if the \Pristine\ metallicity estimate are in fact correct.

\item We provide carbon abundances for 97 stars and upper limits for the rest of the sample. From the 38 stars with  [Fe/H]~$\leq -2.0$ and carbon measurements, 14 are CEMP stars following the criterion of \citet{Aoki2007}, which sets the C-enrichment threshold at [C/Fe]=+0.7. This  results in a global frequency of CEMP stars at 37\%, which is consistent with other studies.

\item However, we almost completely miss the C-rich stars in very metal-poor giants, with only 11\% of CEMP stars compared to 22\% in the sample of \citet{Placco2014}. This is a clear sign for \Pristine\ selection bias  against carbon-rich giants, which will be analysed in a future work.

\item Looking at the abundances in Ba, a few very metal-poor stars stand out:  Pr\_245.5747+6.8844  at [Fe/H]=$-3.17$ is a CEMP-no star.  Pr\_180.2206+9.5683 at
  [Fe/H]=$-2.96$ is another CEMP-no candidate. Pr\_214.5557+7.4670 ([Fe/H] =
  $-2.14$) is most probably a CEMP-s star.  

\item While most our sample is a good match to the known Sr content of the Milky
  Way population, one star, Pr\_210.7513+12.7744 at [Fe/H]=$-2.12$ has a
  particularly low [Sr/Fe]=$-1.10$ for its metallicity. This is typical of the
  abundance ratios found in most of the UFDs, making it a possible fossil of
  accretion in the Milky Way halo. The orbit of Pr\_210.7513+12.7744 is
  perpendicular to the Milky Way plane. Its kinematical parameters are not far
  from those of the older wraps of LMS-1 stream.

\end{itemize}

This work clearly shows the enormous potential of the {\em Pristine} survey and its spectroscopic follow-ups at low- and high-resolution. Many open issues in modern astrophysics and cosmology can be tackled thanks to the accurate chemical tagging of the EMPs identified by the Pristine photometry and studied with spectroscopy, especially when abundances are combined with the information provided by Gaia.
For example, large samples of stars with a chemo-dynamical characterisation, like the one presented in this paper, can be used to check for possible associations with streams in the halo \citep[e.g.][]{Venn2020,Kielty2021}. Along the same lines, it would be of interest to obtain high-resolution observations covering larger wavelength ranges (and spectral features of a wider variety of elements) for the two stars that chemo-dynamically stand out from the others (e.g., Pr\_210.7513+12.7744 and Pr\_255.8043+10.8443; see~\S~\ref{orbits}) to investigate for possible associations with known 
structures/accretion events.

Finally, upcoming spectroscopic surveys with high multiplex capabilities --e.g. WAVE \citep{Dalton2012}, 4MOST \citep{DeJong2019}; will provide us with even larger and more representative samples of metal deficient stars to investigate in fine details the first stages of chemical enrichment of the Milky Way.

\section*{Acknowledgements}

We warmly thank Ian Roederer who very kindly sent us his compilation of Sr and Ba
measurements in Milky Way halo stars. This work is based on observations obtained with \Espadons\ as well as MegaPrime/MegaCam, a joint project of CFHT and CEA/DAPNIA, at the Canada-France-Hawaii Telescope (CFHT) which is operated by the National Research Council (NRC) of Canada, the Institut National des Science de l'Univers of the Centre National de la Recherche Scientifique (CNRS) of France, and the University of Hawaii. The observations at the Canada-France-Hawaii Telescope were performed with care and respect from the summit of Maunakea which is a significant cultural and historic site. NFM gratefully acknowledge support from the French National Research Agency (ANR) funded project “Pristine” (ANR-18-CE31-0017) along with funding from CNRS/INSU through the Programme National Galaxies et Cosmologie and through the CNRS grant PICS07708 and from the European Research Council (ERC) under the European Unions Horizon 2020 research and innovation programme (grant agreement No. 834148). The authors thank the International Space Science
Institute, Berne, Switzerland for providing financial support and meeting
facilities to the international team ``Pristine''. CL acknowledges funding from
Ministero dell'Università e della Ricerca through the Programme ``Rita Levi
Montalcini'' (grant PGR18YRML1). 
ES acknowledges funding through VIDI grant "Pushing Galactic Archaeology to its limits" (with project number VI.Vidi.193.093) which is funded by the Dutch Research Council (NWO).
KAV is grateful for funding through the National Science and Engineering Research Council Discovery Grants program. This work has made use of data from the
European Space Agency (ESA) mission {\it Gaia}
(\url{https://www.cosmos.esa.int/gaia}), processed by the {\it Gaia} Data
Processing and Analysis Consortium (DPAC,
\url{https://www.cosmos.esa.int/web/gaia/dpac/consortium}). Funding for the DPAC
has been provided by national institutions, in particular the institutions
participating in the {\it Gaia} Multilateral Agreement.

\section*{data availability}

The data underlying this article will be shared on reasonable request to the corresponding author.



\bibliographystyle{mnras}
\bibliography{bibtex/biblio} 

\begin{thebibliography}{}
\makeatletter
\relax
\def\mn@urlcharsother{\let\do\@makeother \do\$\do\&\do\#\do\^\do\_\do\%\do\~}
\def\mn@doi{\begingroup\mn@urlcharsother \@ifnextchar [ {\mn@doi@}
  {\mn@doi@[]}}
\def\mn@doi@[#1]#2{\def\@tempa{#1}\ifx\@tempa\@empty \href
  {http://dx.doi.org/#2} {doi:#2}\else \href {http://dx.doi.org/#2} {#1}\fi
  \endgroup}
\def\mn@eprint#1#2{\mn@eprint@#1:#2::\@nil}
\def\mn@eprint@arXiv#1{\href {http://arxiv.org/abs/#1} {{\tt arXiv:#1}}}
\def\mn@eprint@dblp#1{\href {http://dblp.uni-trier.de/rec/bibtex/#1.xml}
  {dblp:#1}}
\def\mn@eprint@#1:#2:#3:#4\@nil{\def\@tempa {#1}\def\@tempb {#2}\def\@tempc
  {#3}\ifx \@tempc \@empty \let \@tempc \@tempb \let \@tempb \@tempa \fi \ifx
  \@tempb \@empty \def\@tempb {arXiv}\fi \@ifundefined
  {mn@eprint@\@tempb}{\@tempb:\@tempc}{\expandafter \expandafter \csname
  mn@eprint@\@tempb\endcsname \expandafter{\@tempc}}}

\bibitem[\protect\citeauthoryear{{Aguado} et~al.,}{{Aguado}
  et~al.}{2019}]{Aguado2019}
{Aguado} D.~S.,  et~al., 2019, \mn@doi [\mnras] {10.1093/mnras/stz2643}, \href
  {https://ui.adsabs.harvard.edu/abs/2019MNRAS.490.2241A} {490, 2241}

\bibitem[\protect\citeauthoryear{{Angelou}, {Stancliffe}, {Church}, {Lattanzio}
   \& {Smith}}{{Angelou} et~al.}{2012}]{Angelou2012}
{Angelou} G.~C.,  {Stancliffe} R.~J.,  {Church} R.~P.,  {Lattanzio} J.~C.,
  {Smith} G.~H.,  2012, \mn@doi [\apj] {10.1088/0004-637X/749/2/128}, \href
  {https://ui.adsabs.harvard.edu/abs/2012ApJ...749..128A} {749, 128}

\bibitem[\protect\citeauthoryear{{Aoki}, {Beers}, {Christlieb}, {Norris},
  {Ryan}  \& {Tsangarides}}{{Aoki} et~al.}{2007}]{Aoki2007}
{Aoki} W.,  {Beers} T.~C.,  {Christlieb} N.,  {Norris} J.~E.,  {Ryan} S.~G.,
  {Tsangarides} S.,  2007, \mn@doi [\apj] {10.1086/509817}, \href
  {https://ui.adsabs.harvard.edu/abs/2007ApJ...655..492A} {655, 492}

\bibitem[\protect\citeauthoryear{{Arlandini}, {K{\"a}ppeler}, {Wisshak},
  {Gallino}, {Lugaro}, {Busso}  \& {Straniero}}{{Arlandini}
  et~al.}{1999}]{Arlandini1999}
{Arlandini} C.,  {K{\"a}ppeler} F.,  {Wisshak} K.,  {Gallino} R.,  {Lugaro} M.,
   {Busso} M.,   {Straniero} O.,  1999, \mn@doi [\apj] {10.1086/307938}, \href
  {https://ui.adsabs.harvard.edu/abs/1999ApJ...525..886A} {525, 886}

\bibitem[\protect\citeauthoryear{{Asplund}, {Grevesse}, {Sauval}  \&
  {Scott}}{{Asplund} et~al.}{2009}]{Asplund2009}
{Asplund} M.,  {Grevesse} N.,  {Sauval} A.~J.,   {Scott} P.,  2009, \mn@doi
  [\araa] {10.1146/annurev.astro.46.060407.145222}, \href
  {http://adsabs.harvard.edu/abs/2009ARA%26A..47..481A} {47, 481}

\bibitem[\protect\citeauthoryear{Bailer-Jones}{Bailer-Jones}{2015}]{Bailer2015}
Bailer-Jones C. A.~L.,  2015, Publications of the Astronomical Society of the
  Pacific, 127, 994

\bibitem[\protect\citeauthoryear{{Beers} \& {Christlieb}}{{Beers} \&
  {Christlieb}}{2005}]{Beers2005}
{Beers} T.~C.,  {Christlieb} N.,  2005, \mn@doi [\araa]
  {10.1146/annurev.astro.42.053102.134057}, \href
  {https://ui.adsabs.harvard.edu/abs/2005ARA&A..43..531B} {43, 531}

\bibitem[\protect\citeauthoryear{{Beers}, {Preston}  \& {Shectman}}{{Beers}
  et~al.}{1992}]{Beers1992}
{Beers} T.~C.,  {Preston} G.~W.,   {Shectman} S.~A.,  1992, \mn@doi [\aj]
  {10.1086/116207}, \href
  {https://ui.adsabs.harvard.edu/abs/1992AJ....103.1987B} {103, 1987}

\bibitem[\protect\citeauthoryear{{Bensby}, {Feltzing}  \& {Oey}}{{Bensby}
  et~al.}{2014}]{Bensby2014}
{Bensby} T.,  {Feltzing} S.,   {Oey} M.~S.,  2014, \mn@doi [\aap]
  {10.1051/0004-6361/201322631}, \href
  {https://ui.adsabs.harvard.edu/abs/2014A&A...562A..71B} {562, A71}

\bibitem[\protect\citeauthoryear{{Bisterzo}, {Gallino}, {Straniero},
  {Cristallo}  \& {K{\"a}ppeler}}{{Bisterzo} et~al.}{2012}]{Bisterzo2012}
{Bisterzo} S.,  {Gallino} R.,  {Straniero} O.,  {Cristallo} S.,
  {K{\"a}ppeler} F.,  2012, \mn@doi [\mnras]
  {10.1111/j.1365-2966.2012.20670.x}, \href
  {https://ui.adsabs.harvard.edu/abs/2012MNRAS.422..849B} {422, 849}

\bibitem[\protect\citeauthoryear{{Bland-Hawthorn} \&
  {Gerhard}}{{Bland-Hawthorn} \& {Gerhard}}{2016}]{BlandHawthorn2016}
{Bland-Hawthorn} J.,  {Gerhard} O.,  2016, \mn@doi [\araa]
  {10.1146/annurev-astro-081915-023441}, \href
  {https://ui.adsabs.harvard.edu/abs/2016ARA&A..54..529B} {54, 529}

\bibitem[\protect\citeauthoryear{{Blanton} et~al.,}{{Blanton}
  et~al.}{2017}]{Blanton2017}
{Blanton} M.~R.,  et~al., 2017, \mn@doi [\aj] {10.3847/1538-3881/aa7567}, \href
  {http://adsabs.harvard.edu/abs/2017AJ....154...28B} {154, 28}

\bibitem[\protect\citeauthoryear{{Bonifacio} et~al.,}{{Bonifacio}
  et~al.}{2019}]{Bonifacio2019}
{Bonifacio} P.,  et~al., 2019, \mn@doi [MNRAS] {10.1093/mnras/stz1378}, \href
  {https://ui.adsabs.harvard.edu/abs/2019MNRAS.487.3797B} {487, 3797}

\bibitem[\protect\citeauthoryear{{Bovy}}{{Bovy}}{2015}]{Bovy2015}
{Bovy} J.,  2015, \mn@doi [\apjs] {10.1088/0067-0049/216/2/29}, \href
  {https://ui.adsabs.harvard.edu/abs/2015ApJS..216...29B} {216, 29}

\bibitem[\protect\citeauthoryear{{Briley}, {Bell}, {Hoban}  \&
  {Dickens}}{{Briley} et~al.}{1990}]{Briley1990}
{Briley} M.~M.,  {Bell} R.~A.,  {Hoban} S.,   {Dickens} R.~J.,  1990, \mn@doi
  [\apj] {10.1086/169066}, \href
  {https://ui.adsabs.harvard.edu/abs/1990ApJ...359..307B} {359, 307}

\bibitem[\protect\citeauthoryear{{Bromm} \& {Larson}}{{Bromm} \&
  {Larson}}{2004}]{Bromm2004}
{Bromm} V.,  {Larson} R.~B.,  2004, \mn@doi [\araa]
  {10.1146/annurev.astro.42.053102.134034}, \href
  {https://ui.adsabs.harvard.edu/abs/2004ARA&A..42...79B} {42, 79}

\bibitem[\protect\citeauthoryear{{Bromm} \& {Loeb}}{{Bromm} \&
  {Loeb}}{2003}]{Bromm2003}
{Bromm} V.,  {Loeb} A.,  2003, \mn@doi [\nat] {10.1038/nature02071}, \href
  {https://ui.adsabs.harvard.edu/abs/2003Natur.425..812B} {425, 812}

\bibitem[\protect\citeauthoryear{{Busso}, {Gallino}  \& {Wasserburg}}{{Busso}
  et~al.}{1999}]{Busso1999}
{Busso} M.,  {Gallino} R.,   {Wasserburg} G.~J.,  1999, \mn@doi [\araa]
  {10.1146/annurev.astro.37.1.239}, \href
  {https://ui.adsabs.harvard.edu/abs/1999ARA&A..37..239B} {37, 239}

\bibitem[\protect\citeauthoryear{{Caffau} et~al.,}{{Caffau}
  et~al.}{2011}]{Caffau2011}
{Caffau} E.,  et~al., 2011, \mn@doi [Nature] {10.1038/nature10377}, \href
  {http://adsabs.harvard.edu/abs/2011Natur.477...67C} {477, 67}

\bibitem[\protect\citeauthoryear{{Caffau} et~al.,}{{Caffau}
  et~al.}{2013}]{Caffau2013}
{Caffau} E.,  et~al., 2013, \mn@doi [\aap] {10.1051/0004-6361/201322488}, \href
  {https://ui.adsabs.harvard.edu/abs/2013A&A...560A..71C} {560, A71}

\bibitem[\protect\citeauthoryear{{Caffau} et~al.,}{{Caffau}
  et~al.}{2017}]{Caffau2017}
{Caffau} E.,  et~al., 2017, \mn@doi [Astronomische Nachrichten]
  {10.1002/asna.201713368}, \href
  {https://ui.adsabs.harvard.edu/abs/2017AN....338..686C} {338, 686}

\bibitem[\protect\citeauthoryear{{Caffau} et~al.,}{{Caffau}
  et~al.}{2020}]{Caffau2020}
{Caffau} E.,  et~al., 2020, \mn@doi [\mnras] {10.1093/mnras/staa589}, \href
  {https://ui.adsabs.harvard.edu/abs/2020MNRAS.493.4677C} {493, 4677}

\bibitem[\protect\citeauthoryear{{Cayrel} et~al.,}{{Cayrel}
  et~al.}{2004}]{Cayrel2004}
{Cayrel} R.,  et~al., 2004, \mn@doi [\aap] {10.1051/0004-6361:20034074}, \href
  {https://ui.adsabs.harvard.edu/abs/2004A&A...416.1117C} {416, 1117}

\bibitem[\protect\citeauthoryear{{Charbonnel}}{{Charbonnel}}{1995}]{Charbonnel1995}
{Charbonnel} C.,  1995, \mn@doi [\apjl] {10.1086/309744}, \href
  {https://ui.adsabs.harvard.edu/abs/1995ApJ...453L..41C} {453, L41}

\bibitem[\protect\citeauthoryear{{Choi}, {Dotter}, {Conroy}, {Cantiello},
  {Paxton}  \& {Johnson}}{{Choi} et~al.}{2016}]{Choi2016}
{Choi} J.,  {Dotter} A.,  {Conroy} C.,  {Cantiello} M.,  {Paxton} B.,
  {Johnson} B.~D.,  2016, \mn@doi [\apj] {10.3847/0004-637X/823/2/102}, \href
  {https://ui.adsabs.harvard.edu/abs/2016ApJ...823..102C} {823, 102}

\bibitem[\protect\citeauthoryear{{Christlieb}, {Sch{\"o}rck}, {Frebel},
  {Beers}, {Wisotzki}  \& {Reimers}}{{Christlieb}
  et~al.}{2008}]{Christlieb2008}
{Christlieb} N.,  {Sch{\"o}rck} T.,  {Frebel} A.,  {Beers} T.~C.,  {Wisotzki}
  L.,   {Reimers} D.,  2008, \mn@doi [\aap] {10.1051/0004-6361:20078748}, \href
  {https://ui.adsabs.harvard.edu/abs/2008A&A...484..721C} {484, 721}

\bibitem[\protect\citeauthoryear{{Cohen}, {Christlieb}, {Thompson},
  {McWilliam}, {Shectman}, {Reimers}, {Wisotzki}  \& {Kirby}}{{Cohen}
  et~al.}{2013}]{Cohen2013}
{Cohen} J.~G.,  {Christlieb} N.,  {Thompson} I.,  {McWilliam} A.,  {Shectman}
  S.,  {Reimers} D.,  {Wisotzki} L.,   {Kirby} E.,  2013, \mn@doi [\apj]
  {10.1088/0004-637X/778/1/56}, \href
  {http://adsabs.harvard.edu/abs/2013ApJ...778...56C} {778, 56}

\bibitem[\protect\citeauthoryear{{Dalton} et~al.,}{{Dalton}
  et~al.}{2012}]{Dalton2012}
{Dalton} G.,  et~al., 2012, in {McLean} I.~S.,  {Ramsay} S.~K.,   {Takami} H.,
  eds,  Society of Photo-Optical Instrumentation Engineers (SPIE) Conference
  Series Vol. 8446, Ground-based and Airborne Instrumentation for Astronomy IV.
  p. 84460P, \mn@doi{10.1117/12.925950}

\bibitem[\protect\citeauthoryear{{Demarque}, {Woo}, {Kim}  \& {Yi}}{{Demarque}
  et~al.}{2004}]{Demarque2004}
{Demarque} P.,  {Woo} J.-H.,  {Kim} Y.-C.,   {Yi} S.~K.,  2004, \mn@doi [\apjs]
  {10.1086/424966}, \href
  {https://ui.adsabs.harvard.edu/abs/2004ApJS..155..667D} {155, 667}

\bibitem[\protect\citeauthoryear{{Deng} et~al.,}{{Deng}
  et~al.}{2012}]{Deng2012}
{Deng} L.-C.,  et~al., 2012, \mn@doi [Research in Astronomy and Astrophysics]
  {10.1088/1674-4527/12/7/003}, \href
  {https://ui.adsabs.harvard.edu/abs/2012RAA....12..735D} {12, 735}

\bibitem[\protect\citeauthoryear{{Donati}, {Catala}, {Landstreet}  \&
  {Petit}}{{Donati} et~al.}{2006}]{Donati2006}
{Donati} J.~F.,  {Catala} C.,  {Landstreet} J.~D.,   {Petit} P.,  2006, in
  {Casini} R.,  {Lites} B.~W.,  eds,  Astronomical Society of the Pacific
  Conference Series Vol. 358, Solar Polarization 4. p.~362

\bibitem[\protect\citeauthoryear{{Dotter}}{{Dotter}}{2016}]{Dotter2016}
{Dotter} A.,  2016, \mn@doi [\apjs] {10.3847/0067-0049/222/1/8}, \href
  {https://ui.adsabs.harvard.edu/abs/2016ApJS..222....8D} {222, 8}

\bibitem[\protect\citeauthoryear{{Eisenstein} et~al.,}{{Eisenstein}
  et~al.}{2011}]{Eisenstein2011}
{Eisenstein} D.~J.,  et~al., 2011, \mn@doi [\aj] {10.1088/0004-6256/142/3/72},
  \href {https://ui.adsabs.harvard.edu/abs/2011AJ....142...72E} {142, 72}

\bibitem[\protect\citeauthoryear{{Fern{\'a}ndez-Trincado}
  et~al.,}{{Fern{\'a}ndez-Trincado} et~al.}{2017}]{Fernandez2017}
{Fern{\'a}ndez-Trincado} J.~G.,  et~al., 2017, \mn@doi [\apjl]
  {10.3847/2041-8213/aa8032}, \href
  {https://ui.adsabs.harvard.edu/abs/2017ApJ...846L...2F} {846, L2}

\bibitem[\protect\citeauthoryear{{Frebel}}{{Frebel}}{2018}]{Frebel2018}
{Frebel} A.,  2018, \mn@doi [Annual Review of Nuclear and Particle Science]
  {10.1146/annurev-nucl-101917-021141}, \href
  {https://ui.adsabs.harvard.edu/abs/2018ARNPS..68..237F} {68, 237}

\bibitem[\protect\citeauthoryear{{Frebel} \& {Norris}}{{Frebel} \&
  {Norris}}{2015}]{Frebel2015}
{Frebel} A.,  {Norris} J.~E.,  2015, \mn@doi [\araa]
  {10.1146/annurev-astro-082214-122423}, \href
  {https://ui.adsabs.harvard.edu/abs/2015ARA&A..53..631F} {53, 631}

\bibitem[\protect\citeauthoryear{{Freiburghaus}, {Rosswog}  \&
  {Thielemann}}{{Freiburghaus} et~al.}{1999}]{Freiburghaus1999}
{Freiburghaus} C.,  {Rosswog} S.,   {Thielemann} F.~K.,  1999, \mn@doi [\apjl]
  {10.1086/312343}, \href
  {https://ui.adsabs.harvard.edu/abs/1999ApJ...525L.121F} {525, L121}

\bibitem[\protect\citeauthoryear{{Gaia Collaboration} et~al.,}{{Gaia
  Collaboration} et~al.}{2016}]{GaiaCollaboration2016}
{Gaia Collaboration} et~al., 2016, \mn@doi [\aap]
  {10.1051/0004-6361/201629272}, \href
  {https://ui.adsabs.harvard.edu/abs/2016A&A...595A...1G} {595, A1}

\bibitem[\protect\citeauthoryear{{Gaia Collaboration} et~al.,}{{Gaia
  Collaboration} et~al.}{2018}]{GaiaCollaboration2018}
{Gaia Collaboration} et~al., 2018, \mn@doi [\aap]
  {10.1051/0004-6361/201833051}, \href
  {https://ui.adsabs.harvard.edu/abs/2018A&A...616A...1G} {616, A1}

\bibitem[\protect\citeauthoryear{{Gaia Collaboration} et~al.,}{{Gaia
  Collaboration} et~al.}{2021}]{GaiaCollaboration2021}
{Gaia Collaboration} et~al., 2021, \mn@doi [\aap]
  {10.1051/0004-6361/202039657}, \href
  {https://ui.adsabs.harvard.edu/abs/2021A&A...649A...1G} {649, A1}

\bibitem[\protect\citeauthoryear{{Gerber}, {Briley}  \& {Smith}}{{Gerber}
  et~al.}{2019}]{Gerber2019}
{Gerber} J.~M.,  {Briley} M.~M.,   {Smith} G.~H.,  2019, \mn@doi [\aj]
  {10.3847/1538-3881/ab0b3f}, \href
  {https://ui.adsabs.harvard.edu/abs/2019AJ....157..154G} {157, 154}

\bibitem[\protect\citeauthoryear{Goswami, Singh~Rathour  \& Goswami}{Goswami
  et~al.}{2021}]{Goswami2021}
Goswami P.~P.,  Singh~Rathour R.,   Goswami A.,  2021, arXiv.org, p.
  arXiv:2101.09518

\bibitem[\protect\citeauthoryear{{Gratton}, {Sneden}, {Carretta}  \&
  {Bragaglia}}{{Gratton} et~al.}{2000}]{Gratton2000}
{Gratton} R.~G.,  {Sneden} C.,  {Carretta} E.,   {Bragaglia} A.,  2000, A\&A,
  \href {http://adsabs.harvard.edu/abs/2000A%26A...354..169G} {354, 169}

\bibitem[\protect\citeauthoryear{{Hansen}, {Andersen}, {Nordstr{\"o}m},
  {Beers}, {Placco}, {Yoon}  \& {Buchhave}}{{Hansen} et~al.}{2016}]{Hansen2016}
{Hansen} T.~T.,  {Andersen} J.,  {Nordstr{\"o}m} B.,  {Beers} T.~C.,  {Placco}
  V.~M.,  {Yoon} J.,   {Buchhave} L.~A.,  2016, \mn@doi [\aap]
  {10.1051/0004-6361/201527409}, \href
  {https://ui.adsabs.harvard.edu/abs/2016A&A...588A...3H} {588, A3}

\bibitem[\protect\citeauthoryear{Hansen, Hansen, Koch, Beers, Nordstr{\"o}m,
  Placco  \& Andersen}{Hansen et~al.}{2019}]{Hansen2019}
Hansen C.~J.,  Hansen T.~T.,  Koch A.,  Beers T.~C.,  Nordstr{\"o}m B.,  Placco
  V.~M.,   Andersen J.,  2019, Astronomy and Astrophysics, 623, A128

\bibitem[\protect\citeauthoryear{{Heger} \& {Woosley}}{{Heger} \&
  {Woosley}}{2010}]{Heger2010}
{Heger} A.,  {Woosley} S.~E.,  2010, \mn@doi [\apj]
  {10.1088/0004-637X/724/1/341}, \href
  {https://ui.adsabs.harvard.edu/abs/2010ApJ...724..341H} {724, 341}

\bibitem[\protect\citeauthoryear{{Hill} et~al.,}{{Hill}
  et~al.}{2019}]{Hill2019}
{Hill} V.,  et~al., 2019, \mn@doi [\aap] {10.1051/0004-6361/201833950}, \href
  {https://ui.adsabs.harvard.edu/abs/2019A&A...626A..15H} {626, A15}

\bibitem[\protect\citeauthoryear{{Hillebrandt}, {Takahashi}  \&
  {Kodama}}{{Hillebrandt} et~al.}{1976}]{Hillebrandt1976}
{Hillebrandt} W.,  {Takahashi} K.,   {Kodama} T.,  1976, \aap, \href
  {https://ui.adsabs.harvard.edu/abs/1976A&A....52...63H} {52, 63}

\bibitem[\protect\citeauthoryear{{Iben}}{{Iben}}{1964}]{Iben1964}
{Iben} Icko J.,  1964, \mn@doi [\apj] {10.1086/148077}, \href
  {https://ui.adsabs.harvard.edu/abs/1964ApJ...140.1631I} {140, 1631}

\bibitem[\protect\citeauthoryear{{Ivans}, {Sneden}, {James}, {Preston},
  {Fulbright}, {H{\"o}flich}, {Carney}  \& {Wheeler}}{{Ivans}
  et~al.}{2003}]{Ivans2003}
{Ivans} I.~I.,  {Sneden} C.,  {James} C.~R.,  {Preston} G.~W.,  {Fulbright}
  J.~P.,  {H{\"o}flich} P.~A.,  {Carney} B.~W.,   {Wheeler} J.~C.,  2003,
  \mn@doi [\apj] {10.1086/375812}, \href
  {https://ui.adsabs.harvard.edu/abs/2003ApJ...592..906I} {592, 906}

\bibitem[\protect\citeauthoryear{{Jablonka} et~al.,}{{Jablonka}
  et~al.}{2015}]{Jablonka2015}
{Jablonka} P.,  et~al., 2015, \mn@doi [\aap] {10.1051/0004-6361/201525661},
  \href {http://adsabs.harvard.edu/abs/2015A%26A...583A..67J} {583, A67}

\bibitem[\protect\citeauthoryear{{Jacobson} et~al.,}{{Jacobson}
  et~al.}{2015}]{Jacobson2015}
{Jacobson} H.~R.,  et~al., 2015, \mn@doi [\apj] {10.1088/0004-637X/807/2/171},
  \href {https://ui.adsabs.harvard.edu/abs/2015ApJ...807..171J} {807, 171}

\bibitem[\protect\citeauthoryear{{Ji}, {Simon}, {Frebel}, {Venn}  \&
  {Hansen}}{{Ji} et~al.}{2019}]{Ji2019}
{Ji} A.~P.,  {Simon} J.~D.,  {Frebel} A.,  {Venn} K.~A.,   {Hansen} T.~T.,
  2019, \mn@doi [\apj] {10.3847/1538-4357/aaf3bb}, \href
  {https://ui.adsabs.harvard.edu/abs/2019ApJ...870...83J} {870, 83}

\bibitem[\protect\citeauthoryear{{Keller} et~al.,}{{Keller}
  et~al.}{2007}]{Keller2007}
{Keller} S.~C.,  et~al., 2007, \mn@doi [\pasa] {10.1071/AS07001}, \href
  {https://ui.adsabs.harvard.edu/abs/2007PASA...24....1K} {24, 1}

\bibitem[\protect\citeauthoryear{{Khoperskov}, {Haywood}, {Snaith}, {Di
  Matteo}, {Lehnert}, {Vasiliev}, {Naroenkov}  \& {Berczik}}{{Khoperskov}
  et~al.}{2021}]{Khoperskov2021}
{Khoperskov} S.,  {Haywood} M.,  {Snaith} O.,  {Di Matteo} P.,  {Lehnert} M.,
  {Vasiliev} E.,  {Naroenkov} S.,   {Berczik} P.,  2021, \mn@doi [\mnras]
  {10.1093/mnras/staa3996}, \href
  {https://ui.adsabs.harvard.edu/abs/2021MNRAS.501.5176K} {501, 5176}

\bibitem[\protect\citeauthoryear{{Kielty} et~al.,}{{Kielty}
  et~al.}{2021}]{Kielty2021}
{Kielty} C.~L.,  et~al., 2021, \mn@doi [\mnras] {10.1093/mnras/stab1783}, \href
  {https://ui.adsabs.harvard.edu/abs/2021MNRAS.506.1438K} {506, 1438}

\bibitem[\protect\citeauthoryear{{Kratz}, {Farouqi}  \& {M{\"o}ller}}{{Kratz}
  et~al.}{2014}]{Kratz2014}
{Kratz} K.-L.,  {Farouqi} K.,   {M{\"o}ller} P.,  2014, \mn@doi [\apj]
  {10.1088/0004-637X/792/1/6}, \href
  {https://ui.adsabs.harvard.edu/abs/2014ApJ...792....6K} {792, 6}

\bibitem[\protect\citeauthoryear{{Kupka}, {Ryabchikova}, {Piskunov}, {Stempels}
   \& {Weiss}}{{Kupka} et~al.}{2000}]{Kupka2000}
{Kupka} F.~G.,  {Ryabchikova} T.~A.,  {Piskunov} N.~E.,  {Stempels} H.~C.,
  {Weiss} W.~W.,  2000, \mn@doi [Baltic Astronomy] {10.1515/astro-2000-0420},
  \href {http://adsabs.harvard.edu/abs/2000BaltA...9..590K} {9, 590}

\bibitem[\protect\citeauthoryear{{Lattimer} \& {Schramm}}{{Lattimer} \&
  {Schramm}}{1974}]{Lattimer1974}
{Lattimer} J.~M.,  {Schramm} D.~N.,  1974, \mn@doi [\apjl] {10.1086/181612},
  \href {https://ui.adsabs.harvard.edu/abs/1974ApJ...192L.145L} {192, L145}

\bibitem[\protect\citeauthoryear{{Lind}, {Primas}, {Charbonnel}, {Grundahl}  \&
  {Asplund}}{{Lind} et~al.}{2009}]{Lind2009}
{Lind} K.,  {Primas} F.,  {Charbonnel} C.,  {Grundahl} F.,   {Asplund} M.,
  2009, \mn@doi [\aap] {10.1051/0004-6361/200912524}, \href
  {https://ui.adsabs.harvard.edu/abs/2009A&A...503..545L} {503, 545}

\bibitem[\protect\citeauthoryear{{Lindegren} et~al.,}{{Lindegren}
  et~al.}{2021}]{Lindegren2020}
{Lindegren} L.,  et~al., 2021, \mn@doi [\aap] {10.1051/0004-6361/202039653},
  \href {https://ui.adsabs.harvard.edu/abs/2021A&A...649A...4L} {649, A4}

\bibitem[\protect\citeauthoryear{{Lucchesi} et~al.,}{{Lucchesi}
  et~al.}{2020}]{Lucchesi2020}
{Lucchesi} R.,  et~al., 2020, \mn@doi [\aap] {10.1051/0004-6361/202037534},
  \href {https://ui.adsabs.harvard.edu/abs/2020A&A...644A..75L} {644, A75}

\bibitem[\protect\citeauthoryear{{Mackereth} et~al.,}{{Mackereth}
  et~al.}{2019}]{Mackereth2019}
{Mackereth} J.~T.,  et~al., 2019, \mn@doi [\mnras] {10.1093/mnras/sty2955},
  \href {https://ui.adsabs.harvard.edu/abs/2019MNRAS.482.3426M} {482, 3426}

\bibitem[\protect\citeauthoryear{{Majewski}, {APOGEE Team}  \& {APOGEE-2
  Team}}{{Majewski} et~al.}{2016}]{Majewski2016}
{Majewski} S.~R.,  {APOGEE Team}  {APOGEE-2 Team} 2016, \mn@doi [Astronomische
  Nachrichten] {10.1002/asna.201612387}, \href
  {https://ui.adsabs.harvard.edu/abs/2016AN....337..863M} {337, 863}

\bibitem[\protect\citeauthoryear{{Malhan}, {Yuan}, {Ibata}, {Arentsen},
  {Bellazzini}  \& {Martin}}{{Malhan} et~al.}{2021}]{Malhan2021}
{Malhan} K.,  {Yuan} Z.,  {Ibata} R.,  {Arentsen} A.,  {Bellazzini} M.,
  {Martin} N.~F.,  2021, arXiv e-prints, \href
  {https://ui.adsabs.harvard.edu/abs/2021arXiv210409523M} {p. arXiv:2104.09523}

\bibitem[\protect\citeauthoryear{{Martell}, {Smith}  \& {Briley}}{{Martell}
  et~al.}{2008}]{Martell2008}
{Martell} S.~L.,  {Smith} G.~H.,   {Briley} M.~M.,  2008, \mn@doi [\pasp]
  {10.1086/525060}, \href {http://adsabs.harvard.edu/abs/2008PASP..120....7M}
  {120, 7}

\bibitem[\protect\citeauthoryear{{Mashonkina}, {Jablonka}, {Sitnova},
  {Pakhomov}  \& {North}}{{Mashonkina} et~al.}{2017}]{Mashonkina2017}
{Mashonkina} L.,  {Jablonka} P.,  {Sitnova} T.,  {Pakhomov} Y.,   {North} P.,
  2017, \mn@doi [\aap] {10.1051/0004-6361/201731582}, \href
  {https://ui.adsabs.harvard.edu/abs/2017A&A...608A..89M} {608, A89}

\bibitem[\protect\citeauthoryear{Masseron, Johnson, Plez, van Eck, Primas,
  Goriely  \& Jorissen}{Masseron et~al.}{2010}]{Masseron2010}
Masseron T.,  Johnson J.~A.,  Plez B.,  van Eck S.,  Primas F.,  Goriely S.,
  Jorissen A.,  2010, Astronomy and Astrophysics, 509, A93

\bibitem[\protect\citeauthoryear{Matsuno, Aoki, Suda  \& Li}{Matsuno
  et~al.}{2017}]{Matsuno2017}
Matsuno T.,  Aoki W.,  Suda T.,   Li H.,  2017, Publications of the
  Astronomical Society of Japan, 69, 24

\bibitem[\protect\citeauthoryear{{Mucciarelli}}{{Mucciarelli}}{2013}]{Mucciarelli2013}
{Mucciarelli} A.,  2013, arXiv e-prints, \href
  {https://ui.adsabs.harvard.edu/abs/2013arXiv1311.1403M} {p. arXiv:1311.1403}

\bibitem[\protect\citeauthoryear{{Nishimura}, {Kotake}, {Hashimoto}, {Yamada},
  {Nishimura}, {Fujimoto}  \& {Sato}}{{Nishimura} et~al.}{2006}]{Nishimura2006}
{Nishimura} S.,  {Kotake} K.,  {Hashimoto} M.-a.,  {Yamada} S.,  {Nishimura}
  N.,  {Fujimoto} S.,   {Sato} K.,  2006, \mn@doi [\apj] {10.1086/500786},
  \href {https://ui.adsabs.harvard.edu/abs/2006ApJ...642..410N} {642, 410}

\bibitem[\protect\citeauthoryear{{Nissen} \& {Schuster}}{{Nissen} \&
  {Schuster}}{2010}]{Nissen2010}
{Nissen} P.~E.,  {Schuster} W.~J.,  2010, \mn@doi [\aap]
  {10.1051/0004-6361/200913877}, \href
  {https://ui.adsabs.harvard.edu/abs/2010A&A...511L..10N} {511, L10}

\bibitem[\protect\citeauthoryear{{Norris} \& {Yong}}{{Norris} \&
  {Yong}}{2019}]{Norris2019}
{Norris} J.~E.,  {Yong} D.,  2019, \mn@doi [\apj] {10.3847/1538-4357/ab1f84},
  \href {https://ui.adsabs.harvard.edu/abs/2019ApJ...879...37N} {879, 37}

\bibitem[\protect\citeauthoryear{{Pagel}}{{Pagel}}{1997}]{Pagel1997}
{Pagel} B. E.~J.,  1997, {Nucleosynthesis and Chemical Evolution of Galaxies}

\bibitem[\protect\citeauthoryear{{Paxton}, {Bildsten}, {Dotter}, {Herwig},
  {Lesaffre}  \& {Timmes}}{{Paxton} et~al.}{2011}]{Paxton2011}
{Paxton} B.,  {Bildsten} L.,  {Dotter} A.,  {Herwig} F.,  {Lesaffre} P.,
  {Timmes} F.,  2011, \mn@doi [\apjs] {10.1088/0067-0049/192/1/3}, \href
  {https://ui.adsabs.harvard.edu/abs/2011ApJS..192....3P} {192, 3}

\bibitem[\protect\citeauthoryear{{Piskunov}, {Kupka}, {Ryabchikova}, {Weiss}
  \& {Jeffery}}{{Piskunov} et~al.}{1995}]{Piskunov1995}
{Piskunov} N.~E.,  {Kupka} F.,  {Ryabchikova} T.~A.,  {Weiss} W.~W.,
  {Jeffery} C.~S.,  1995, \aaps, \href
  {http://adsabs.harvard.edu/abs/1995A%26AS..112..525P} {112, 525}

\bibitem[\protect\citeauthoryear{{Placco}, {Frebel}, {Beers}  \&
  {Stancliffe}}{{Placco} et~al.}{2014}]{Placco2014}
{Placco} V.~M.,  {Frebel} A.,  {Beers} T.~C.,   {Stancliffe} R.~J.,  2014,
  \mn@doi [\apj] {10.1088/0004-637X/797/1/21}, \href
  {https://ui.adsabs.harvard.edu/abs/2014ApJ...797...21P} {797, 21}

\bibitem[\protect\citeauthoryear{{Plez}}{{Plez}}{2012}]{Plez2012}
{Plez} B.,  2012, {Turbospectrum: Code for spectral synthesis}, Astrophysics
  Source Code Library (\mn@eprint {ascl} {1205.004})

\bibitem[\protect\citeauthoryear{{Prochaska} \& {McWilliam}}{{Prochaska} \&
  {McWilliam}}{2000}]{Prochaska2000}
{Prochaska} J.~X.,  {McWilliam} A.,  2000, \mn@doi [\apjl] {10.1086/312749},
  \href {https://ui.adsabs.harvard.edu/abs/2000ApJ...537L..57P} {537, L57}

\bibitem[\protect\citeauthoryear{{Roederer}}{{Roederer}}{2013}]{Roederer2013}
{Roederer} I.~U.,  2013, \mn@doi [\aj] {10.1088/0004-6256/145/1/26}, \href
  {https://ui.adsabs.harvard.edu/abs/2013AJ....145...26R} {145, 26}

\bibitem[\protect\citeauthoryear{{Roederer}, {Preston}, {Thompson}, {Shectman},
  {Sneden}, {Burley}  \& {Kelson}}{{Roederer} et~al.}{2014}]{Roederer2014}
{Roederer} I.~U.,  {Preston} G.~W.,  {Thompson} I.~B.,  {Shectman} S.~A.,
  {Sneden} C.,  {Burley} G.~S.,   {Kelson} D.~D.,  2014, \mn@doi [\aj]
  {10.1088/0004-6256/147/6/136}, \href
  {https://ui.adsabs.harvard.edu/abs/2014AJ....147..136R} {147, 136}

\bibitem[\protect\citeauthoryear{{Rosswog}, {Davies}, {Thielemann}  \&
  {Piran}}{{Rosswog} et~al.}{2000}]{Rosswog2000}
{Rosswog} S.,  {Davies} M.~B.,  {Thielemann} F.~K.,   {Piran} T.,  2000, \aap,
  \href {https://ui.adsabs.harvard.edu/abs/2000A&A...360..171R} {360, 171}

\bibitem[\protect\citeauthoryear{{Ryabchikova}, {Piskunov}, {Kupka}  \&
  {Weiss}}{{Ryabchikova} et~al.}{1997}]{Ryabchikova1997}
{Ryabchikova} T.~A.,  {Piskunov} N.~E.,  {Kupka} F.,   {Weiss} W.~W.,  1997,
  \mn@doi [Baltic Astronomy] {10.1515/astro-1997-0216}, \href
  {http://adsabs.harvard.edu/abs/1997BaltA...6..244R} {6, 244}

\bibitem[\protect\citeauthoryear{{Sakari} et~al.,}{{Sakari}
  et~al.}{2019}]{Sakari2019}
{Sakari} C.~M.,  et~al., 2019, \mn@doi [\apj] {10.3847/1538-4357/ab0c02}, \href
  {https://ui.adsabs.harvard.edu/abs/2019ApJ...874..148S} {874, 148}

\bibitem[\protect\citeauthoryear{{Schlegel}, {Finkbeiner}  \&
  {Davis}}{{Schlegel} et~al.}{1998}]{Schlegel1998}
{Schlegel} D.~J.,  {Finkbeiner} D.~P.,   {Davis} M.,  1998, \mn@doi [\apj]
  {10.1086/305772}, \href
  {https://ui.adsabs.harvard.edu/abs/1998ApJ...500..525S} {500, 525}

\bibitem[\protect\citeauthoryear{{Sestito} et~al.,}{{Sestito}
  et~al.}{2019}]{Sestito2019}
{Sestito} F.,  et~al., 2019, \mn@doi [\mnras] {10.1093/mnras/stz043}, \href
  {https://ui.adsabs.harvard.edu/abs/2019MNRAS.484.2166S} {484, 2166}

\bibitem[\protect\citeauthoryear{{Sestito} et~al.,}{{Sestito}
  et~al.}{2020}]{Sestito2020}
{Sestito} F.,  et~al., 2020, \mn@doi [\mnras] {10.1093/mnrasl/slaa022}, \href
  {https://ui.adsabs.harvard.edu/abs/2020MNRAS.497L...7S} {497, L7}

\bibitem[\protect\citeauthoryear{{Shetrone} et~al.,}{{Shetrone}
  et~al.}{2019}]{Shetrone2019}
{Shetrone} M.,  et~al., 2019, \mn@doi [\apj] {10.3847/1538-4357/aaff66}, \href
  {https://ui.adsabs.harvard.edu/abs/2019ApJ...872..137S} {872, 137}

\bibitem[\protect\citeauthoryear{{Sitnova} et~al.,}{{Sitnova}
  et~al.}{2021}]{Sitnova2021}
{Sitnova} T.~M.,  et~al., 2021, \mn@doi [\mnras] {10.1093/mnras/stab786}, \href
  {https://ui.adsabs.harvard.edu/abs/2021MNRAS.504.1183S} {504, 1183}

\bibitem[\protect\citeauthoryear{{Spite} et~al.,}{{Spite}
  et~al.}{2005}]{Spite2005}
{Spite} M.,  et~al., 2005, \mn@doi [\aap] {10.1051/0004-6361:20041274}, \href
  {https://ui.adsabs.harvard.edu/abs/2005A&A...430..655S} {430, 655}

\bibitem[\protect\citeauthoryear{{Spite} et~al.,}{{Spite}
  et~al.}{2006}]{Spite2006}
{Spite} M.,  et~al., 2006, \mn@doi [\aap] {10.1051/0004-6361:20065209}, \href
  {https://ui.adsabs.harvard.edu/abs/2006A&A...455..291S} {455, 291}

\bibitem[\protect\citeauthoryear{Spite, Caffau, Bonifacio, Spite, Ludwig, Plez
  \& Christlieb}{Spite et~al.}{2013}]{Spite2013}
Spite M.,  Caffau E.,  Bonifacio P.,  Spite F.,  Ludwig H.~G.,  Plez B.,
  Christlieb N.,  2013, Astronomy and Astrophysics, 552, A107

\bibitem[\protect\citeauthoryear{{Starkenburg} et~al.,}{{Starkenburg}
  et~al.}{2017}]{Starkenburg2017}
{Starkenburg} E.,  et~al., 2017, \mn@doi [MNRAS] {10.1093/mnras/stx1068}, \href
  {http://adsabs.harvard.edu/abs/2017MNRAS.471.2587S} {471, 2587}

\bibitem[\protect\citeauthoryear{{Starkenburg} et~al.,}{{Starkenburg}
  et~al.}{2018}]{Starkenburg2018}
{Starkenburg} E.,  et~al., 2018, \mn@doi [\mnras] {10.1093/mnras/sty2276},
  \href {https://ui.adsabs.harvard.edu/abs/2018MNRAS.481.3838S} {481, 3838}

\bibitem[\protect\citeauthoryear{{Steinmetz} et~al.,}{{Steinmetz}
  et~al.}{2006}]{Steinmetz2006}
{Steinmetz} M.,  et~al., 2006, \mn@doi [\aj] {10.1086/506564}, \href
  {https://ui.adsabs.harvard.edu/abs/2006AJ....132.1645S} {132, 1645}

\bibitem[\protect\citeauthoryear{{Stetson} \& {Pancino}}{{Stetson} \&
  {Pancino}}{2008}]{Stetson2008}
{Stetson} P.~B.,  {Pancino} E.,  2008, \mn@doi [\pasp] {10.1086/596126}, \href
  {http://adsabs.harvard.edu/abs/2008PASP..120.1332S} {120, 1332}

\bibitem[\protect\citeauthoryear{{Sweigart} \& {Mengel}}{{Sweigart} \&
  {Mengel}}{1979}]{Sweigart1979}
{Sweigart} A.~V.,  {Mengel} J.~G.,  1979, \mn@doi [\apj] {10.1086/156996},
  \href {https://ui.adsabs.harvard.edu/abs/1979ApJ...229..624S} {229, 624}

\bibitem[\protect\citeauthoryear{{Tafelmeyer} et~al.,}{{Tafelmeyer}
  et~al.}{2010}]{Tafelmeyer2010}
{Tafelmeyer} M.,  et~al., 2010, \mn@doi [\aap] {10.1051/0004-6361/201014733},
  \href {http://adsabs.harvard.edu/abs/2010A%26A...524A..58T} {524, A58}

\bibitem[\protect\citeauthoryear{{Theler} et~al.,}{{Theler}
  et~al.}{2020}]{Theler2020}
{Theler} R.,  et~al., 2020, \mn@doi [\aap] {10.1051/0004-6361/201937146}, \href
  {https://ui.adsabs.harvard.edu/abs/2020A&A...642A.176T} {642, A176}

\bibitem[\protect\citeauthoryear{{Thielemann}, {Eichler}, {Panov}  \&
  {Wehmeyer}}{{Thielemann} et~al.}{2017}]{Thielemann2017}
{Thielemann} F.~K.,  {Eichler} M.,  {Panov} I.~V.,   {Wehmeyer} B.,  2017,
  \mn@doi [Annual Review of Nuclear and Particle Science]
  {10.1146/annurev-nucl-101916-123246}, \href
  {https://ui.adsabs.harvard.edu/abs/2017ARNPS..67..253T} {67, 253}

\bibitem[\protect\citeauthoryear{{Van der Swaelmen}, {Hill}, {Primas}  \&
  {Cole}}{{Van der Swaelmen} et~al.}{2013}]{VanderSwaelmen2013}
{Van der Swaelmen} M.,  {Hill} V.,  {Primas} F.,   {Cole} A.~A.,  2013, \mn@doi
  [\aap] {10.1051/0004-6361/201321109}, \href
  {http://adsabs.harvard.edu/abs/2013A%26A...560A..44V} {560, A44}

\bibitem[\protect\citeauthoryear{{Venn} et~al.,}{{Venn}
  et~al.}{2020}]{Venn2020}
{Venn} K.~A.,  et~al., 2020, \mn@doi [\mnras] {10.1093/mnras/stz3546}, \href
  {https://ui.adsabs.harvard.edu/abs/2020MNRAS.492.3241V} {492, 3241}

\bibitem[\protect\citeauthoryear{{Wanajo}}{{Wanajo}}{2013}]{Wanajo2013}
{Wanajo} S.,  2013, \mn@doi [\apjl] {10.1088/2041-8205/770/2/L22}, \href
  {https://ui.adsabs.harvard.edu/abs/2013ApJ...770L..22W} {770, L22}

\bibitem[\protect\citeauthoryear{{Wanajo}, {Kajino}, {Mathews}  \&
  {Otsuki}}{{Wanajo} et~al.}{2001}]{Wanajo2001}
{Wanajo} S.,  {Kajino} T.,  {Mathews} G.~J.,   {Otsuki} K.,  2001, \mn@doi
  [\apj] {10.1086/321339}, \href
  {https://ui.adsabs.harvard.edu/abs/2001ApJ...554..578W} {554, 578}

\bibitem[\protect\citeauthoryear{{Woosley} \& {Weaver}}{{Woosley} \&
  {Weaver}}{1986}]{Woosley1986}
{Woosley} S.~E.,  {Weaver} T.~A.,  1986, \mn@doi [\araa]
  {10.1146/annurev.aa.24.090186.001225}, \href
  {https://ui.adsabs.harvard.edu/abs/1986ARA&A..24..205W} {24, 205}

\bibitem[\protect\citeauthoryear{{Woosley}, {Wilson}, {Mathews}, {Hoffman}  \&
  {Meyer}}{{Woosley} et~al.}{1994}]{Woosley1994}
{Woosley} S.~E.,  {Wilson} J.~R.,  {Mathews} G.~J.,  {Hoffman} R.~D.,   {Meyer}
  B.~S.,  1994, \mn@doi [\apj] {10.1086/174638}, \href
  {https://ui.adsabs.harvard.edu/abs/1994ApJ...433..229W} {433, 229}

\bibitem[\protect\citeauthoryear{{Xing}, {Zhao}, {Aoki}, {Honda}, {Li},
  {Ishigaki}  \& {Matsuno}}{{Xing} et~al.}{2019}]{Xing2019}
{Xing} Q.-F.,  {Zhao} G.,  {Aoki} W.,  {Honda} S.,  {Li} H.-N.,  {Ishigaki}
  M.~N.,   {Matsuno} T.,  2019, \mn@doi [Nature Astronomy]
  {10.1038/s41550-019-0764-5}, \href
  {https://ui.adsabs.harvard.edu/abs/2019NatAs...3..631X} {3, 631}

\bibitem[\protect\citeauthoryear{{Yanny} et~al.,}{{Yanny}
  et~al.}{2009}]{Yanny2009}
{Yanny} B.,  et~al., 2009, \mn@doi [\aj] {10.1088/0004-6256/137/5/4377}, \href
  {https://ui.adsabs.harvard.edu/abs/2009AJ....137.4377Y} {137, 4377}

\bibitem[\protect\citeauthoryear{{Yong} et~al.,}{{Yong}
  et~al.}{2013}]{Yong2013}
{Yong} D.,  et~al., 2013, \mn@doi [\apj] {10.1088/0004-637X/762/1/26}, \href
  {http://adsabs.harvard.edu/abs/2013ApJ...762...26Y} {762, 26}

\bibitem[\protect\citeauthoryear{{York} et~al.,}{{York}
  et~al.}{2000}]{York2000}
{York} D.~G.,  et~al., 2000, \mn@doi [\aj] {10.1086/301513}, \href
  {https://ui.adsabs.harvard.edu/abs/2000AJ....120.1579Y} {120, 1579}

\bibitem[\protect\citeauthoryear{{Youakim} et~al.,}{{Youakim}
  et~al.}{2017}]{Youakim2017}
{Youakim} K.,  et~al., 2017, \mn@doi [MNRAS] {10.1093/mnras/stx2005}, \href
  {http://adsabs.harvard.edu/abs/2017MNRAS.472.2963Y} {472, 2963}

\bibitem[\protect\citeauthoryear{{Yuan}, {Chang}, {Beers}  \& {Huang}}{{Yuan}
  et~al.}{2020}]{Yuan2020}
{Yuan} Z.,  {Chang} J.,  {Beers} T.~C.,   {Huang} Y.,  2020, \mn@doi [\apjl]
  {10.3847/2041-8213/aba49f}, \href
  {https://ui.adsabs.harvard.edu/abs/2020ApJ...898L..37Y} {898, L37}

\bibitem[\protect\citeauthoryear{{de Jong} et~al.,}{{de Jong}
  et~al.}{2019}]{DeJong2019}
{de Jong} R.~S.,  et~al., 2019, \mn@doi [The Messenger]
  {10.18727/0722-6691/5117}, \href
  {https://ui.adsabs.harvard.edu/abs/2019Msngr.175....3D} {175, 3}

\makeatother
\end{thebibliography}



\bsp	
\label{lastpage}
\end{document}